\RequirePackage{letltxmacro}
\LetLtxMacro{\LaTeXtextbf}{\textbf}
\documentclass{ieeeaccess}
\LetLtxMacro{\textbf}{\LaTeXtextbf}

\usepackage{cite}
\usepackage{amsmath,amssymb,amsfonts}
\usepackage[ruled,linesnumbered]{algorithm2e}
\usepackage{algpseudocode}

\algdef{SE}[DOWHILE]{Do}{doWhile}{\algorithmicdo}[1]{\algorithmicwhile\ #1}%
\algnewcommand\algorithmicinput{\textbf{Input:}}
\algnewcommand\algorithmicoutput{\textbf{Output:}}
\algnewcommand\Input{\item[\algorithmicinput]}
\algnewcommand\Output{\item[\algorithmicoutput]}
\usepackage{graphicx}
\usepackage{textcomp}
\usepackage{caption,setspace}
\usepackage[labelsep=period]{caption}
\usepackage{longtable}
\usepackage[mathscr]{eucal}
\usepackage[normal]{subfigure}
\usepackage{comment}
\usepackage{paralist}
\usepackage{boxedminipage}
\usepackage{array}
\usepackage{multirow}
\usepackage{float}
\usepackage{color}
\usepackage{soul}
\usepackage{mwe}
\usepackage{stackengine}
\usepackage{mathtools}
\usepackage{bm}
\def\BibTeX{{\rm B\kern-.05em{\sc i\kern-.025em b}\kern-.08em
    T\kern-.1667em\lower.7ex\hbox{E}\kern-.125emX}}

\DeclareMathOperator{\arctantwo}{arctan2}

\begin{document}
\history{Date of publication xxxx 00, 0000, date of current version xxxx 00, 0000.}
\doi{10.1109/ACCESS.2017.DOI}

\title{Design and Implementation of 5.8 GHz RF Wireless Power Transfer System}
\author{\uppercase{Je Hyeon Park},
\IEEEmembership{Student Member, IEEE},
\uppercase{Nguyen Minh Tran},
\IEEEmembership{Student Member, IEEE}
\uppercase{Sa Il Hwang},
\uppercase{Dong In Kim},
\IEEEmembership{Fellow, IEEE}, AND
\uppercase{Kae Won Choi},
\IEEEmembership{Senior Member, IEEE}}
\address{Department of Electrical and Computer Engineering, College of Information and Communication Engineering, Sungkyunkwan University, Suwon 16419, South Korea}

\tfootnote{This research was partly supported by the MSIT(Ministry of Science and ICT), Korea, under the ICT Creative Consilience program (IITP-2021-2020-0-01821) supervised by the IITP(Institute for Information {\&} communications Technology Planning {\&} Evaluation), and partly supported by the MSIT(Ministry of Science and ICT), Korea, under the ITRC(Information Technology Research Center) support program(IITP-2021-0-02046) supervised by the IITP(Institute of Information {\&} Communications Technology Planning {\&} Evaluation)}

\markboth
{J. H. Park, \emph{et al.}: Design and Implementation of 5.8 GHz RF Wireless Power Transfer System}
{J. H. Park, \emph{et al.}: Design and Implementation of 5.8 GHz RF Wireless Power Transfer System}

\corresp{Corresponding author: Kae Won Choi (kaewonchoi@skku.edu)}

\begin{abstract}
In this paper, we present a 5.8 GHz radio-frequency (RF) wireless power transfer (WPT) system that consists of 64 transmit antennas and 16 receive antennas. Unlike the inductive or resonant coupling-based near-field WPT, RF WPT has a great advantage in powering low-power internet of things (IoT) devices with its capability of long-range wireless power transfer. We also propose a beam scanning algorithm that can effectively transfer the power no matter whether the receiver is located in the radiative near-field zone or far-field zone. The proposed beam scanning algorithm is verified with a real-life WPT testbed implemented by ourselves. By experiments, we confirm that the implemented 5.8 GHz RF WPT system is able to transfer 3.67 mW at a distance of 25 meters with the proposed beam scanning algorithm. Moreover, the results show that the proposed algorithm can effectively cover radiative near-field region differently from the conventional scanning schemes which are designed under the assumption of the far-field WPT.
\end{abstract}


\titlepgskip=-15pt

\maketitle

\section{Introduction}\label{section:introduction}
\PARstart{T}{he} Internet of Things (IoT) has been regarded as a representative technology of the next industrial revolution in terms of hyper-connected society. According to the forecast from Transform Insights TAM Forecast Database, at the end of 2019, there were 7.6 billion active IoT devices, which is expected to grow to 24.1 billion in 2030. Along the way, with the upcoming industrial wave, supplying electrical power to a tremendous number of IoT devices will become a great challenge. Conventional ways to supply power to devices (e.g., connecting power cords or periodically replacing batteries) are a huge loss in various aspects not only for the costs and efforts but also for the quality of service when it comes to large-scale IoT connectivity.

Over decades, there have been various approaches for wirelessly charging electrical devices. Near-field wireless power transfer (WPT) based on resonant or inductive coupling methods has shown great technical progress and even introduced commercialized products which are able to charge mobile devices\cite{Hui:2014}. However, despite the capability of high-efficiency power transfer of near-field WPT, the application area is very restricted due to the short charging range. Although the wire is not connected to the target device directly, it still demands the devices be located near the power source, even on a specific spot. 

On the other hand, radio frequency (RF) WPT, based on the electromagnetic (EM) wave, is capable of transferring wireless power enough to operate IoT devices up to hundreds of meters \cite{Clerckx:2018}. Moreover, RF WPT has a powerful advantage in that it can be extended to simultaneous wireless information and power transfer (SWIPT) \cite{Choi:2020}. However, the required power for operating an IoT device is around 1 mW \cite{Huang:2015}, which is not an easy performance target to achieve for the RF WPT system. It is because most of the radiated RF power dissipates into the air, and then only a small fraction of the transmitted EM wave can be captured at the receiver due to the dispersive nature of free-space EM wave. This is the reason  why the foremost obstacle of RF WPT is low power transfer efficiency.

High-efficiency RF WPT can be achieved by synthesizing the beam of an EM wave and steering it to the desired direction, which means focusing wireless power on the receiver \cite{Shinohara:2021}. Moreover, to maximize available energy with received power, many studies have been conducted to attain high RF-to-DC conversion efficiency of rectifiers \cite{khan:2020}.

To make RF WPT technology much more reliable for real-life applications, we have to focus on a systematic and integrated view of the overall system.
A number of papers provide details of practical implementation of RF WPT together with some associated theories on beamforming and protocol design, such as \cite{Setiawan:2017, Shen:2020, Kim:2017, Clerckx:2018_2, Hui:2019, Choi:2019, Koo:2020, Bae:2020, Pabbisetty:2019, Arai:2021,  Belo:2019, Gowda:2016, Xianjin:2019}. They experimentally verified their own proposed beamforming scheme with a prototype testbed setup. Especially, the authors of \cite{Hui:2019, Choi:2019, Koo:2020,  Bae:2020, Pabbisetty:2019, Arai:2021, Belo:2019, Gowda:2016, Xianjin:2019} have built the RF WPT prototype system of their own. 

The authors of \cite{Hui:2019} and \cite{Choi:2019} have fabricated a transmitter module consisting of 64 phased array antennas that operate at around 920 MHz. In \cite{Hui:2019}, the maximum output power of transmitter reaches 100 W, and the power transfer efficiency at 15 meters is roughly calculated to be 10 {\%}. And \cite{Choi:2019}, which is our previous work, demonstrated the outdoor test results up to 50 meters and verified the sensor device is successfully kept alive with only the received power. At the distance of 50 meters, the sensor node received 1 mW with total transmitted power of 11 W.

Recently, the research works on the RF WPT system operating at the frequency from 5 GHz to 6 GHz have gained great momentum (e.g., \cite{Koo:2020, Bae:2020, Pabbisetty:2019, Belo:2019, Arai:2021}). These works introduce their own WPT prototype based on a phased array antenna transmitter with different operation frequencies from 5.2 GHz to 5.8 GHz, respectively. Both  \cite{Koo:2020} and \cite{Bae:2020} use the same hardware architecture of the RF WPT system, which operates at 5.2 GHz. While \cite{Koo:2020} used 8-by-8 transmitter elements, \cite{Bae:2020} only used 4-by-8 elements. The receiver consists of one antenna to report the received RF power and five rectennas to harvest energy. The received dc power of 100 mW was reported from the total transmitted power of 32 W at a distance of 4 meters in \cite{Koo:2020}. In comparison, received dc power was 191.1 mW at a distance of 1 meter with 16 W of transmitted power in \cite{Bae:2020}.

In \cite{Pabbisetty:2019}, the authors have built an RF WPT system at a frequency of 5.745 GHz. They proposed a time-sharing beamforming algorithm and experimentally verified it. Four nodes located at 2 meters apart from the transmitter with different azimuthal angles have received the power of 23 mW, 23.38 mW, 31.4 mW, and 27.92 mW with equivalent isotropically radiated power (EIRP) of 10 kW. The work \cite{Arai:2021} proposed a tile-based 8-by-8 triangular grid array transmitter for a 5.7 GHz RF WPT system. The proposed transmitter achieves up to 68.6 dBm EIRP at 5.75 GHz. 

There have been several works about the 5.8 GHz RF WPT system demonstration (e.g., \cite{Belo:2019, Gowda:2016, Xianjin:2019}), which has the same target frequency as the proposed system in this paper.
In \cite{Belo:2019}, the proposed RF WPT system consists of 4-by-4 transmit antenna elements and 4-by-4 receive antenna elements. At a distance of 0.5 meters, around 7 mW was reported at the receiver with 1.3 W of transmit power.
The authors of \cite{Gowda:2016} and \cite{Xianjin:2019} have built the transmitter, which is capable of controlling the phase of each element by using different lengths of connected transmission lines at 5.8 GHz frequency band. They both experimentally verified a focused antenna array method with their own proposed system. The work  \cite{Gowda:2016} used 8-by-8 transmit antenna array, and 4-by-4 receive antenna array. The received power at a distance of 0.4 meters was reported as  33.2 mW with a total transmitted power of 100 mW. In \cite{Xianjin:2019}, the authors used a very large-sized (1 m $\times$ 1 m) transmitter and receiver. With 500 W of transmitted power, 209.26 W of RF power was reported at the receiver at a distance of 10 meters.

In this paper, we have designed and implemented a full-fledged 5.8 GHz RF WPT system consisting of an 8-by-8 transmit antenna array and a 4-by-4 receive rectenna array. Overall hardware components in this paper have been developed with the aim to make the RF WPT system compact and simple. The transmitter is comprised of a phase control board, amplifying board, and antenna board. By stacking up three boards with connectors as a sandwich structure, we have a single module of phased antenna array with 16 RF paths since each board has 4-by-4 elements. With this configuration, it is easy to increase the number of elements for the transmitter, and we have combined 4 modules to implement 8-by-8 transmit antenna elements. The receiver has 16 rectifiers in a square form, and each rectifier is designed with a one-stage Dickson charge pump structure. The rectifier is directly connected to the antenna element, and we will call this structure a rectenna. The harvested DC energy from each rectenna is combined  in parallel.

We also propose an elaborated beam scanning method that is able to cover the far-field and near-field region over two scanning iterations.
In the phased array-based wireless communications systems, the codebook-based beam scanning schemes have been studied in many ways in terms of pencil beam, wide beam, and arbitrarily shaped beam (e.g., \cite{Liu:2018, Sun:2018, Aslan:2019, Morabito:2012, Yoon:2021}). However, most beam synthesis methods are proposed based on the far-field assumption since, in communications systems, the distance between the base station and user equipment is actually very far. On the other hand, in the RF WPT system, the far-field assumption does not hold anymore. 
For example, the primary potential application scenario of the RF WPT system is a smart home. In this scenario, it is not assured that IoT devices are located in the far-field all the time. Besides, for operating an IoT device, the RF WPT system demands much higher transfer efficiency because the required power is around 0 dBm, whereas the data communication is possible unless the received power is lower than the noise floor (i.e., $-120$ dBm). So the beam scanning algorithm should work within the radiative near-field region to achieve a reasonable power transfer efficiency.

The proposed beam scanning method, in this paper, divides the whole far-field region into a specific size of the grid for the first scanning phase. Each point of considered u-v coordinate represents a certain direction of the beam from the transmitter. 
In the first scanning phase, the transmitter scans the whole far-field region with a given size and the number of the grid points. Based on the measured power at the receiver according to the first scanning, the transmitter determines the optimal direction. 
After that, the second scanning phase is conducted to cover radiative near-field zone.
In the second scanning phase, the transmitter synthesizes the beams correspond to the different distances, which are decided with a given step size up to the end of the radiative near-field region towards the obtained direction from the first scanning phase.

The rest of the paper is organized as follows. First, we present the proposed beam scanning method in Sections \ref{section:system model} and  \ref{section:beam scanning}. A detailed description of the implemented RF WPT system is provided in \ref{section:system}. Testbed setup and experimental results are presented in Section \ref{section:Experiments}, and the paper is concluded in Section \ref{section:Conclusion}. 

\section{System Model}\label{section:system model}

\subsection{Coordinate system}\label{section:coordinate system}

\begin{figure}
\centering
        \includegraphics[width=0.8\linewidth] {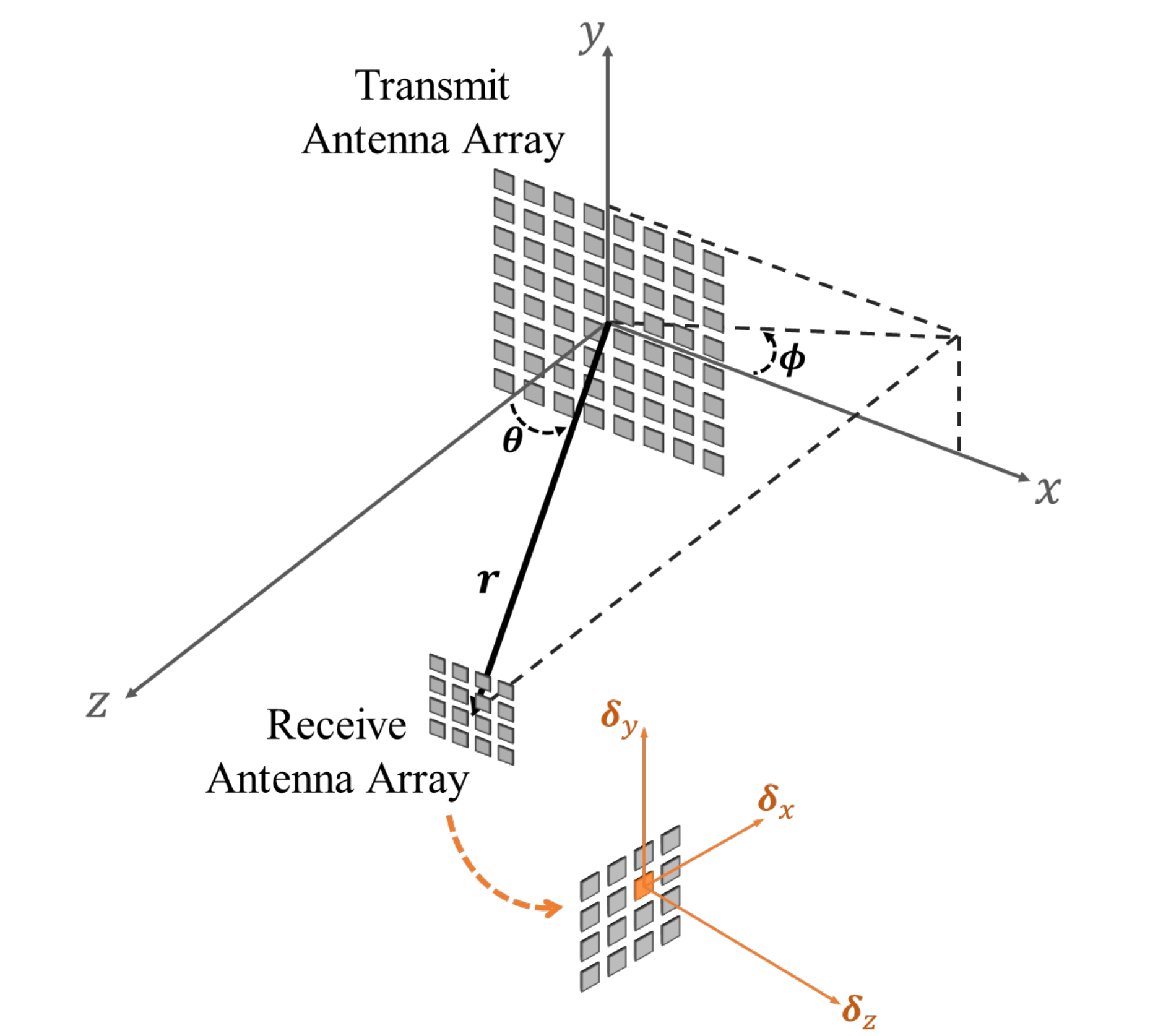}
        \caption{Cartesian and spherical coordinate system model}
        \label{fig:coordinate}
\end{figure}

Fig. \ref{fig:coordinate} shows RF WPT system in consideration, which consists of a phased antenna array transmitter and rectenna array receiver. The transmitter is located on the x-y plane in the Cartesian coordinate system, and the EM wave is radiated towards the positive z-axis. 
The transmitter is a rectangular planar antenna array with $N^\text{Tx}_\text{row}\times N^\text{Tx}_\text{col}$ antenna elements. Let $N^\text{Tx}_\text{row}$ and $N^\text{Tx}_\text{col}$ denote antenna elements along x-axis and y-axis, also we denote the total number of antenna elements as $N^\text{Tx}$ ($=N^\text{Tx}_\text{row} \times N^\text{Tx}_\text{col} $) . Each antenna element is indexed by $n$ ($=1,\ldots,N^\text{Tx}$), and then $n$ th antenna element is denoted by antenna $n$.

In the global Cartesian coordinate system (GCS) with the x, y, and z axes, the position of transmit antenna $n$ is denoted by 
\begin{align}
{\mathbf a}^\text{GCS}_n = (a^x_n,a^y_n,a^z_n)^T.
\end{align}
And the distances between two neighboring rows and columns of antennas, which is called antenna spacing, are denoted by $l^\text{Tx}_\text{row}$ and $l^\text{Tx}_\text{col}$, respectively. Then, the position of antenna $n$ is given by 
\begin{align}
\begin{split}
&a^x_n = l^\text{Tx}_\text{col}  \cdot\left(\left({\frac{N^\text{Tx}_\text{col}+1}{2}}\right)-\tau^\text{Tx}_\text{col}\right),\\
&a^y_n = l^\text{Tx}_\text{row}\cdot\left(\tau^\text{Tx}_\text{row}- \left({\frac{N^\text{Tx}_\text{row}+1}{2}}\right)\right),\\
&a^z_n = 0.
\end{split}
\end{align}
where $\tau^\text{Tx}_\text{row}$ and $\tau^\text{Tx}_\text{col}$ are, respectively, row and column indices of antenna $n$ such that 
\begin{align}
\begin{split}
\tau^\text{Tx}_\text{row}&=\lfloor(n-1)/N^\text{Tx}_\text{row}\rfloor+1,\\
\tau^\text{Tx}_\text{col}&=((n-1)\,\bmod\,N^\text{Tx}_\text{col}) +1.
\end{split}
\end{align}
Similarly, the receive antenna array can be defined as follows: the receiver is composed of $M^\text{Rx}_\text{col}\times M^\text{Rx}_\text{row}$ array of antenna elements with the spacing between neighboring antenna array elements denoted by $l^\text{Rx}_\text{col}$ and $l^\text{Rx}_\text{row}$. The total number of antenna elements is denoted as $M^\text{Rx} (= M^\text{Rx}_\text{col}\times M^\text{Rx}_\text{row})$. And the position of the receive antenna $m$ in GCS is denoted by
\begin{align}
{\bold s}^\text{GCS}_m = (s^x_m,s^y_m,s^z_m)^T.
\end{align}

In our system, the position and attitude of the transmit antenna array are fixed on the x-y plane with the center of the array at the origin (i.e., $(0,\ 0,\ 0)$) of the GCS. However, the receiver can be freely located in a GCS with arbitrary attitudes. 
To describe the location of the receive antenna, including its attitude, we introduce the receiver's local coordinate system (LCS), which is represented by three orthogonal axes $\delta_x$, $\delta_y$, and $\delta_z$. The origin of the receiver's LCS is ${\bold s}^\text{LCS}_{\text{o}} = (0, 0, 0)^T$ and the position of receive antenna $m$ in LCS is denoted by
\begin{align}
{\bold{s}}^\text{LCS}_m = (s^{\delta_x}_m,s^{\delta_y}_m,s^{\delta_z}_m)^T.
\end{align}

In the viewpoint of GCS, LCS can be seen as the rotation matrix that rotates the receiver with their own attitude, whose axes are aligned with the x, y, and z axes at first.
We represent the rotation of LCS by the Euler angle. The rotation matrices that represent the rotation of $\alpha$, $\beta$, and $\gamma$ around x, y, and z axes of GCS are, respectively, given by
\begin{align}\label{eq:euler}
\begin{split}
\centering
{\bold R}_x(\alpha)&=
\left[\begin{smallmatrix}
1 & 0 & 0 \\
0 & \cos\alpha & -\sin\alpha \\
0 & \sin\alpha & \cos\alpha
\end{smallmatrix}\right],\
{\bold R}_y(\beta)=
\left[\begin{smallmatrix}
\cos\beta & 0 & \sin\beta \\
0 & 1 & 0 \\
-\sin\beta & 0 & \cos\beta
\end{smallmatrix}\right],\\
{\bold R}_z(\gamma)&=
\left[\begin{smallmatrix}
\cos\gamma & -\sin\gamma & 0 \\
\sin\gamma & \cos\gamma & 0 \\
0 & 0 & 1
\end{smallmatrix}\right].
\end{split}
\end{align}
Then the rotation matrix of receive antenna ${\bold R}_\text{Rx}$ which is rotated by an Euler angle $(\alpha, \beta, \gamma)$ with the x-y'-z'' intrinsic rotation is given by 
\begin{align}
{\bold R}_\text{Rx} = {\bold R}_x(\alpha){\bold R}_y(\beta){\bold R}_z(\gamma).
\end{align}
The three axes of the LCS of the receiver (i.e., $\delta_x$, $\delta_y$, and $\delta_z$) correspond to the columns of rotation matrix ${\bold R}_\text{Rx}$.
If we present the origin of the receiver's LCS in GCS as
\begin{align}\label{eq:originLCS}
{\bold{s}}^\text{GCS}_{\text{o}} = (s^x_{\text{o}},s^y_{\text{o}},s^z_{\text{o}})^T,
\end{align}
we can fully describe the position and attitude of the receive antenna element $m$ in GCS as
\begin{align}\label{eq:globalrx_o}
&{\bold s}^\text{GCS}_m ={\bold s}^\text{GCS}_{\text{o}} + {\bold R}_\text{Rx} {\bold s}^\text{LCS}_m .
\end{align}
The position of transmit antenna $n$ ${\bold a}_n^\text{Tx,LCS}$ in LCS is the same as ${\bold a}^\text{Tx,GCS}_n$ since the center of transmit antenna array is located at the origin of GCS.

In this paper, we designate one antenna element $m^*$, which is located at the center of the receive antenna array, as an anchor antenna. We set the position of the anchor antenna in LCS as the origin such that
\begin{align}\label{eq:o_LCS}
{\bold{s}}^\text{LCS}_{m^*} = {\bold{s}}^\text{LCS}_{\text{o}} = (0,\ 0,\ 0)^T,
\end{align}
and the position of the anchor antenna in GCS is denoted as
\begin{align}\label{eq:s}
{\bold s}^\text{GCS}_{m^*} = (s^x_{m^*},s^y_{m^*},s^z_{m^*})^T.
\end{align}
From \eqref{eq:o_LCS} and \eqref{eq:s}, \eqref{eq:globalrx_o} is rewritten as
\begin{align}\label{eq:globalrx}
&{\bold s}^\text{GCS}_m ={\bold s}^\text{GCS}_{m^*} + {\bold R}_\text{Rx} {\bold s}^\text{LCS}_m .
\end{align}

The position of ${\bold s}^\text{GCS}_{m^*}$ can also be denoted in the spherical coordinate system as
${\bold s}^\text{GCS}_{m^*} = (r^\text{Rx},\theta^\text{Rx},\phi^\text{Rx})^T$ where $r^\text{Rx}$, $\theta^\text{Rx}$, and $\phi^\text{Rx}$ denote the radius, elevation, and azimuth of the anchor antenna $m^*$.
Consequently, we can derive the position of antenna $m$ in the Cartesian coordinate system and spherical coordinate system both in terms of $r^\text{Rx}$, $\theta^\text{Rx}$, and $\phi^\text{Rx}$ by
\begin{align}
&s^x_{m^*} = r^\text{Rx}\sin\theta^\text{Rx}\cos\phi^\text{Rx},\label{eq:sx}\\
&s^y_{m^*} = r^\text{Rx}\sin\theta^\text{Rx}\sin\phi^\text{Rx},\label{eq:sy}\\
&s^z_{m^*} = r^\text{Rx}\cos\theta^\text{Rx}.\label{eq:sz}
\end{align}

\subsection{Near-field EM wave propagation model}\label{section:propagation model}
In this subsection, we model the EM wave propagation between transmit antenna array and receive antenna array in the radiative near field.
The frequency of the EM wave is denoted by $f$, and the free-space wavelength of the EM wave is denoted by $\lambda$. The EM wave from the transmitter is radiated towards the positive direction along the z-axis. The direction from the transmitter to the receiver is defined as elevation $\theta$ and azimuth $\phi$. We assume that $ -\pi/2 \le \theta \le \pi/2 $ , $-\pi \le \phi \le \pi $  which means the receiver is located  within the range of the EM wave radiation from the transmitter. 

In our system, only the phase can be controlled with a phase shifter for hardware simplicity, and hence each element of transmit antenna array radiates equal power $\rho^\text{Tx}$.
The transmitted power wave at the port of antenna $n$, denoted by $x_n$, is defined as 
\begin{align}\label{eq:x}
x_n =\sqrt{2\rho^\text{Tx}}\exp(-j\omega_n), 
\end{align}
where $\omega_n$ is the phase of $x_n$. 
Here we define the transmitted power wave vector as
\begin{align}\label{eq:xvec}
{\bold x} = (x_{n})_{n=1,\ldots,N^\text{Tx}} .
\end{align}
If we denote $y_m$ as the received power wave from the receive antenna $m$, $y_m$ is given by
\begin{align}\label{eq:rxp}
\begin{split}
&y_m = \sum_{n=1}^{N^\text{Tx}} h_{n,m}x_{n}.
\end{split}
\end{align}
where $h_{n,m}$ is the channel gain from transmit antenna $n$ to receive antenna $m$. 

In the free space, the channel gain $h$ between the transmit and receive antennas is given by
\begin{align}\label{eq:h}
h = \frac{\lambda}{4\pi d}\sqrt{G^\text{Tx} G^\text{Rx}}\exp\bigg(-j\frac{2\pi}{\lambda}d\bigg),
\end{align}
where $d$ is the distance between two antennas, $G^\text{Tx}$ is the gain of transmit antenna, and $G^\text{Rx}$ is the gain of the receive antenna. Based on \eqref{eq:h}, we can derive the channel gain between transmit antenna $n$ to receive antenna $m$.

If we consider LCS and GCS in \eqref{eq:globalrx}, the distance between antenna $n$ of the transmitter and antenna $m$ of the receiver, which is denoted by $d_{n,m}$, is represented by 
\begin{align}\label{eq:distnm}
\begin{split}
&d_{n,m} = \|{\mathbf a}^\text{GCS}_n - ({\bold s}^\text{GCS}_{m^*} + {\bold R}_\text{Rx} {\bold s}^\text{LCS}_m)\|\\
&= \sqrt{\|{{\bold s}^\text{GCS}_{m^*}}\|^2 + 2(-{\bold s}^\text{GCS}_{m^*})^T\bm{\kappa}_{n,m} + \|\bm{\kappa}_{n,m}\|^2} ,
\end{split}
\end{align}
where
\begin{align}
\bm{\kappa}_{n,m} &= {\mathbf a}^\text{GCS}_n - {\bold R}_\text{Rx}{\bold s}^\text{LCS}_m.
\end{align}
By using the first-order approximation with the Taylor expansion $\sqrt{x^2+y} \simeq x+\frac{1}{2x}y$, \eqref{eq:distnm} can be rewritten as 
\begin{align}\label{eq:distgcs}
\begin{split}
d_{n,m} &= \|{\bold s}^\text{GCS}_{m^*}\| - \frac{({\bold s}^\text{GCS}_{m^*})^T}{\|{\bold s}^\text{GCS}_{m^*}\|}{\bm{\kappa}_{n,m}} + \frac{\|\bm{\kappa}_{n,m}\|^2}{2\|{\bold s}^\text{GCS}_{m^*}\|}.\\
\end{split}
\end{align}

Based on \eqref{eq:h} and \eqref{eq:distgcs}, the channel gain $h_{n,m}$ between transmit antenna $n$ and receive antenna $m$ is given by
\begin{align}\label{eq:hnm}
\begin{split}
h_{n,m} &= \frac{\lambda}{4\pi d_{n,m}}\sqrt{G^\text{Tx} G^\text{Rx}}\exp\bigg(-j\frac{2\pi}{\lambda}d_{n,m}\bigg)\\
&= \frac{\lambda\sqrt{G^\text{Tx} G^\text{Rx}}}{4\pi d_{n,m}}
\exp\bigg(-j\frac{2\pi}{\lambda} \|{\bold s}^\text{GCS}_{m^*}\| \bigg)\\
&\quad\times\exp\bigg(j\frac{2\pi}{\lambda}\frac{({\bold s}^\text{GCS}_{m^*})^T}{\|{\bold s}^\text{GCS}_{m^*}\|}{\bm{\kappa}_{n,m}}\bigg)\\
&\quad\times\exp\bigg(-j\frac{2\pi}{\lambda}\frac{\|{\bm{\kappa}_{n,m}}\|^2}{2\|{\bold s}^\text{GCS}_{m^*}\|}\bigg).
\end{split}
\end{align}
In \eqref{eq:hnm}, the magnitude of ${\bold s}^\text{GCS}_{m^*}$ is equal to the distance between the center point of the transmit antenna array and anchor antenna $m^*$, which is denoted as below
\begin{align}\label{eq:r}
r^\text{Rx} = \|{\bold s}^\text{GCS}_{m^*}\| = \sqrt{(s^x_{m^*})^2 + (s^y_{m^*})^2 + (s^z_{m^*})^2}.
\end{align}
Furthermore, in \eqref{eq:hnm}, ${({\bold s}^\text{GCS}_{m^*})^T} / {\|{\bold s}^\text{GCS}_{m^*}\|}$ is interpreted as the unit direction vector from the center of the transmit antenna array to the anchor antenna of the receive antenna array, which is denoted by
\begin{align}\label{eq:unitdr}
\frac{({\bold s}^\text{GCS}_{m^*})^T}{\|{\bold s}^\text{GCS}_{m^*}\|}
= (\sin\theta^\text{Rx}\cos\phi^\text{Rx}, \sin\theta^\text{Rx}\sin\phi^\text{Rx}, \cos\theta^\text{Rx})^T.
\end{align}

Based on \eqref{eq:r} and \eqref{eq:unitdr}, we can simplify \eqref{eq:hnm} as 
\begin{align}\label{eq:hnm2}
\begin{split}
h_{n,m} = \Phi (r^\text{Rx}) \Omega_{n,m} (\theta^\text{Rx}, \phi^\text{Rx}) \Lambda_{n,m} (r^\text{Rx}).
\end{split}
\end{align}
In \eqref{eq:hnm2}, $\Phi (r^\text{Rx})$ represents the magnitude and phase rotation of the channel gain related to the distance between the center of the transmit antenna array and the anchor antenna of the receiver, which is defined as 
\begin{align}\label{eq:Phi}
\Phi (r^\text{Rx}) = \frac{\lambda\sqrt{G^\text{Tx} G^\text{Rx}}}{4\pi r^\text{Rx}}
\exp\bigg(-j\frac{2\pi}{\lambda} r^\text{Rx} \bigg),
\end{align}
where $d_{n,m}$ is approximated to $r^\text{Rx}$. In \eqref{eq:hnm2}, $\Phi (r^\text{Rx})$ has the same value for all channel gains (i.e., $(h_{n,m})_{n=1,\ldots,N^\text{Tx} \atop m=1,\ldots,M^\text{Rx}}$) regardless of the indices of transmit and receive antennas.

In addition, in \eqref{eq:hnm2}, $\Omega_{n,m} (\theta^\text{Rx}, \phi^\text{Rx})$ represents the phase rotation in the far-field region caused by the direction from the transmitter to the anchor antenna $m^*$, which is defined as
\begin{align}\label{eq:Omega}
\begin{split}
&\Omega_{n,m} (\theta^\text{Rx}, \phi^\text{Rx}) = \\
& \exp\bigg(j\frac{2\pi}{\lambda} (\sin\theta^\text{Rx}\cos\phi^\text{Rx}, \sin\theta^\text{Rx}\sin\phi^\text{Rx}, \cos\theta^\text{Rx})^T {\bm{\kappa}_{n,m}} \bigg).
\end{split}
\end{align}

Furthermore, in \eqref{eq:hnm2}, $\Lambda_{n,m} (r^\text{Rx})$ is the near-field region related term that denotes the phase rotation of $h_{n,m}$ related to the distance $r^{\text{Rx}}$, which is denoted by
\begin{align}\label{eq:Lambda}
\Lambda_{n,m} (r^\text{Rx}) =
\exp\bigg(-j\frac{2\pi}{\lambda}\frac{\|{\bm{\kappa}_{n,m}}\|^2}{2r^\text{Rx}}\bigg).
\end{align}
Most of the research works on the beam scanning algorithm have not considered $\Lambda_{n,m} (r^\text{Rx})$
since under the far-field assumption, $\Lambda_{n,m} (r^\text{Rx})$ becomes negligibly small.
However, without $\Lambda_{n,m} (r^\text{Rx})$, the transmitter is not able to focus the EM wave beam on to a small spot within the near-field region. The convex lens-like phase distribution of the transmit antenna array can be formed by $\Lambda_{n,m} (r^\text{Rx})$. We have addressed $\Lambda_{n,m} (r^\text{Rx})$ to propose the elaborated beam scanning algorithm which is capable of covering the near-field region. 

In this paper, since the single element of the receiver is made up of a rectenna, the received RF power at the antenna is directly converted to DC power. The converted DC power from each element is combined together at the end of the rectifier. With a slight abuse of definition, we call this technique as DC combining method to represent the total DC power from multiple rectenna elements.
We define $\epsilon_m$ as the RF-to-DC conversion efficiency of the rectifier which is connected to receive antenna $m$.
If $y_{m}$ denotes the received power wave from antenna array $m$ of the receiver, the total received DC power $p^\text{Rx}_\text{DC}$ is given by
\begin{align}\label{eq:totpow}
p^\text{Rx}_\text{DC} = \sum_{m=1}^{M^\text{Rx}}  \epsilon_m \frac{|y_{m}|^2}{2}.
\end{align}

\section{Beam scanning method}\label{section:beam scanning}
In this section, we introduce the beam scanning method for maximizing the received power with the proposed RF WPT system.
We have described the coordinate system model and EM wave propagation model in the previous section. However, when it comes to radiative RF WPT, derivation of EM wave propagation between antenna elements is not very practical. In order to achieve high transfer efficiency, a phased array transmitter forms the power beam to the receiver by controlling antenna weights (i.e., magnitude and phase).

We use the anchor antenna $m^*$ as a sensor antenna which measures the received power for each scanning beam. 
In this paper, we prefer to use u-v coordinates to represent the direction. With $\theta$ and $\phi$, we can derive the direction control parameter $\boldsymbol \xi$ as u-v coordinates of $\theta$ and $\phi$ as
\begin{align}\label{eq:uv}
{\boldsymbol \xi} = (u,v)^T = (\sin\theta\cos\phi, \sin\theta\sin\phi)^T. 
\end{align}
Based on \eqref{eq:uv}, we define the position of the anchor antenna $m^*$ as 
$({\boldsymbol \xi}^\text{Rx},r^\text{Rx})$.

From \eqref{eq:o_LCS}, \eqref{eq:hnm}, and \eqref{eq:hnm2}, the channel gain from transmit antenna $n$ to the anchor antenna $m^*$ is derived as
\begin{align}\label{eq:hnmfn}
\begin{split}
&h_{n,m^*} = \frac{\lambda}{4\pi d_{n,m^*}}\sqrt{G^\text{Tx} G^\text{Rx}}\exp\bigg(-j\frac{2\pi}{\lambda}d_{n,m^*}\bigg)\\
&= \Phi (r^\text{Rx})\exp\bigg(j\frac{2\pi}{\lambda} (\boldsymbol \xi^\text{Rx})^T {\bold u}^\text{Tx}_{n}\bigg)\exp\bigg(-j\frac{2\pi}{\lambda}\frac{\|{\bold a}^\text{Tx}_n\|^2}{2r^\text{Rx}}\bigg),
\end{split}
\end{align}
where ${\bold u}^\text{Tx}_{n} = (a^x_n,a^y_n)$ is the position of antenna $n$ on the x-y plane. 

From \eqref{eq:rxp} and \eqref{eq:hnmfn}, the received power wave at sensor antenna $m^*$ is denoted by
\begin{align}\label{eq:ym}
\begin{split}
&y_{m^*} = \sum_{n=1}^{N^\text{Tx}} h_{n,m^*}x_{n}\\
&= \sum_{n=1}^{N^\text{Tx}} \sqrt{2\rho^\text{Tx}}\Phi (r^\text{Rx}) \exp\bigg(j\Big(\big(\zeta_n({\boldsymbol \xi}^\text{Rx}) + \eta_n(r^\text{Rx})\big) - \omega_n \Big)\bigg),
\end{split}
\end{align}
where
\begin{align}
 \zeta_n({\boldsymbol \xi}^\text{Rx}) &= \frac{2\pi}{\lambda} ({\boldsymbol \xi}^\text{Rx})^T {\bold u}^\text{Tx}_{n},\\
\eta_n(r^\text{Rx}) &= -\frac{2\pi}{\lambda} \frac{\|{\bold a}^\text{Tx}_n\|^2}{2r^\text{Rx}}.
\end{align}
The basic concept of beamforming for radiative RF WPT is  pursuing transmitted power waves from each antenna element to be combined in-phase at the receive antenna.
The received power at the sensor antenna is maximized when the phases $\big(\big(\zeta_n({\boldsymbol \xi}^\text{Rx}) + \eta_n(r^\text{Rx})\big) - \omega_n\big)$ are aligned for all $n = 1,\ldots,N^\text{Tx}$. Then the optimal power wave of antenna $n$ is
\begin{align}\label{eq:txopt}
x_n^\text{opt} =\sqrt{2\rho^\text{Tx}} \exp(-j\omega^*_n),
\end{align}
where
\begin{align}\label{eq:optphase}
\begin{split}
\omega^*_n &= \zeta_n({\boldsymbol \xi}^\text{Rx}) + \eta_n(r^\text{Rx})\\
&= \frac{2\pi}{\lambda} \bigg( ({\boldsymbol \xi}^\text{Rx})^T{\bold u}^\text{Tx}_{n} - \frac{\|{\bold a}^\text{Tx}_n\|^2}{2r^\text{Rx}} \bigg).
\end{split}
\end{align}
Based on \eqref{eq:optphase}, we can see the optimal phase for each antenna element is determined by ${\boldsymbol \xi}^\text{Rx}$, which represents the direction of the receiver and the distance between transmitter and receiver $r^\text{Rx}$. 
In the far-field region, $\eta_n(r^\text{Rx})$ is approximated to zero since the distance $r^\text{Rx}$ is very large. However, the WPT system should work within the radiative near-field region to guarantee a reasonable power transfer efficiency and in order to do that, $\eta_n(r^\text{Rx})$ should be addressed to cover the radiative near-field region.

From \eqref{eq:ym}, the received power at the sensor antenna $P_{m^*}$ can be defined as the function of the direction and the distance from the transmitter as
\begin{align}\label{eq:senpow}
\begin{split}
&P_{m^*} = \frac{|y_{m^*}|^2}{2}\\
&=\frac{\bigg|\sum_{n=1}^{N^\text{Tx}} I(r^\text{Rx}) \exp\bigg(j\Big(\big(\zeta_n({\boldsymbol \xi}^\text{Rx}) + \eta_n(r^\text{Rx})\big) - \omega_n \Big)\bigg)\bigg|^2}
{2},
\end{split}
\end{align}
where $I(r^\text{Rx}) = \sqrt{2\rho^\text{Tx}}\Phi (r^\text{Rx}). $

\begin{figure}
\centering
        \includegraphics[width=\linewidth] {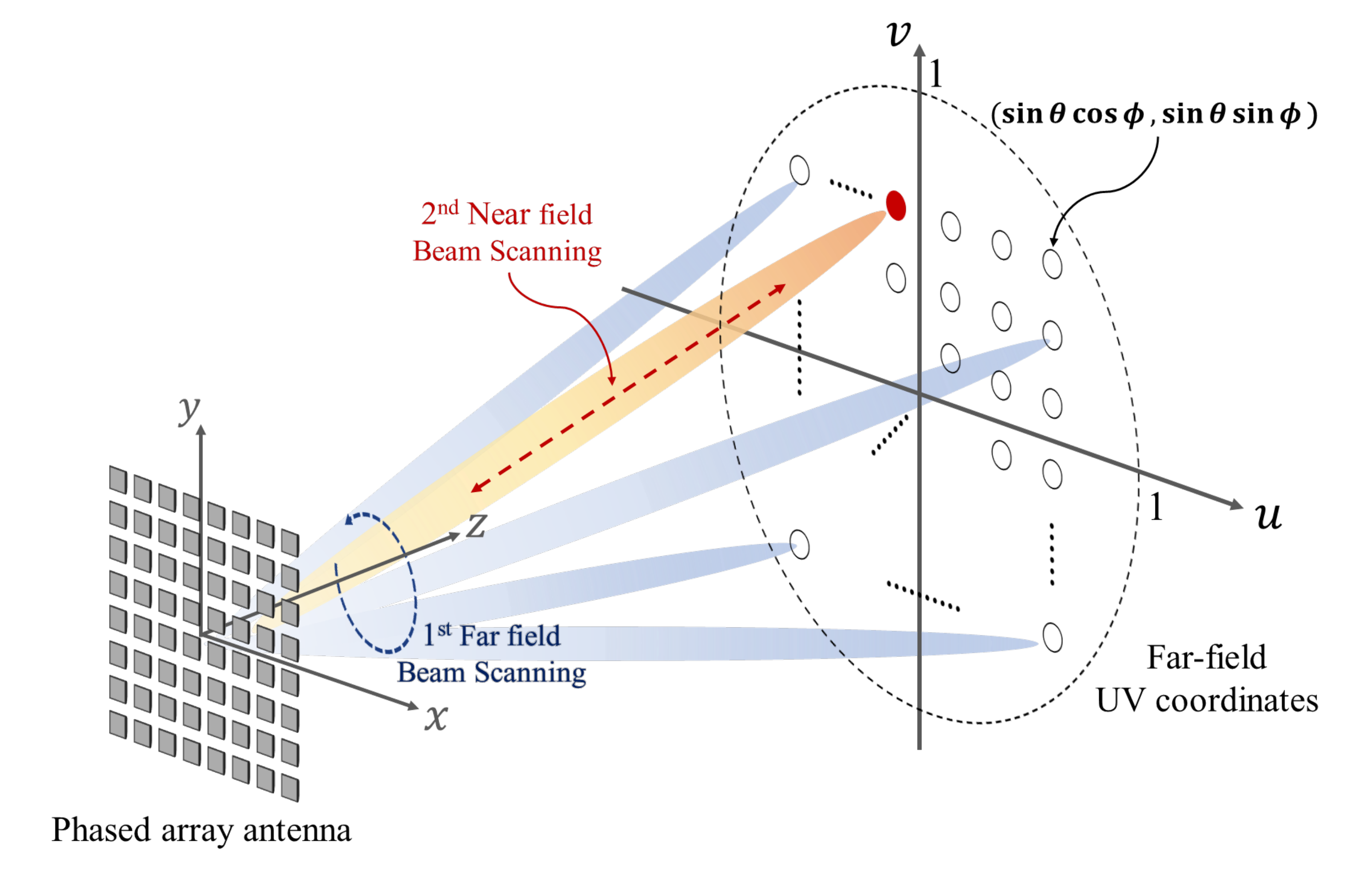}
        \caption{Proposed beam scanning scheme}
        \label{fig:scanning}
\end{figure}

Fig.~\ref{fig:scanning} shows a conceptual diagram of the proposed beam scanning method, which is able to cover far-field and radiative near-field both sequentially.
Firstly, we assume that the receiver is located within the far-field region of the transmitter. 
The far-field region means the distance between transmitter and receiver $r$ is longer than the Fraunhofer distance of the transmitter. Even though there is no clear boundary between far-field and radiative near-field, if we define the boundary as $r_\text{b}$, each field is derived as
\begin{align}\label{eq:rb}
&r_\text{b} =  \frac{2 L^\text{2}_\text{Tx}}{\lambda},\\
&\text{Radiative near-field} < r_\text{b},\\
&\text{Far-field} > r_\text{b},
\end{align}
where the maximum linear dimension of the transmit antenna array $L_\text{Tx}$ is denoted by
\begin{align}
L_\text{Tx} = \sqrt{
(N^\text{Tx}_\text{row}\times l^\text{Tx}_\text{row})^2 + (N^\text{Tx}_\text{col}\times l^\text{Tx}_\text{col})^2
}.
\end{align}

Under the far-field assumption, the term $\eta_n(r^\text{Rx})$ which is related to the distance becomes negligibly small and it means the direction $\boldsymbol \xi$ is the only control parameter to steer the beam.
When the direction control parameter ${\boldsymbol \xi}$ is given, the transmitted power wave from transmit antenna element $n$ in \eqref{eq:x} is rewritten as 
\begin{align}\label{eq:xdir}
\begin{split}
x^\text{far}_{n}({\boldsymbol \xi}) = \sqrt{2\rho^\text{Tx}} \exp\Big(-j\frac{2\pi}{\lambda}{\boldsymbol \xi}^T {\bold u}^\text{Tx}_{n} \Big).
\end{split}
\end{align}
We also rewrite the transmitter excitation vector in \eqref{eq:xvec} as
\begin{align}\label{eq:xfvec}
{\bold x}^\text{far}({\boldsymbol \xi}) = (x^\text{far}_{n}({\boldsymbol \xi}))_{n=1,\ldots,N^\text{Tx}}.
\end{align}
Furthermore, from \eqref{eq:optphase}, \eqref{eq:senpow}, and \eqref{eq:xdir}, we can calculate the received power at the sensor antenna $m^*$ corresponds to the far-field beam, which is steered to the direction of $\boldsymbol \xi$ as follows:
\begin{align}\label{eq:farpower}
\begin{split}
&P_{m^*}^\text{far} ({\boldsymbol \xi})=\\
&\frac{\Big|\sum_{n=1}^{N^\text{Tx}} I(r^\text{Rx}) \exp\Big(j\frac{2\pi}{\lambda}\big( ( {\boldsymbol \xi}^\text{Rx} -  {\boldsymbol \xi})^T{\bold u}^\text{Tx}_{n} -\frac{\|{\bold a}^\text{Tx}_n\|^2}{2r^\text{Rx}} \big ) \Big)\Big|^2}
{2}.
\end{split}
\end{align}

We introduce the u-v grid concept for the far-field region from the transmitter.
The far-field region from the transmit antenna array can be represented by a unit disk form in the u-v coordinate. The set of scanning beams for the transmitter is generated by a uniform grid in the u-v coordinate system. Each scanning beam is determined by $\boldsymbol \xi$.
Then, we define $k$th scanning beam ${\boldsymbol \xi}^{(k)}$ as
\begin{align}
{\boldsymbol \xi}^{(k)} = \bigg(\frac{\Delta^\text{u}}{\Psi^\text{u}-1}\chi_{k^u}^{\Psi^\text{u}},
\frac{\Delta^\text{v}}{\Psi^\text{v}-1}\chi_{k^v}^{\Psi^\text{v}}\bigg)^T,
\label{eq:Txscanbeam}
\end{align}
where $\chi_j^J$ is the $j$th grid point in the uniform grid with size $J$ centered at the origin, that is
\begin{align}\label{eq:grid}
    \chi_j^J = j-J/2-1/2,
\end{align}
and $\Delta^\text{u}$ and $\Delta^\text{v}$ are, respectively, the scan widths along the u and v axes, $\Psi^\text{u}$ and $\Psi^\text{v}$ are, respectively, the numbers of scanning beams along the u and v axes, and $k^u = ((k-1) \pmod{\Psi^\text{u})}+1$ and $k^v = \lfloor(k-1)/\Psi^\text{u}\rfloor + 1$ are, respectively, the indices of the scanning beam along the u and v axes.
In this paper, for a $N^\text{Tx}_\text{row}\times N^\text{Tx}_\text{col}$ planar array, we generate $2N^\text{Tx}_\text{row}\times2N^\text{Tx}_\text{col}$ scanning beams with $2N^\text{Tx}_\text{row}$ scanning points in the u-axis and $2N^\text{Tx}_\text{col}$ scanning points in the v-axis.
The number of scanning beams $K$ of the transmitter is 
\begin{align}\label{eq:scannumber}
    K = \Psi^\text{u}\Psi^\text{v} = 2N^\text{Tx}_\text{row}\times2N^\text{Tx}_\text{col}.
\end{align}

During the far-field scanning along the designed u-v grid, the sensor antenna measures the received power for each beam index and then finds out the optimal index $k^*$ which has the highest received power.
With the optimal direction $\boldsymbol \xi^{(k^*)}$ from the first scanning phase, the transmitter conducts the near-field beam scanning with different distance value $r$. From the first scanning phase, the optimal u-v coordinate is obtained as $\boldsymbol \xi_\text{opt} = (u_\text{opt},v_\text{opt})^T$.

Under the radiative near-field assumption, the phase of each transmit antenna is determined not only by the direction but also by the distance. 
The transmitted power wave from the transmitter in the near-field region is defined as
\begin{align}\label{eq:xdis}
\begin{split}
x^\text{near}_{n}({\boldsymbol \xi}, r) = \sqrt{2\rho^\text{Tx}} \exp \bigg(-j\frac{2\pi}{\lambda}  \Big({\boldsymbol \xi}^T{\bold u}^\text{Tx}_{n} - \frac{\|{\bold a}^\text{Tx}_n\|^2}{2r}\Big) \bigg).
\end{split}
\end{align}
The transmitter excitation vector in the near-field region is defined as 
\begin{align}\label{{eq:xnvec}}
{\bold x}^\text{near}({\boldsymbol \xi}, r) = (x^\text{near}_{n}({\boldsymbol \xi}, r))_{n=1,\ldots,N^\text{Tx}}.
\end{align}
In the same manner as \eqref{eq:farpower}, based on \eqref{eq:xdis}, the corresponding received power for the near-field beam is defined with the direction control parameter $\boldsymbol \xi$ and distance control parameter $r$ as
\begin{align}\label{eq:nearpower}
\begin{split}
&P_{m^*}^\text{near} ({\boldsymbol \xi}, r)=\\
&\frac{\Big|\sum_{n=1}^{N^\text{Tx}} I(r^\text{Rx}) \exp\Big(j\frac{2\pi}{\lambda}\big( F(\boldsymbol \xi) -\frac{\|{\bold a}^\text{Tx}_n\|^2}{2}(\frac{1}{r^\text{Rx}} - \frac{1}{r}) \big ) \Big)\Big|^2}
{2},
\end{split}
\end{align}
where $F(\boldsymbol \xi) = ( {\boldsymbol \xi}^\text{Rx} -  {\boldsymbol \xi})^T{\bold u}^\text{Tx}_{n}$.

With the fixed optimal direction $\boldsymbol \xi_\text{opt}$ from the first scanning phase, the set of scanning beams is generated by a uniform distance grid in order to find out the optimal focal distance.
Each scanning beam is indexed by $i$ ($=1,\ldots,\Upsilon^\text{d}$) and defined as 
\begin{align}\label{eq:rgrid}
r^{(i)} = \frac{r_\text{b}}{\Upsilon^\text{d}} i,
\end{align}
where $\Upsilon^\text{d}$ is the number of the grid for the distance. The reason why we set the maximum distance as $r_\text{b}$ is to conduct the second scanning phase within the near-field region. Based on the received power at the sensor antenna corresponding each beam index, we can find out the optimal distance parameter $r_\text{opt}$ with the second scanning phase.

\begin{algorithm}[!ht]
\SetAlgoLined
\KwIn{
\\Direction control parameter ${\boldsymbol \xi}$\\ 
Distance control parameter ${r}$
}
\KwOut{
\\ Optimal direction control parameter ${\boldsymbol \xi}_\text{opt}$\\
Optimal distance control parameter ${r}_\text{opt}$
}
\BlankLine
\For{$k \leftarrow 1$ \KwTo $K$}{
Set the transmitter excitation vector to ${\bold x}^\text{far} ({\boldsymbol \xi}^{(k)})$\\
Measure the receive power at the sensor antenna $P_{m^*}^\text{far} ({\boldsymbol \xi}^{(k)})$\\
}
$k^{*}\leftarrow \arg\max_{k=1,\ldots, K}P_{m^*}^\text{far} ({\boldsymbol \xi}^{(k)})$\\
${\boldsymbol \xi}_\text{opt} \leftarrow {\boldsymbol \xi}^{(k^{*})}$\\
Set the direction control parameter ${\boldsymbol \xi}^{(k^*)}$ \\
\For{$i \leftarrow 1$ \KwTo $\Upsilon^\text{d}$}{
Set the transmitter excitation vector to ${\bold x}^\text{near}({\boldsymbol \xi^{(k^*)}}, r^{(i)})$\\
Measure the receive power at the sensor antenna $P_{m^*}^\text{near} ({\boldsymbol \xi}^{(k^*)}, r^{(i)})$\\
}
$i^*\leftarrow \arg\max_{i=1,\ldots, \Upsilon^\text{d}}P_{m^*}^\text{near} ({\boldsymbol \xi}^{(k^*)}, r^{(i)})$\\
${\bold r}_\text{opt} \leftarrow {\bold r}^{(i^*)}$\\
\Return ${\boldsymbol \xi}_\text{opt}$, ${r}_\text{opt}$
\caption{Beam Scanning Algorithm} \label{Scan}
\end{algorithm}

The overall operation of the proposed beam scanning algorithm is summarized in Algorithm \ref{Scan}.
Algorithm starts by scanning the direction control parameter of the transmitter with $\boldsymbol \xi^{(k)}$ ($k = 1,\ldots,K$) (line 1).
For each scanning beam $k$, the corresponding received power at the sensor antenna $P_{m^*}^\text{far} ({\boldsymbol \xi}^{(k)})$ is measured (line 3) so that the optimal direction control parameter $\boldsymbol \xi_{\text{opt}}$ is found at lines 5--6.
After the first scanning phase, the direction control parameter of the transmitter is fixed to $\boldsymbol \xi_{\text{opt}}$ at line 7, and then we move onto the second scanning phase with different distance values $r^{(i)}$ ($i = 1,\ldots,\Upsilon^\text{d}$) (line 8). 
Then the optimal distance parameter $r_{\text{opt}}$ is returned (line 14) based on the received power for each scanning beam $P_{m^*}^\text{near} ({\boldsymbol \xi}^{(k^*)}, r^{(i)})$ in the second scanning phase at lines 10--13.

\begin{figure}
\centering
        \includegraphics[width=\linewidth] {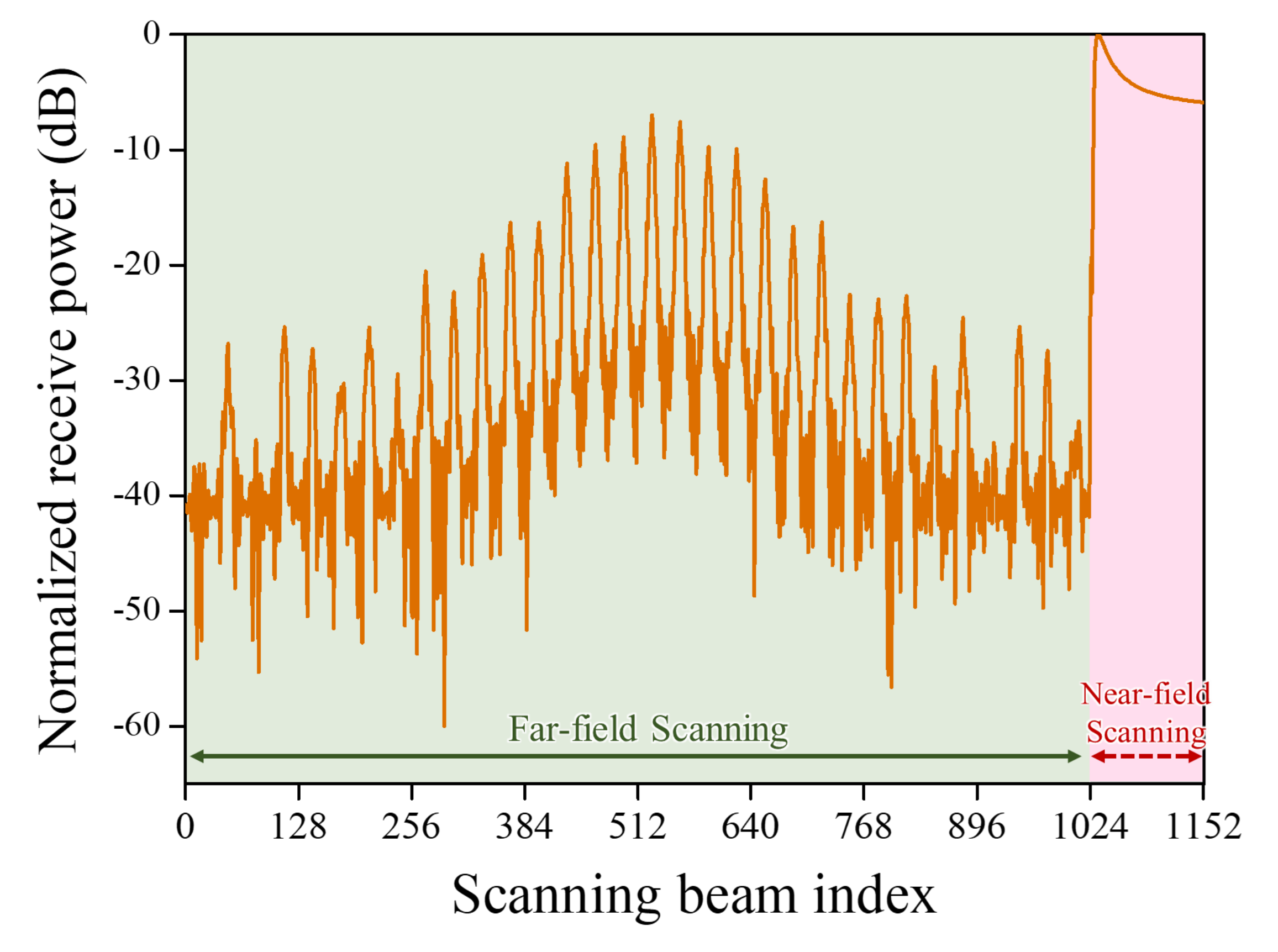}
        \caption{Receive power according to scanning beams}
        \label{fig:scanpat}
\end{figure}

\begin{figure}
    \centering
    \subfigure[Far-field]{
        \label{fig:Txphase_far}\includegraphics[width=0.33\linewidth] {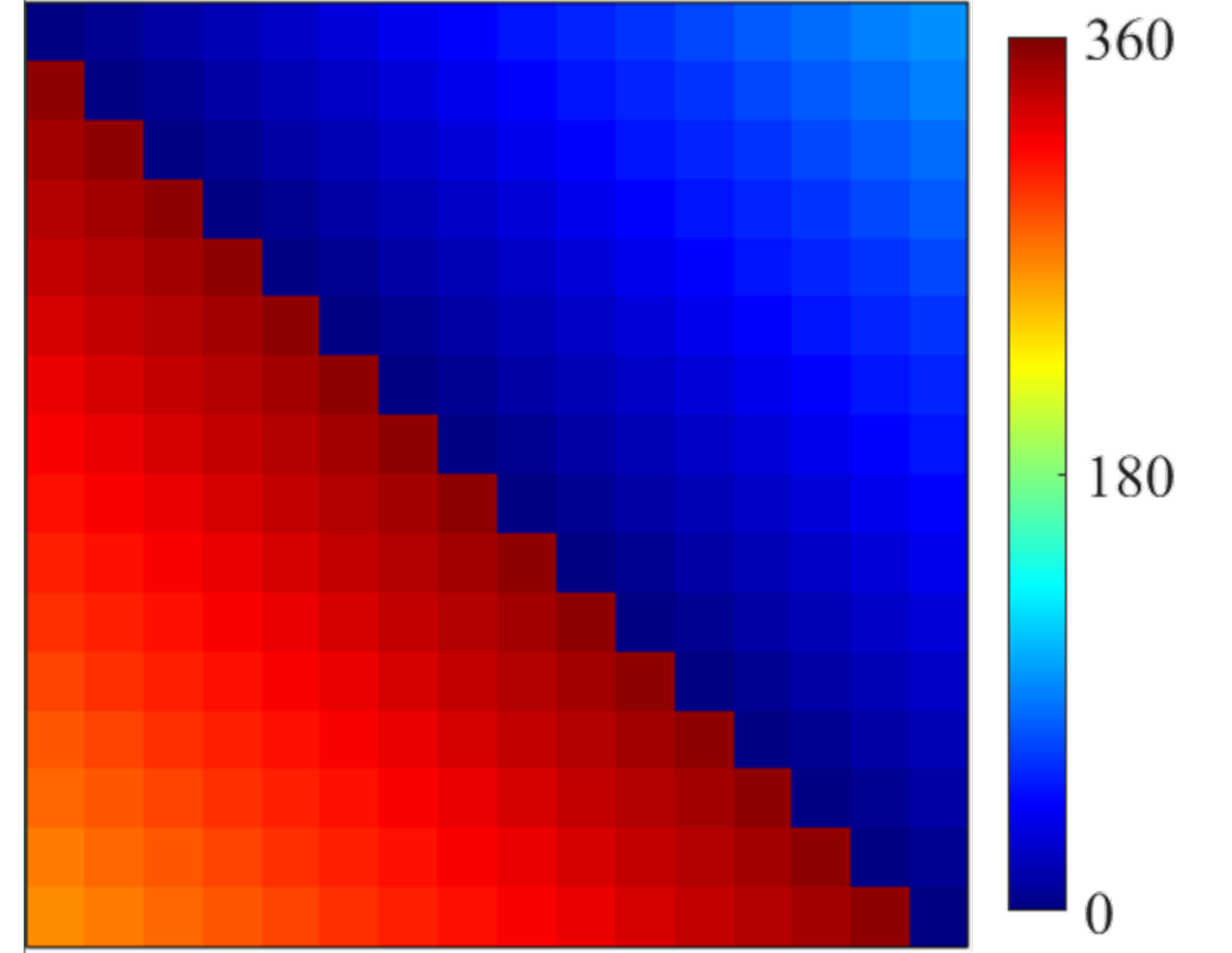}
    }%
    \subfigure[1m]{
        \label{fig:Txphase_1m}\includegraphics[width=0.33\linewidth] {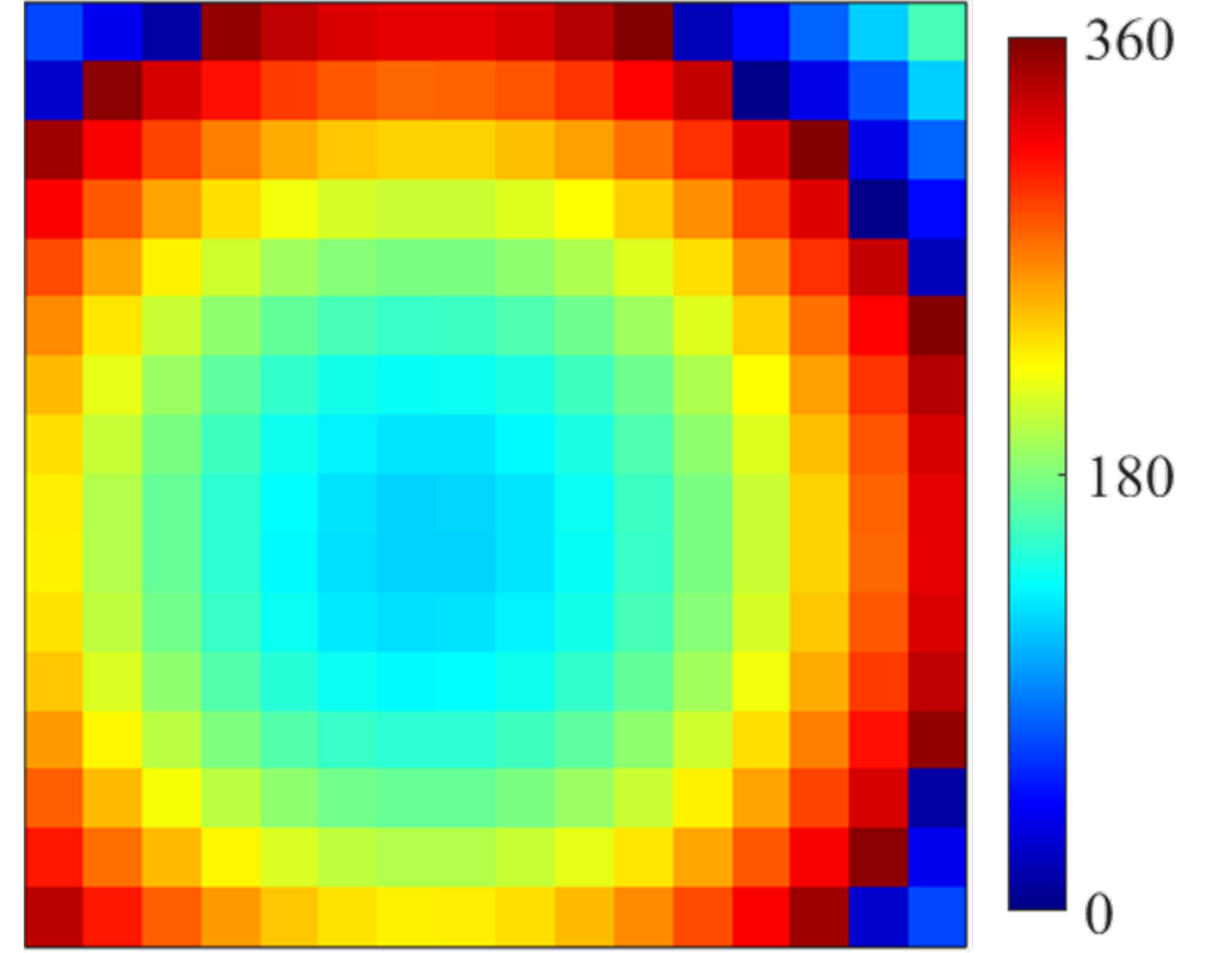}
    }%
        \subfigure[2m]{
        \label{fig:Txphase_2m}\includegraphics[width=0.33\linewidth] {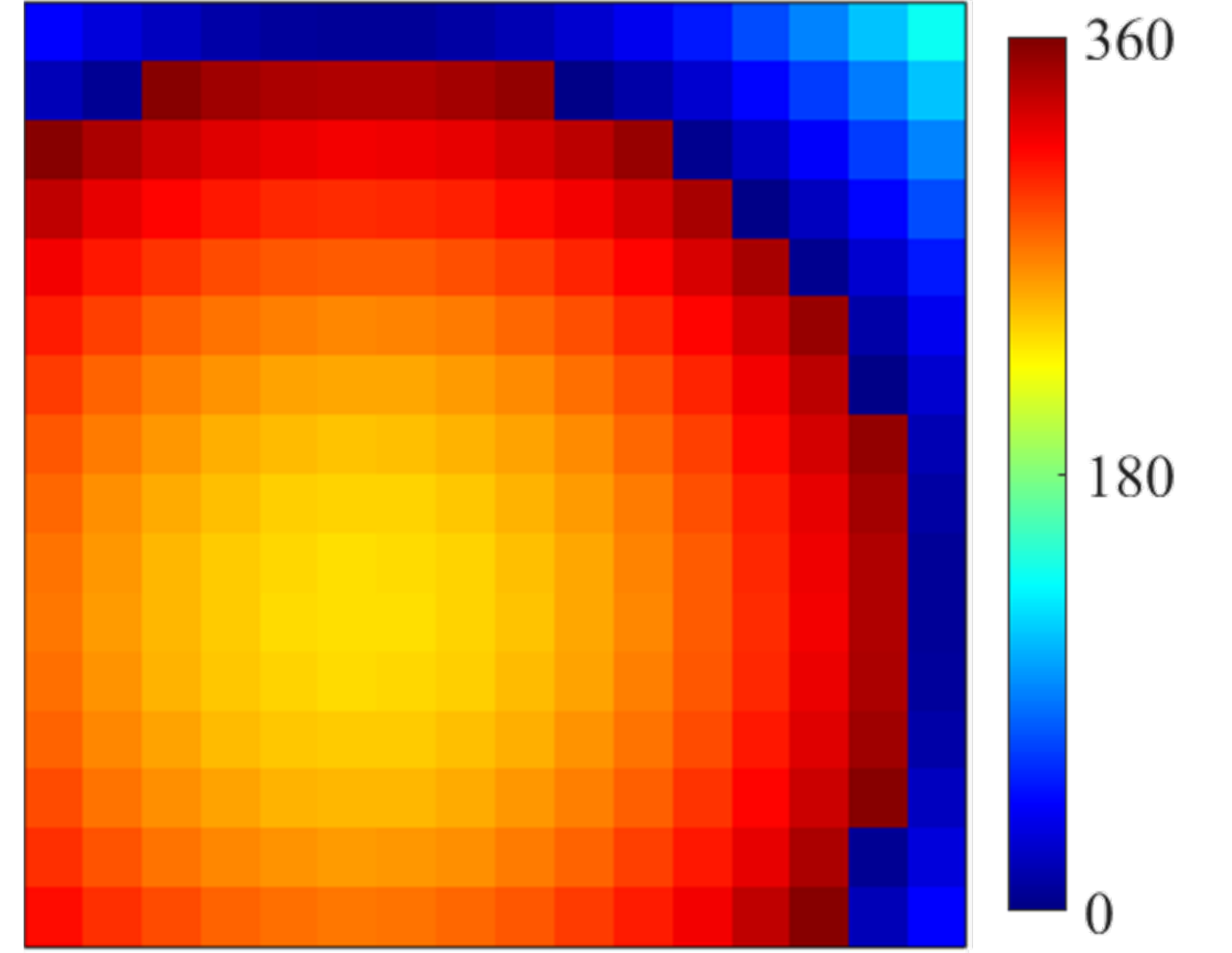}
    }\\
        \subfigure[4m]{
        \label{fig:Txphase_4m}\includegraphics[width=0.33\linewidth] {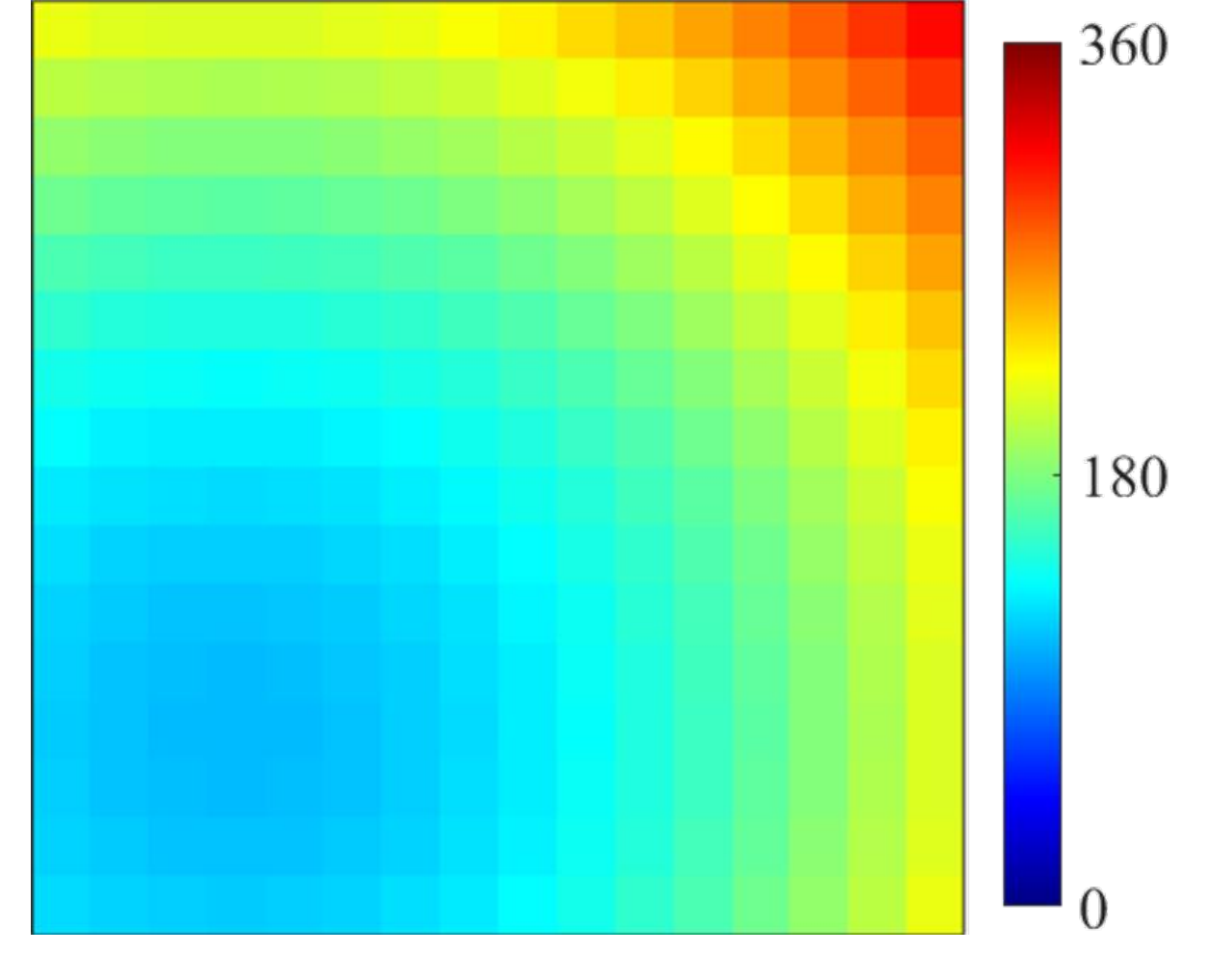}
    }%
    \subfigure[8m]{
        \label{fig:Txphase_8m}\includegraphics[width=0.33\linewidth] {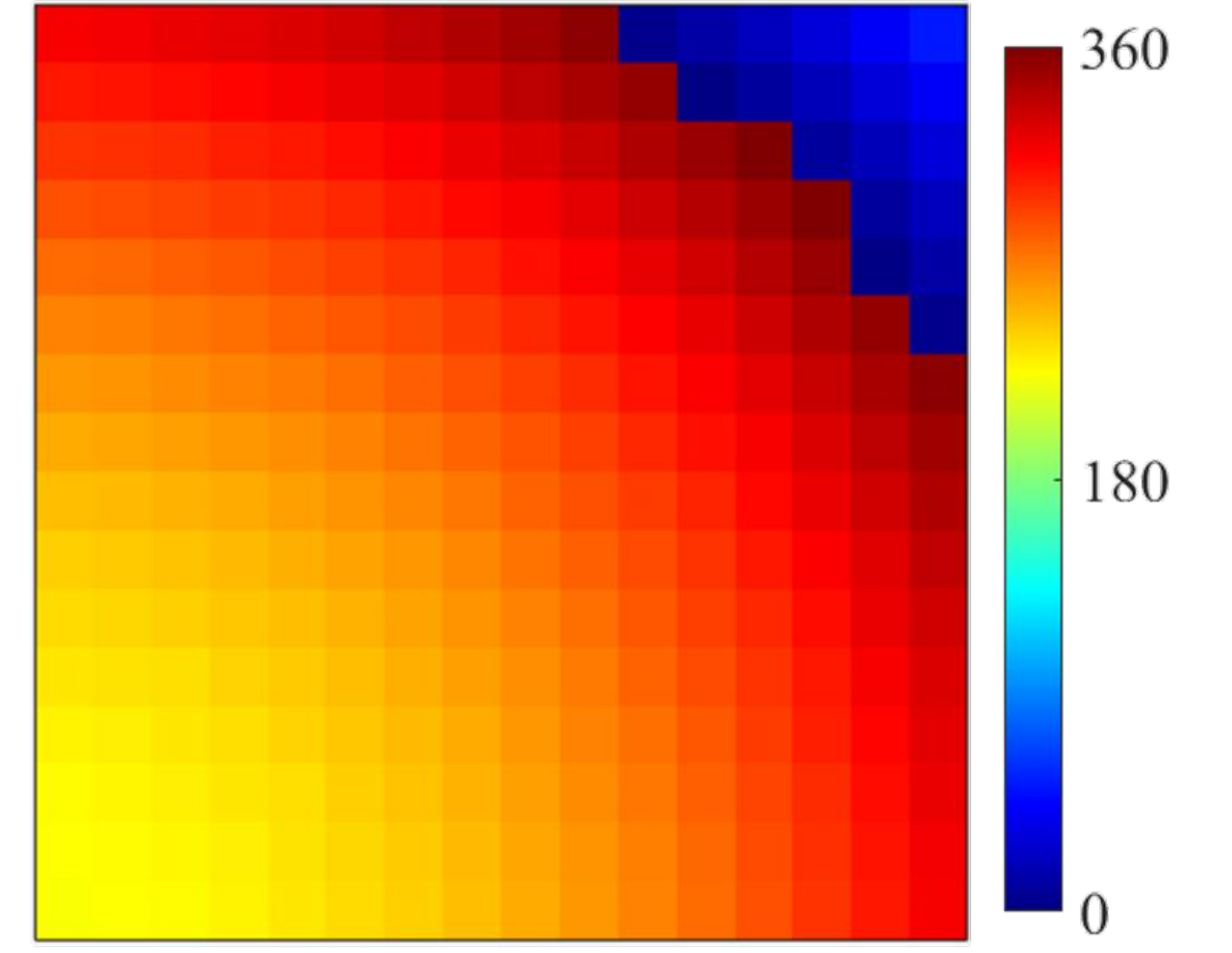}
    }%
        \subfigure[13m]{
        \label{fig:Txphase_13m}\includegraphics[width=0.33\linewidth] {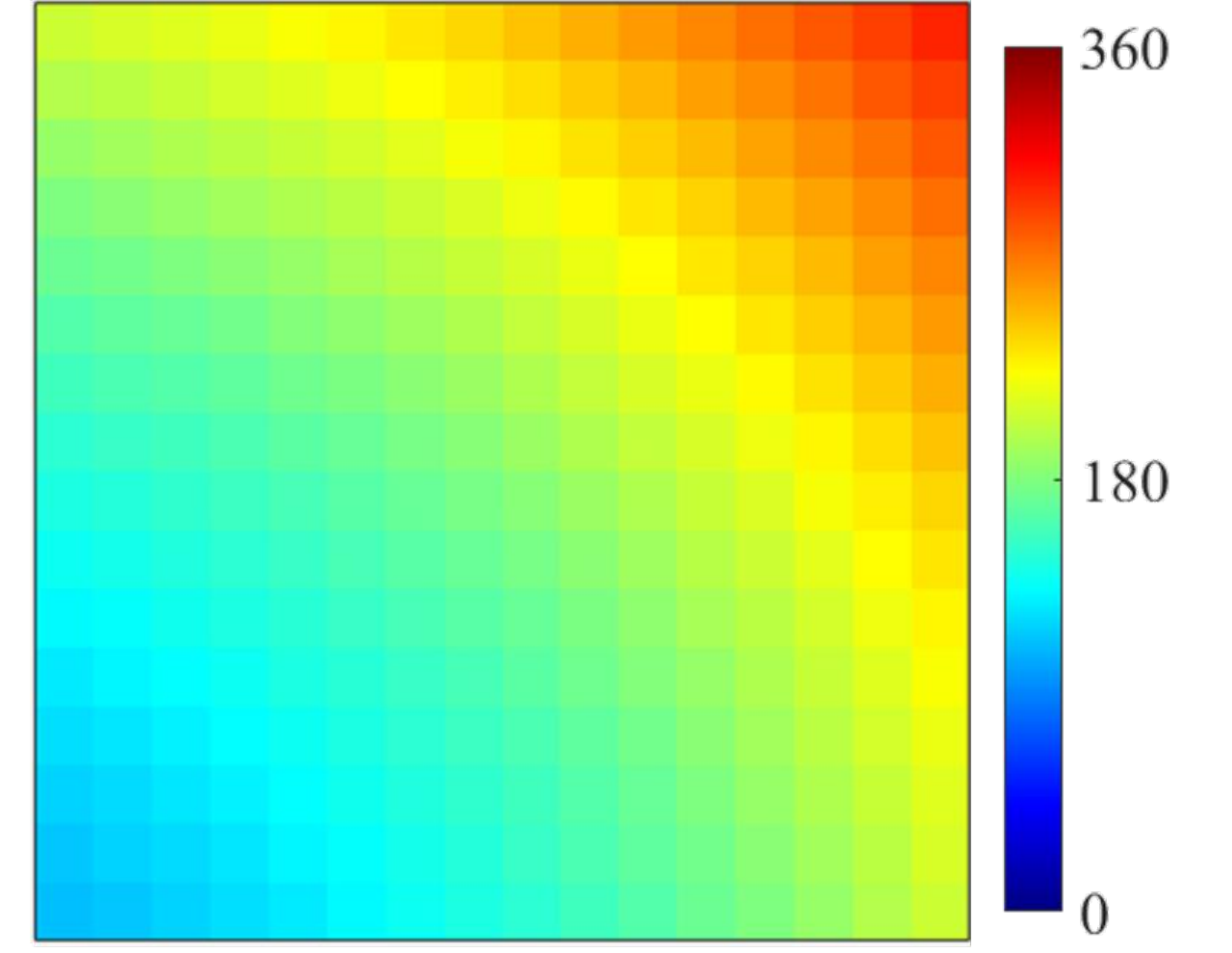}
    }\\
    \caption{Phase of transmit array weights}
    \label{fig:txphase}
\end{figure}

To verify the proposed beam scanning algorithm, we have simulated the WPT system with MATLAB software. In the simulation, the transmitter consists of a 16-by-16 phased array antenna, which is larger compared to the fabricated system in our works. According to \eqref{eq:Txscanbeam} and \eqref{eq:scannumber}, 1024 scanning beams are generated for the far-field scanning.
We set the center points of transmit antenna array and receive antenna array have the same height and increase the distance from 1 m to 13 m since based on \eqref{eq:rb}, up to 13 meters from the transmitter is considered within radiative near-field. Fig.~\ref{fig:scanpat} shows the normalized receive power with the proposed beam scanning algorithm when the distance between transmitter and receiver is 1 meter. The 497th scanning beam achieves the highest receive power, and with direction control parameter ${\xi}^{497}$, the near-field scanning is conducted. For the near-field scanning, we set the number of the grid for the distance $\Upsilon^\text{d}$ to 130 and the step size of the distance grid is 0.1 meter.
Fig.~\ref{fig:txphase} shows the phases of the transmit array weights according to the different distances as heat map distribution. In Fig.~\ref{fig:scanpat}, we can see that around 7 dB improvement in the received power is obtained by the near-field scanning compared to the far-field-scanning-only. Figs.~\ref{fig:Txphase_far} is the phase distribution of the optimal (497th) scanning beam. As we can see in Figs.~\ref{fig:Txphase_1m}-(f), the convex lens-like form to focus the beam at the receiver gradually changes to the linear form as the receiver moves towards the far-field region.

\section{System Design and Implementation}\label{section:system}

\begin{figure}
    \centering
        \subfigure[Transmitter block diagram]{
        \label{fig:transmitter diagram}\includegraphics[width=0.95\linewidth] {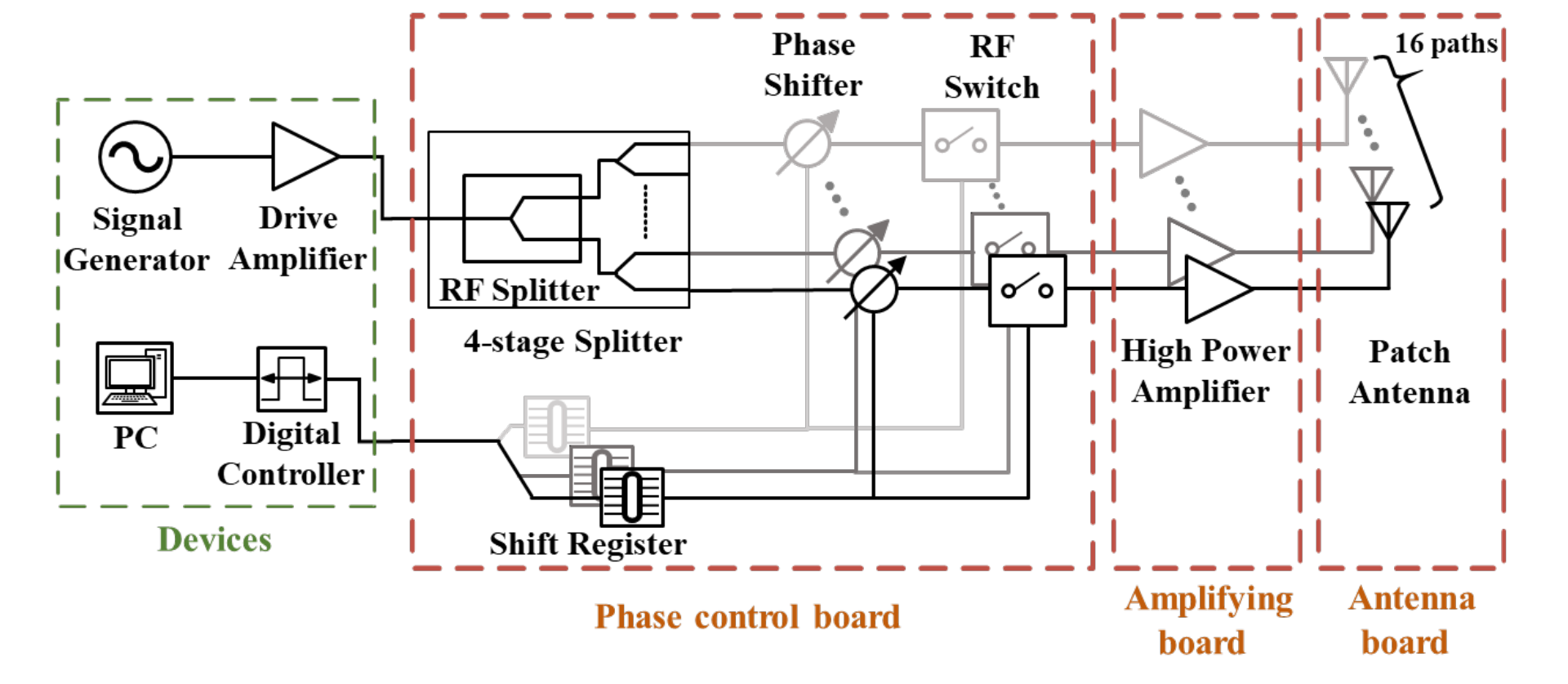}
        }\\
        \subfigure[Receiver block diagram]{
        \label{fig:receiver diagram}\includegraphics[width=0.9\linewidth] {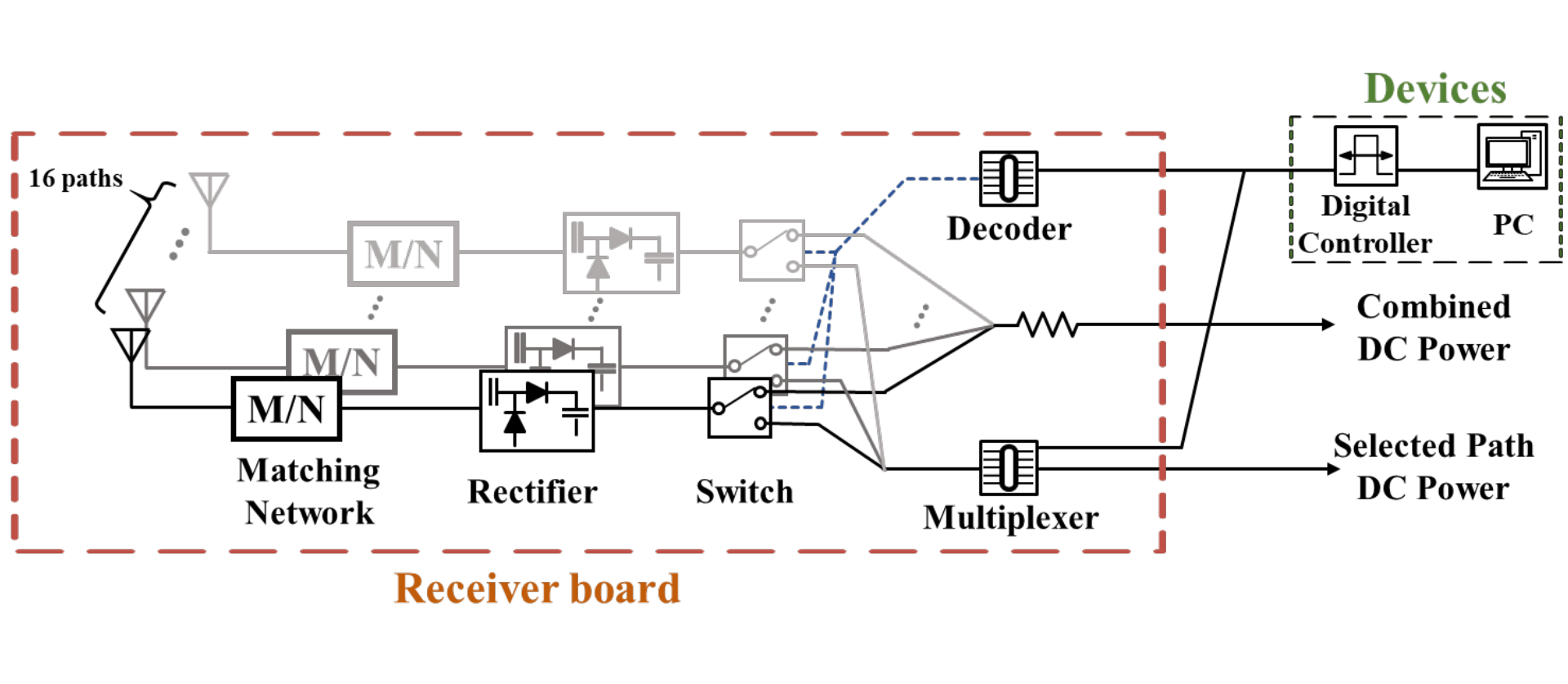}
        }\\
    \caption{System diagram}
    \label{fig:Blcok diagram}
\end{figure}

In this section, we present the design, structure, and fabrication of all hardware components in our system.
Fig.~\ref{fig:Blcok diagram} shows the block diagram of the proposed WPT system. Implemented WPT system consists of an 8-by-8 circularly-polarized phased antenna array transmitter and a 4-by-4 rectenna array receiver that operates at 5.8 GHz.
Selected components for the system are listed in Table \ref{component table}. Technical descriptions of the transmitter and receiver are presented in the following subsections.

\subsection{Phased Array Antenna
Transmitter}\label{section:transmitter}

\begin{figure}
    \centering
    \subfigure[Phase control board]{
        \label{fig:phase board}\includegraphics[width=0.44\linewidth] {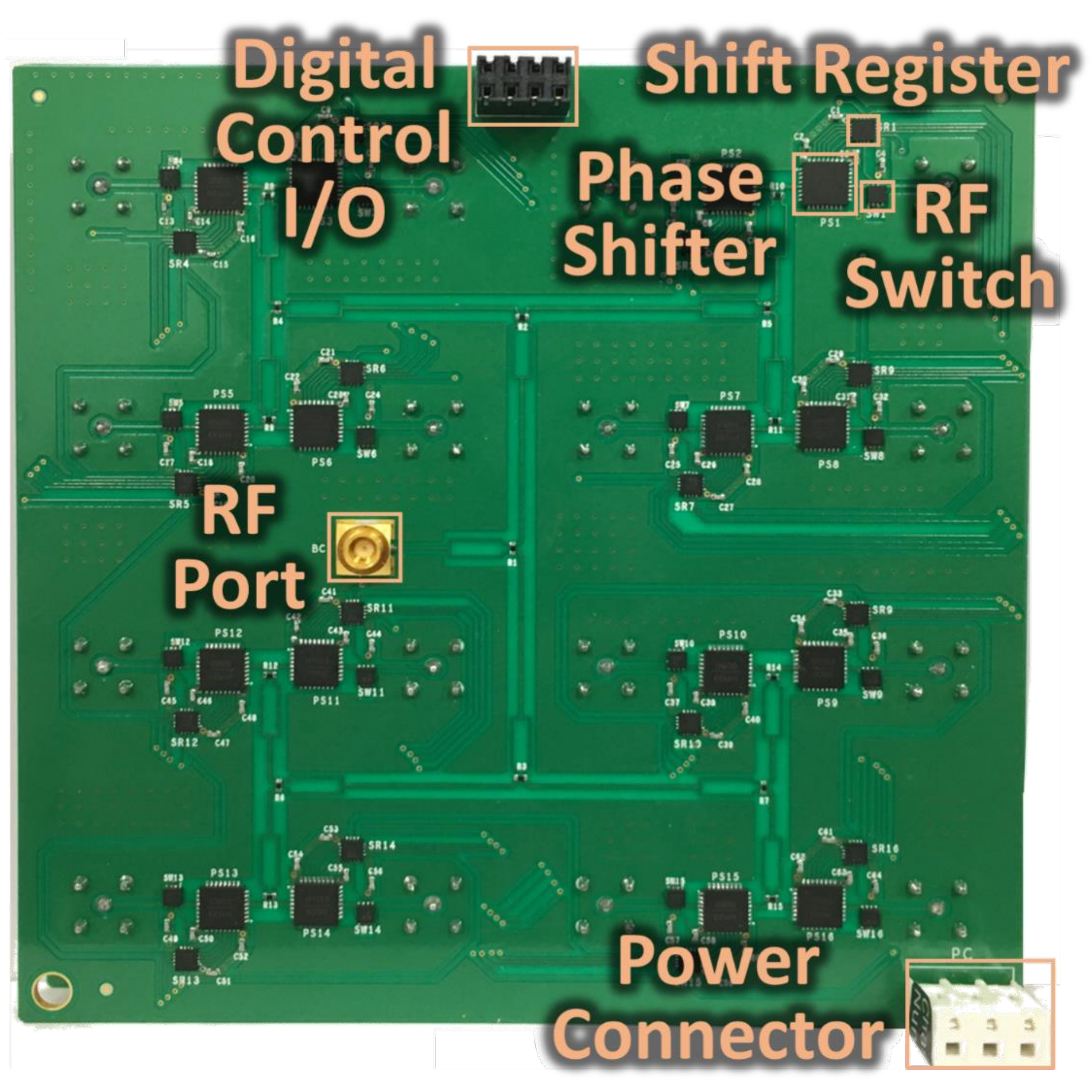}
        }
    \subfigure[Amplifying board]{
        \label{fig:amplifying board}\includegraphics[width=0.42\linewidth] {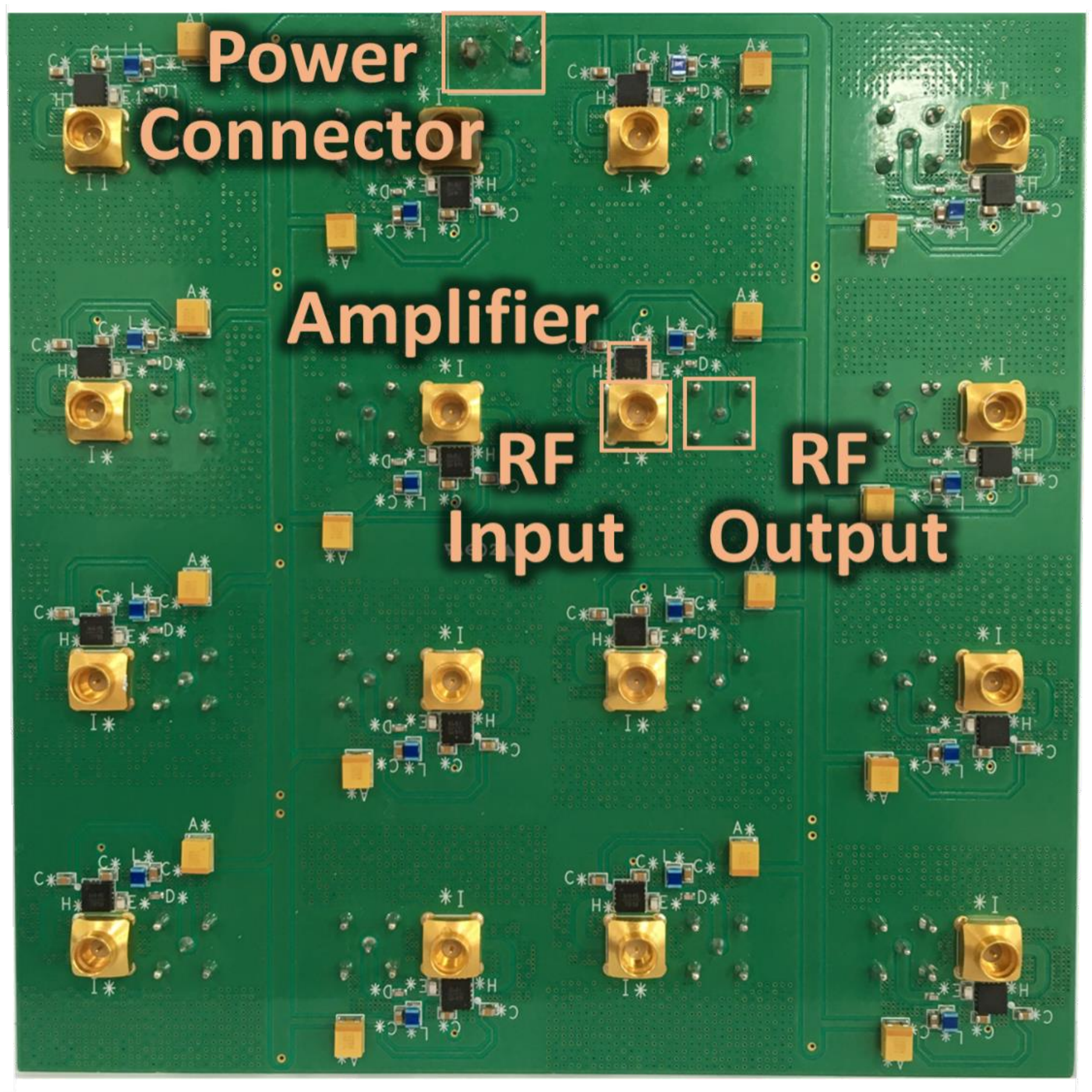}
        }\\
        \subfigure[Module combining board]{
        \label{fig:additional baord}\includegraphics[width=0.55\linewidth] {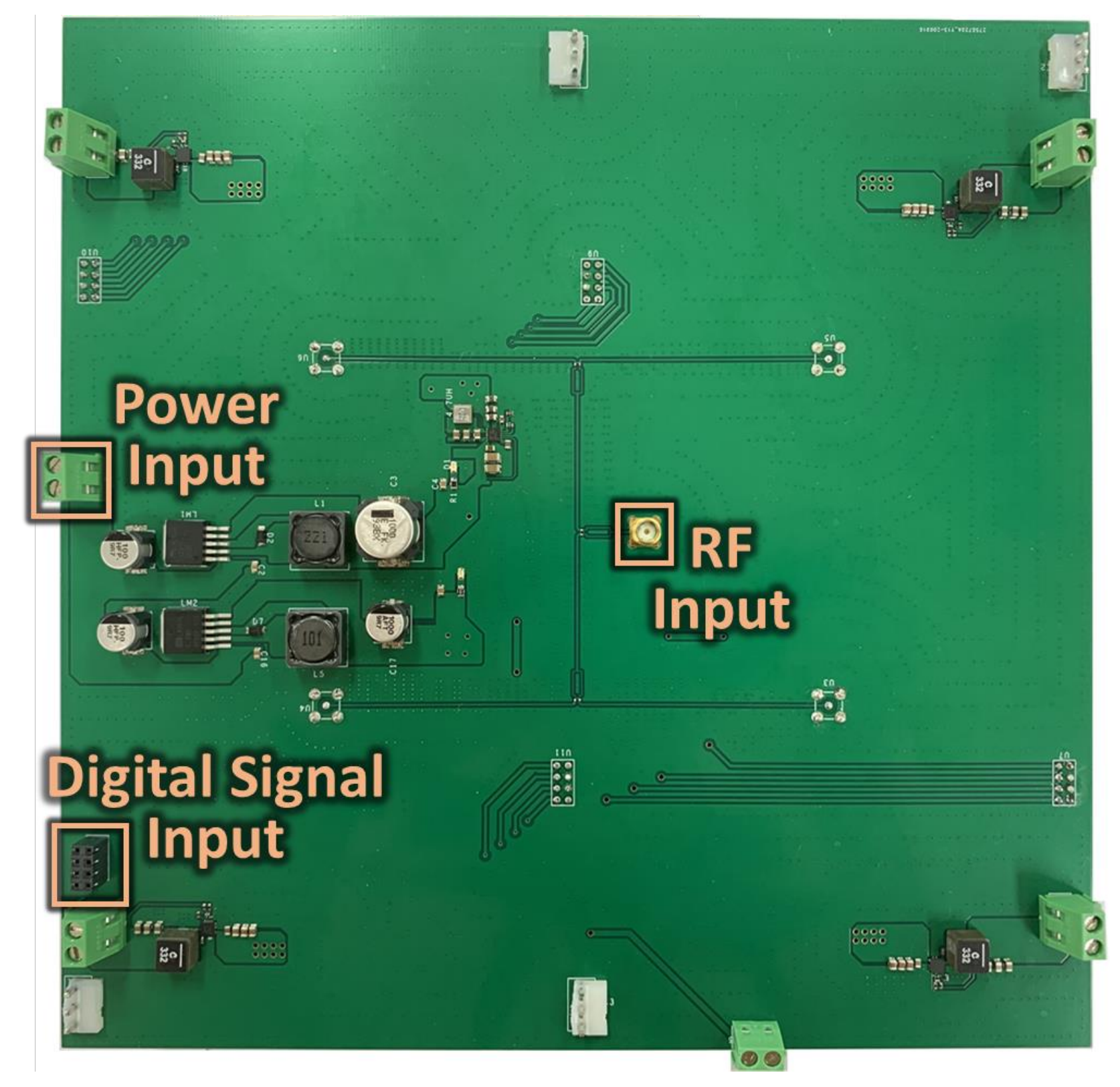}
        }\\
    \caption{Phased array antenna transmitter}
    \label{fig:transmitter}
\end{figure}

The 8-by-8 transmitter array consists of four 4-by-4 unit modules. A 4-by-4 unit module is built by connecting a phased array board, an amplifier board, and an antenna board as a sandwich structure. 
We designed and fabricated a 16-way phase control board that is able to control both the phase of the input RF wave and ON/OFF states for each path.
Firstly, to divide the transmit power into multiple RF paths, we have designed a Wilkinson power divider that has an optimal performance by using the advanced design system (ADS) software. A 16-way power divider is implemented by using four stages of the Wilkinson power divider.
And then we have deployed a phase shifter, RF switch, and shift register every port, as seen in Fig.~\ref{fig:phase board}. The RF switch enables turning on and off the path, and the phase shifter provides a 360-degree coverage of a phase shifting value, with a least significant bit (LSB) of 5.625 degrees. The RF switch and phase shifter in each path are all digitally controlled by a shift register chip.

The power of the output RF signal from the phase control board is not sufficient for WPT due to the insertion loss of phase shifter and RF switch chips. Therefore, in order to increase the output power of the transmitter, we have designed a 4-by-4 amplifier array board, which consists of an amplifier as in Fig.~\ref{fig:amplifying board}. Each output port of the phase control board is directly connected to the input of the amplifier board via a board-to-board connector. The phase control board and amplifier array board are both fabricated on Rogers RO4350B board with a thickness of 1.542 mm.

The more antenna elements the transmitter has, the sharper beam can be formed, which leads to higher power transfer efficiency. To set up the transmitter with 8-by-8 antenna arrays, we have fabricated an integrating board that can combine four phase shifter and amplifier modules in one transmit antenna array. Fig.~\ref{fig:additional baord} shows the module combining board which has one RF input port divided by a 2-stage Wilkinson divider into 4 RF paths. At the end of each path, there is an RF connector that is supposed to be connected to each module. We put all of the dc-dc power converters together in this board which is needed to provide power to the phase control board and amplifying board. Also, we can control shift registers for all RF ports with a single digital I/O port by forming a daisy chain to provide a serial control.

\begin{figure}
    \centering
    \subfigure[Layout with dimensions]{
        \label{fig:antenna layout}\includegraphics[width=0.4\linewidth] {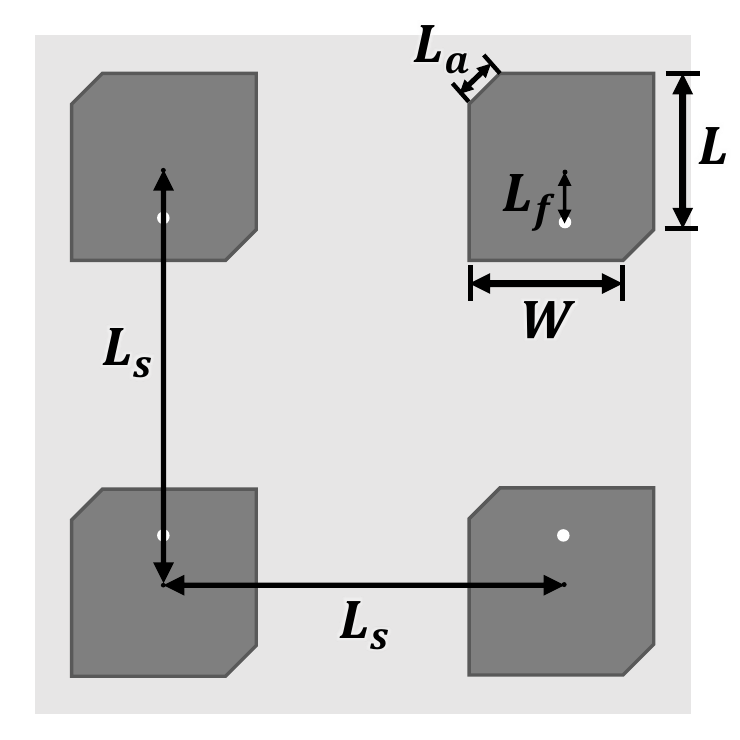}
        }
    \subfigure[Fabricated board]{
        \label{fig:antenna board}\includegraphics[width=0.4\linewidth] {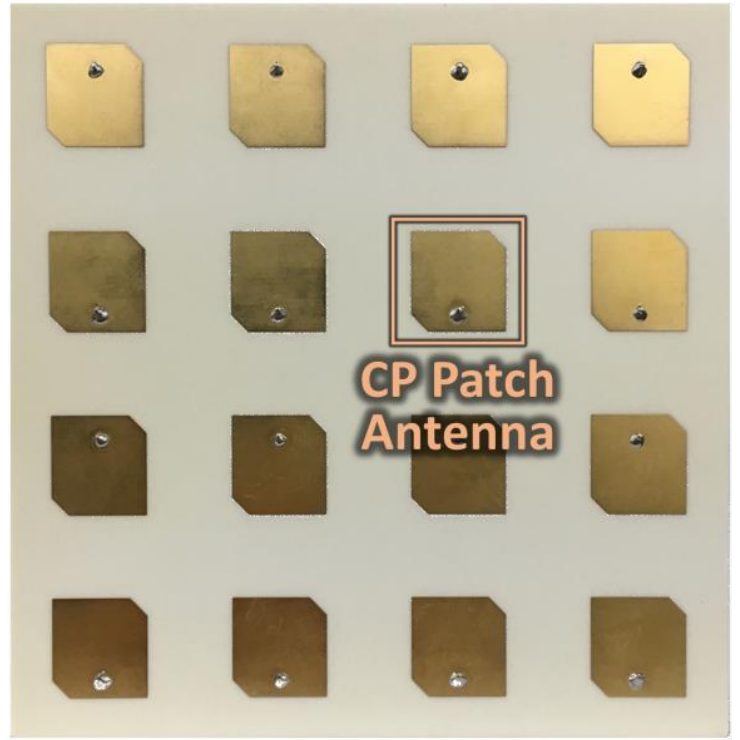}
        }\\
    \caption{Antenna array}
    \label{fig:antenna figure}
\end{figure}

\begin{figure}
    \centering
    \subfigure[xz-plane]{
        \label{fig:xz}\includegraphics[width=0.4\linewidth] {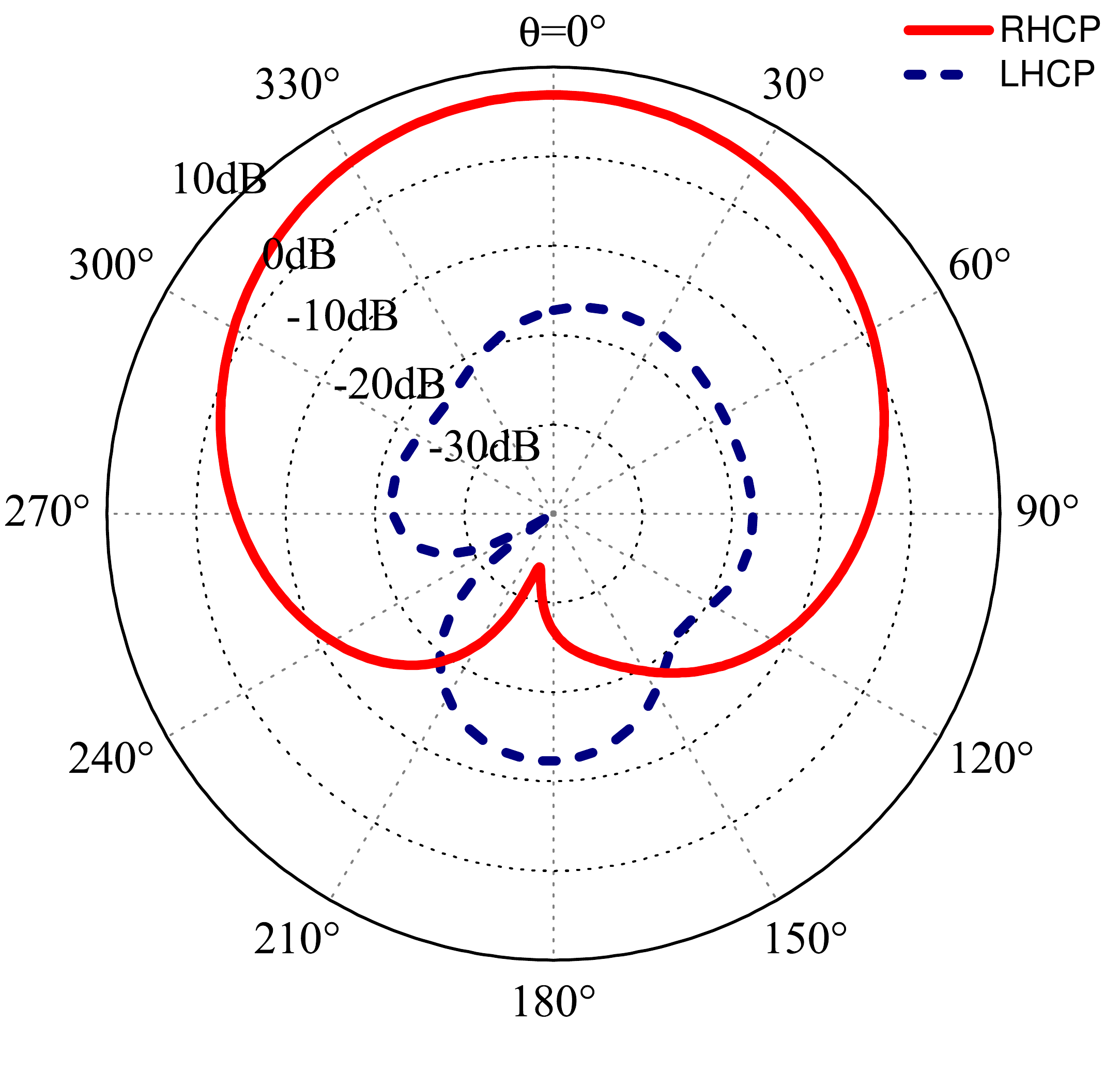}
        }
    \subfigure[yz-plane]{
        \label{fig:yz}\includegraphics[width=0.4\linewidth] {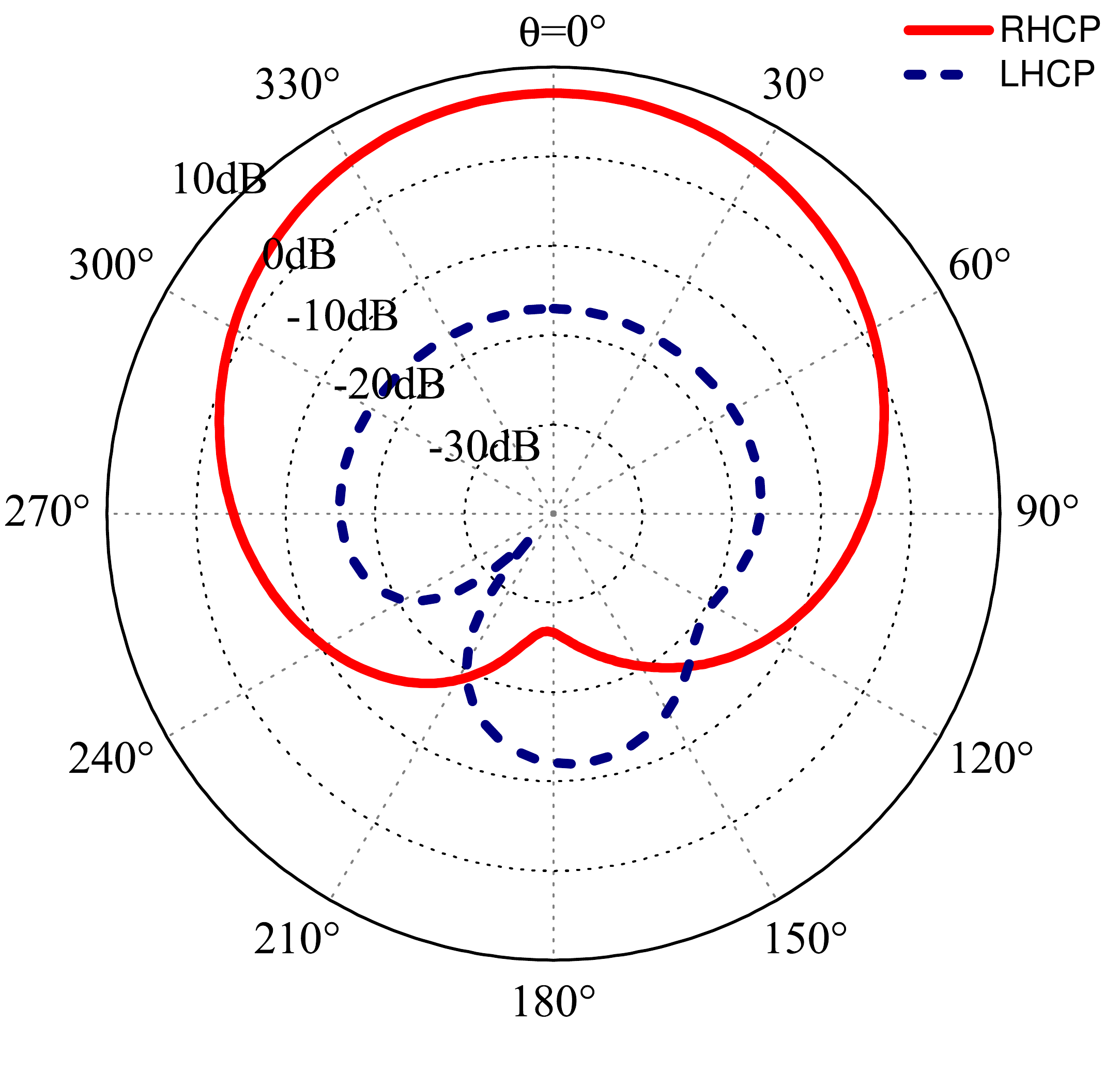}
        }\\
    \caption{Simulated radiation patterns of the patch antenna}
    \label{fig:radiation pattern}
\end{figure}

A circularly polarized (CP) microstrip patch antenna is used for both the transmitter and receiver. In the RF WPT system, in most cases, the receiver is freely deployed with arbitrary attitudes, and it leads to polarization mismatch in terms of the transmitter and receive antennas. 
In order to achieve reasonable power transfer efficiency even if the polarization is not matched due to the different attitudes of the transmitter and receiver, CP antenna is used for transmit and receive antenna arrays.

Fig.~\ref{fig:antenna layout} shows the layout of the CP antenna array with antenna patch dimensions. The distance between the feed point and center point of the patch antenna is $L_f$ (4.398 mm). The dimensions of the patch are $W$ = $12.75$ mm, $L$ = $12.17$ mm, and $L_a$ = $3.82$ mm. The antenna spacing between two antenna elements ($L_s$) is $28.2$ mm. Simulated 2-D radiation patterns along two cutting planes (xz- and yz-planes) are seen in Fig.~\ref{fig:radiation pattern} at the frequency of 5.8 GHz. 
We can see that the designed antenna mainly
radiates unidirectional RHCP waves in the upper hemisphere.

Based on simulation results, we have fabricated a 4-by-4 array antenna on Rogers RO4725JXR board with a thickness of 1.542mm, as seen in Fig.~\ref{fig:antenna board}. The reflection coefficient of the antenna element, which is measured by a network analyzer, was observed to be lower than -10 dB for all ports at 5.8 GHz.
Each patch antenna is fed by a coaxial feed from the backside of the board, which is connected to the transmitter and receiver modules via a board-to-board connector.

\subsection{Rectenna Array Receiver}\label{section:receiver}

\begin{figure}
    \centering
        \subfigure[Schematic diagram]{
        \label{fig:rectifier diagram}\includegraphics[width=0.45\linewidth] {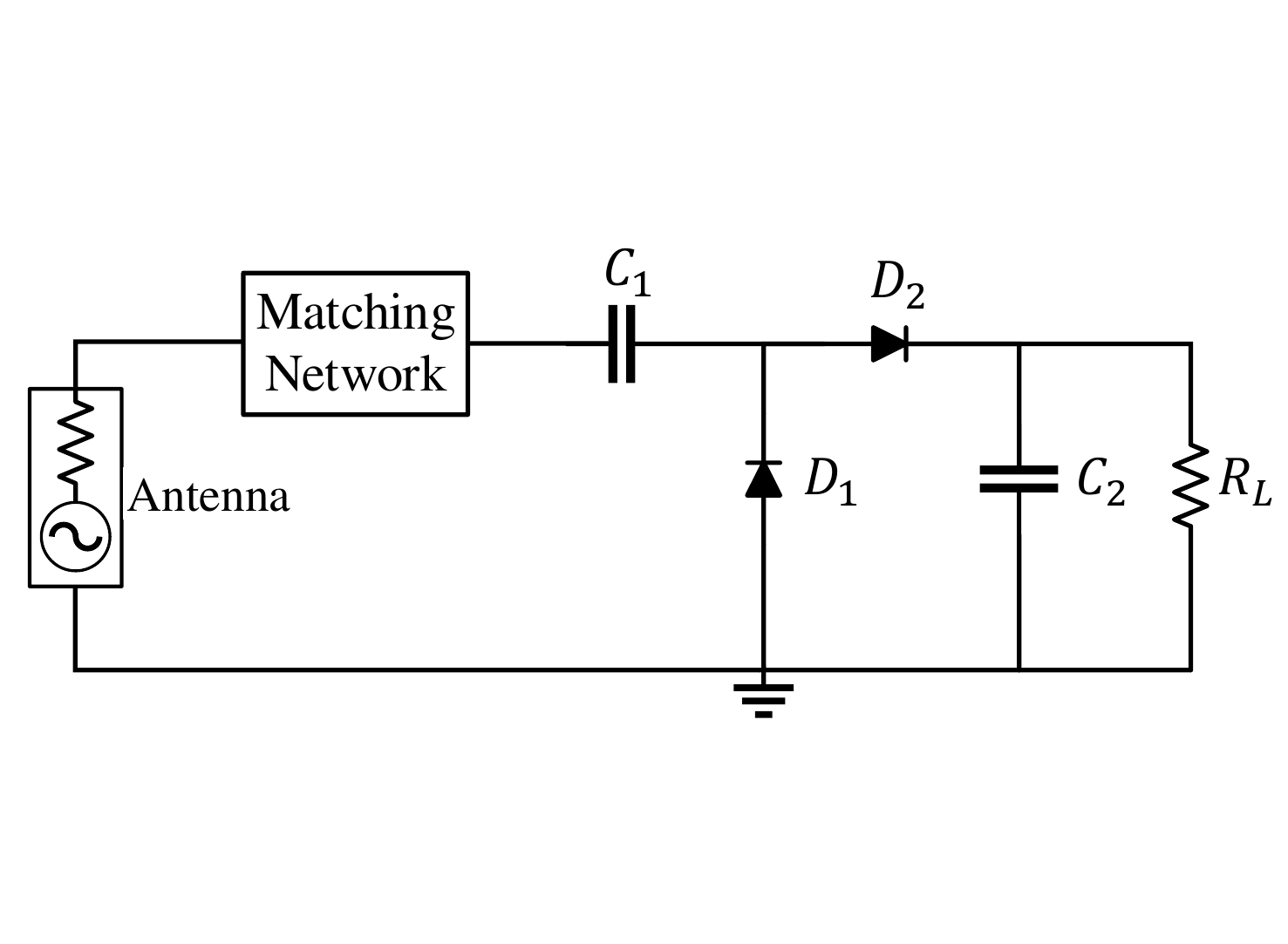}
        }
    \subfigure[Fabricated board]{
        \label{fig:rectifier board}\includegraphics[width=0.45\linewidth] {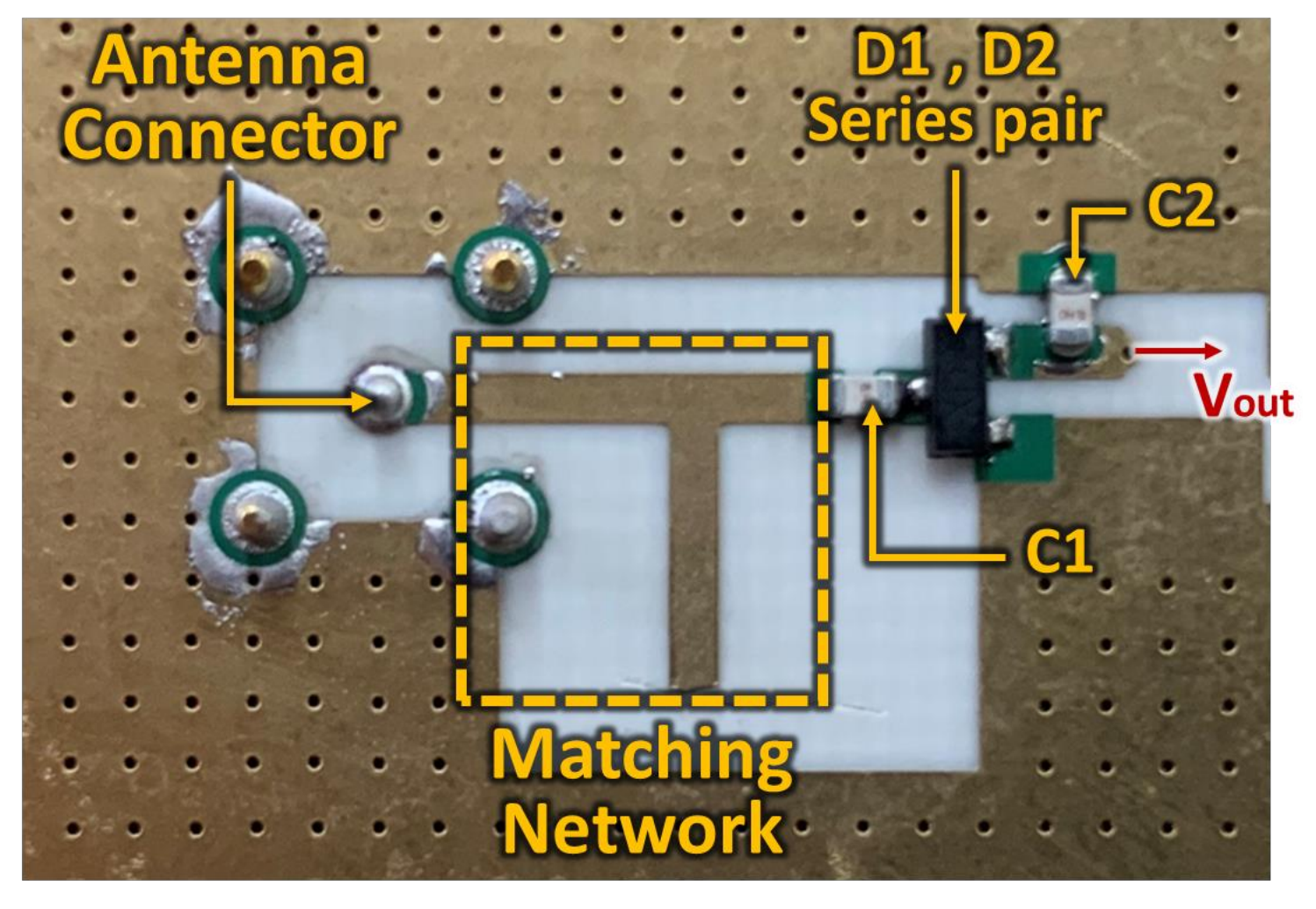}
        }\\
    \caption{Rectifier design}
    \label{fig:rectifier}
\end{figure}

Fig.~\ref{fig:rectifier diagram} shows a circuit diagram of the one-stage Dickson charge pump rectifier circuit that we have designed and fabricated. Since the Dickson charge pump structure has two Schottky diodes in series, the output voltage reaches almost two times the single shunt diode rectifier's output voltage \cite{Bae:2017}.
We have matched the input impedance at the input power of 5 mW and the load of 1 k$\Omega$. In addition, we have used ADS to derive the optimal rectifier performance, including input impedance. The two diodes in the rectifier schematic are integrated as series pairs in one diode package, Skywork SMS7630-005.
The parameters for the capacitors and resistors in
Fig.~\ref{fig:rectifier diagram} are $C_1$ = 2 pF, $C_2$ = 2 pF, and $R_L$ = 1 k$\Omega$.
 In Fig.~\ref{fig:rectifier board}, we show the photograph
of the fabricated rectifier circuit as well. The substrate material is Rogers 4003, with a thickness of 0.508 mm and permittivity of 3.55 F/m.

\begin{figure}
    \centering
        \subfigure[Output voltage]{
        \label{fig:voutrect}\includegraphics[width=0.46\linewidth] {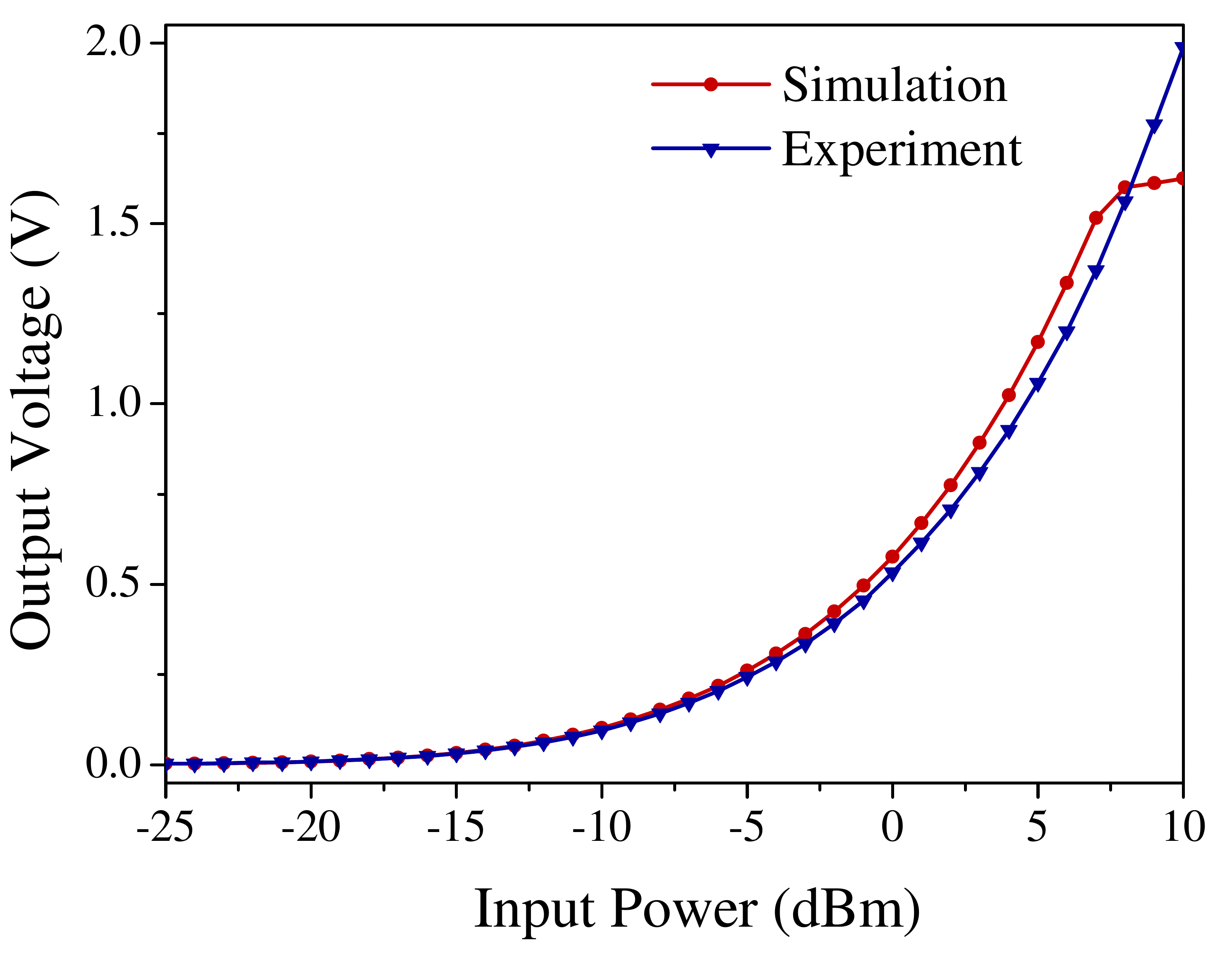}
        }
    \subfigure[Efficiency]{
        \label{fig:effrect}\includegraphics[width=0.46\linewidth] {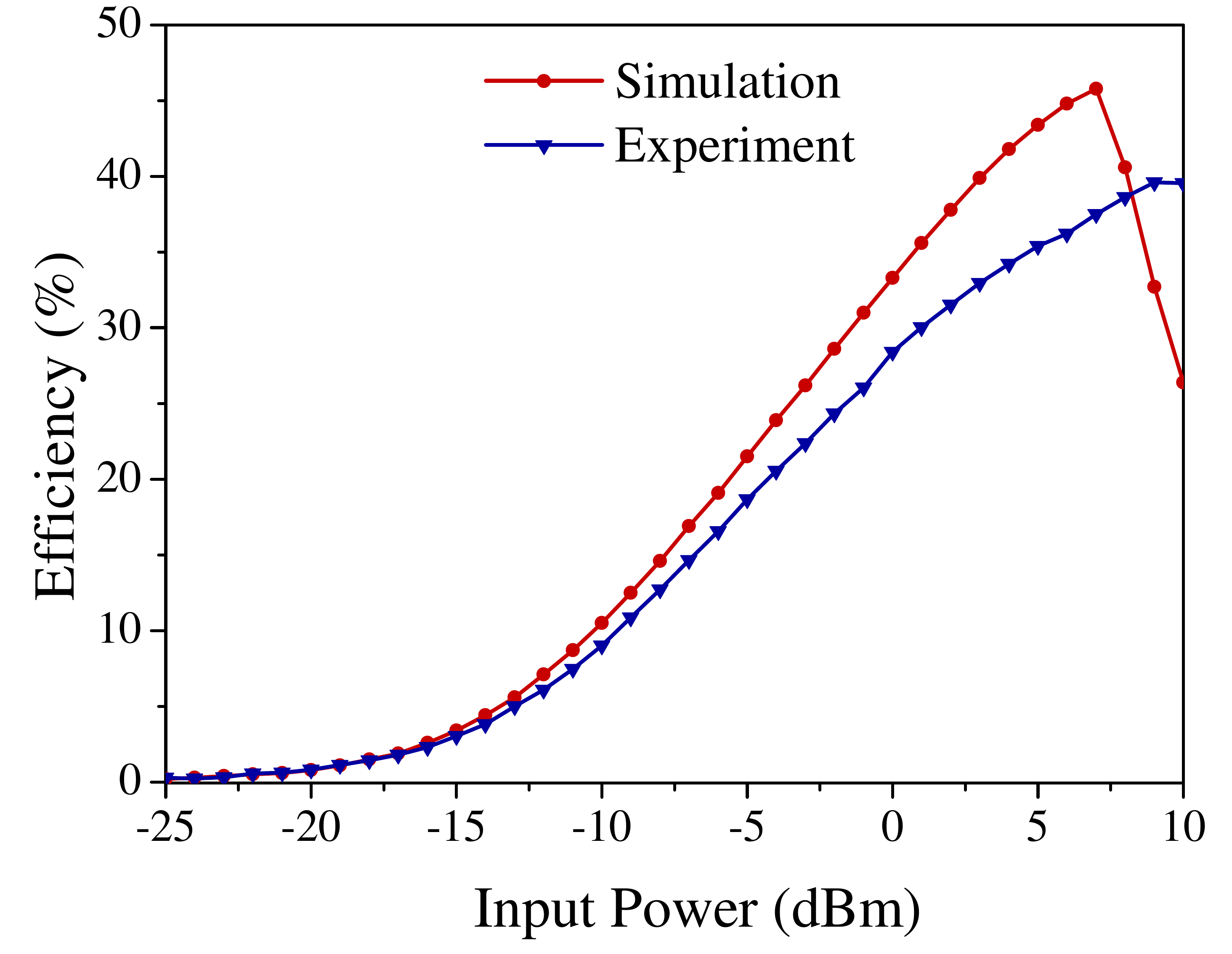}
        }\\
    \caption{Simulated and measured results}
    \label{fig:efficiency}
\end{figure}

\begin{figure}
    \centering
        \subfigure[Front view]{
        \label{fig:receiver front}\includegraphics[width=0.3\linewidth] {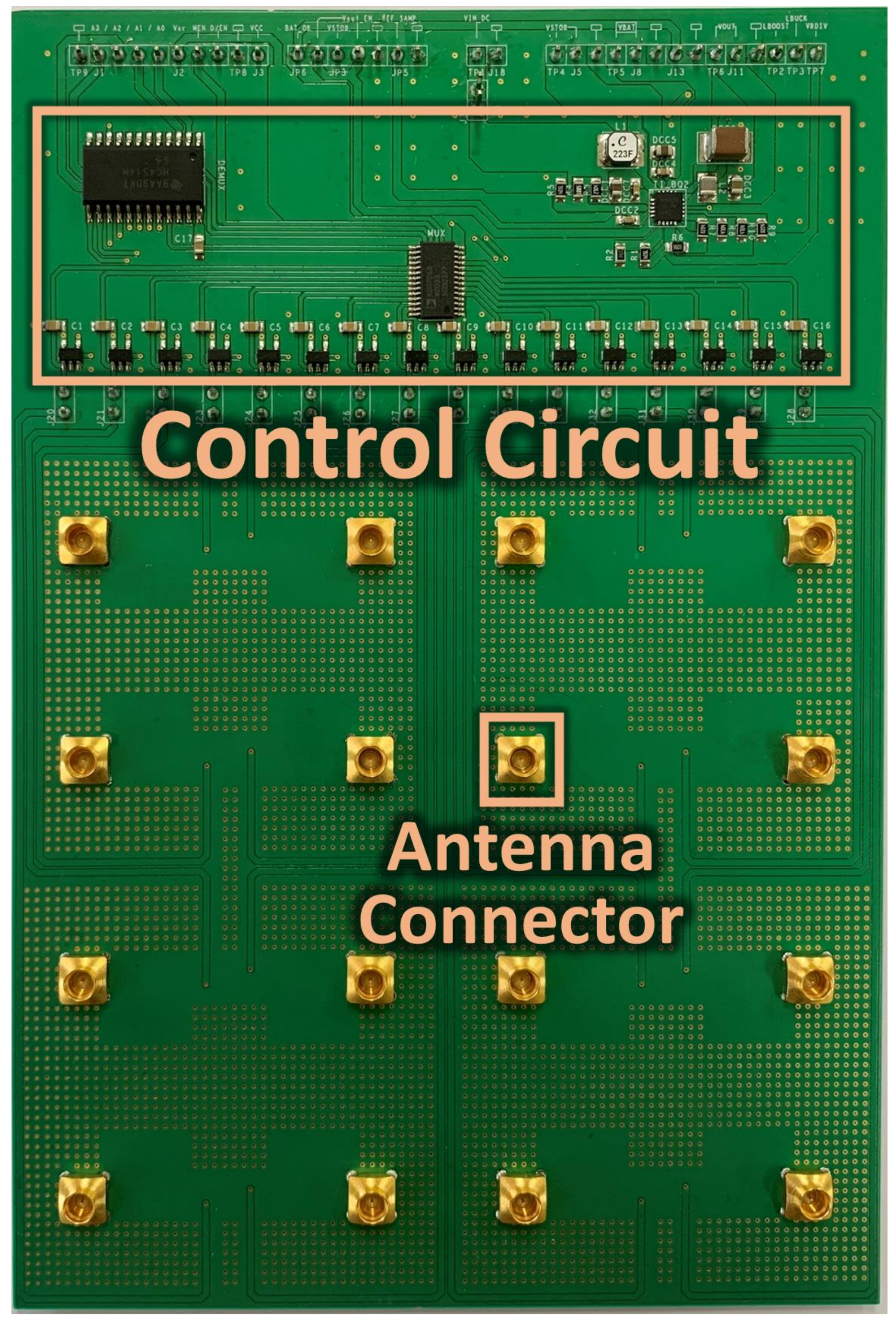}
        }
    \subfigure[Back view]{
        \label{fig:receiver back}\includegraphics[width=0.3\linewidth] {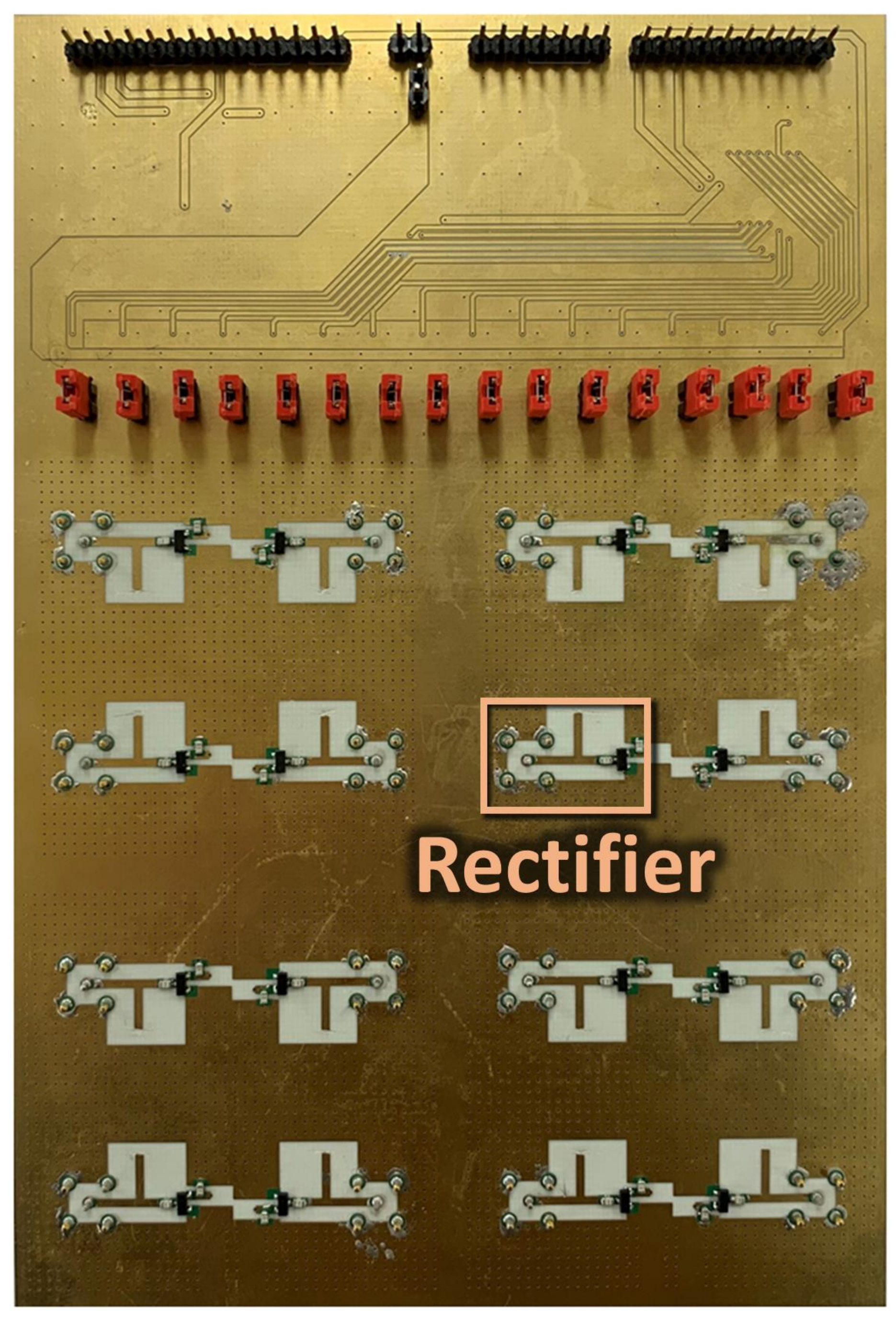}
        }\\
    \caption{Rectenna array receiver}
    \label{fig:receiver board}
\end{figure}

Based on the designed rectifier, we have built a 4-by-4 rectifier array receiver as seen in Fig.~\ref{fig:receiver front} and Fig.~\ref{fig:receiver back}. We can control the ON/OFF states of the rectifier by controlling the deployed analog switch of each path. We can select one path with a multiplexer and decoder to change ON/OFF states. When the switch is in the OFF state, we can measure the open-circuit voltage  ${V_\text{oc}}$ of a rectifier which is used as an indicator of the received power.
On the other hand, in the ON state, the converted dc power is delivered to dc power combining circuit.
In this system, we used the dc combining technique which combines the dc outputs of the 16 rectifiers in parallel.

\begin{table}
\centering
\caption{Selected Components for System}
\label{component table}
\scriptsize
\begin{tabular}{c|cc}
\hline\hline
System Part & Component & Part Number\\
\hline
\multirow{10}{*}{Transmitter}
& Phase Shifter & HMC1133LP5E\\
& RF Switch & QPC6014\\
& Shift Register & SN74HC595B\\
& RF Amplifier & HMC415LP3ETR\\
& RF Connector & SMP-MSSB-PCT\\
& DC/DC Converter (12V) & LM2576-12.0\\
& DC/DC Converter (5V) & LM2576-5.0\\
& DC/DC Converter (3.3V) & TPS56637\\
& DC/DC Converter (-5V) & TPS63710\\
& Board-to-Board RF Connector & SMP-FSBA-645\\
\hline
\multirow{6}{*}{Receiver}
& Diode & SMS7630-005\\
& Capacitor & 600S2R0AT250XT\\
& Switch & TS5A3167\\
& Multiplexer & ADG706BRUZ\\
& Decoder & CD74HC4514M\\
\hline
\hline
\end{tabular}
\end{table}

\section{Experimental Results}\label{section:Experiments}

\begin{figure}
    \centering
    \subfigure[Transmitter and receiver]{
        \label{fig:wpt system}\includegraphics[width=0.95\linewidth] {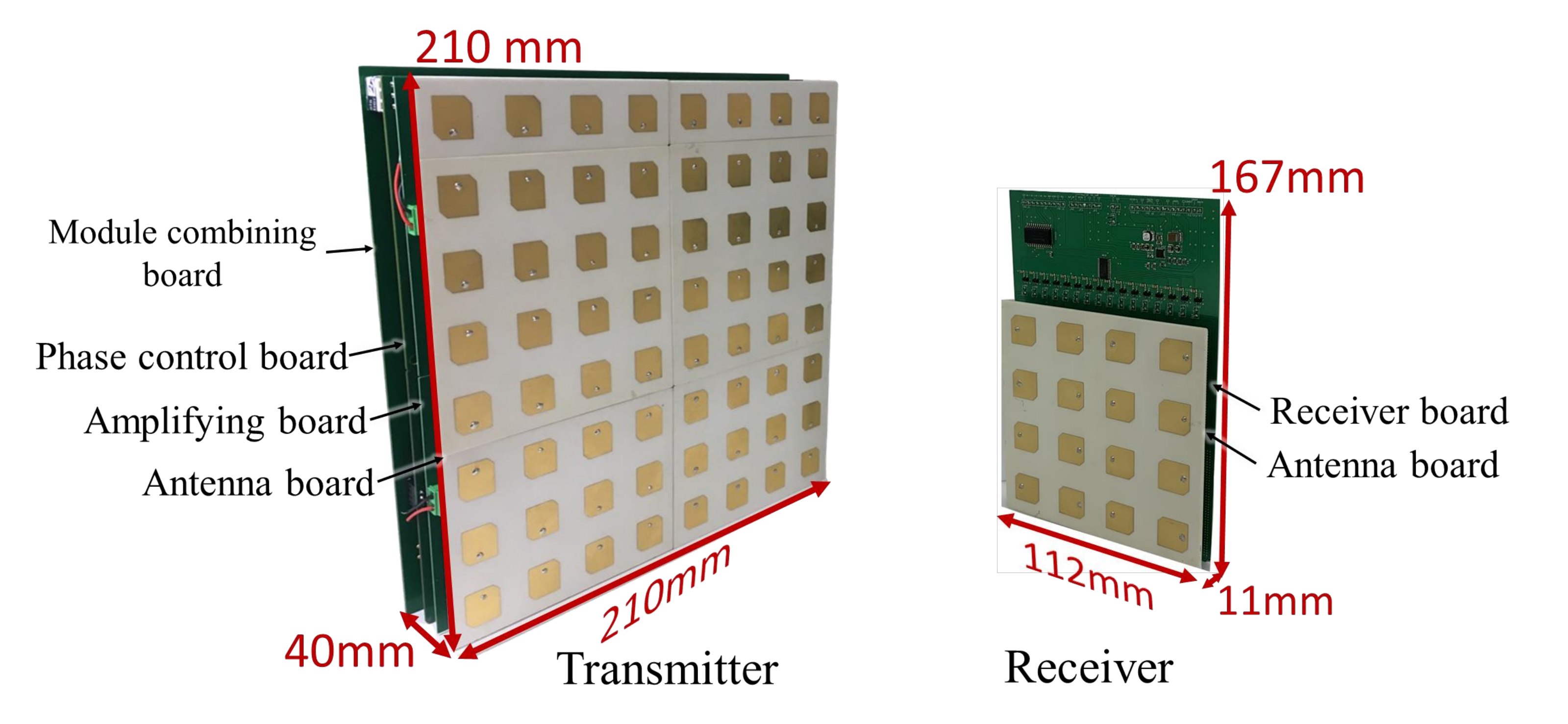}
        }\\
    \subfigure[Testbed setup]{
        \label{fig:testbed setup}\includegraphics[width=0.95\linewidth] {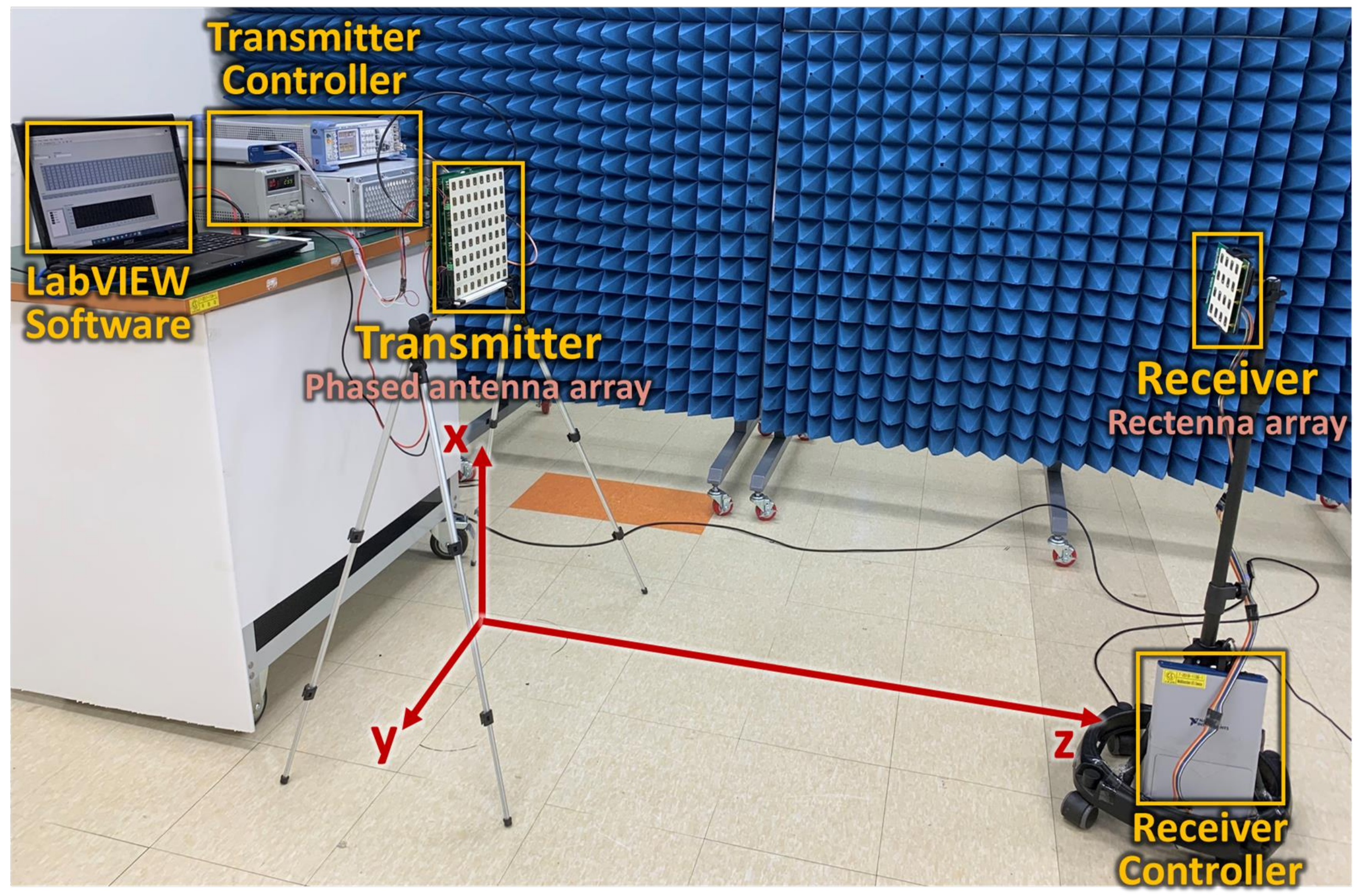}
        }\\
    \caption{Experimental setup}
    \label{fig:tx and rx}
\end{figure}

In this section, we present the experimental results that we have conducted to verify the performance of the proposed WPT system. For the experiments, 8-by-8 transmit antenna array and 4-by-4 receive rectenna array are built by combining and stacking up each component which is introduced in Section~\ref{section:system}. The final prototype is shown in Fig.~\ref{fig:wpt system} with dimensions.
The experimental setup is shown in Fig.~\ref{fig:testbed setup}. 
We have used two data acquisition devices (NI USB-6351) for controlling and measuring the implemented system. On the transmitter side, the continuous RF signal is provided by a microwave signal generator (R{\&}S SMB 100A) and amplified by an RF amplifier (Ophir 5291). The overall system is controlled by LabVIEW software in the PC.

The scanning beams over the u-v coordinate of the transmitter are generated according to \eqref{eq:Txscanbeam}. Therefore, 256 scanning beams are generated for the $8\times8$ phased array transmitter.
The total size of the transmit antenna array is 210.04 mm $\times$ 210.99 mm as seen in Fig.~\ref{fig:wpt system} and the maximum linear dimension of which is 299.12 mm. 
Based on \eqref{eq:rb}, the radiative near-field region is up to 3.45 meters ($r_\text{b}$ = 3.45). In experiments, we set the number of the grid for the distance $\Upsilon^\text{d}$ to 35 so that the step size is 0.1 meters.

For the validation of the proposed scanning algorithm, we have conducted experiments to measure the received power and obtained the power transfer efficiency over distances. In this real test scenario, the transmitter is located on x-y plane, and the receiver directly faces the transmitter.

\begin{figure}
\centering
        \includegraphics[width=\linewidth] {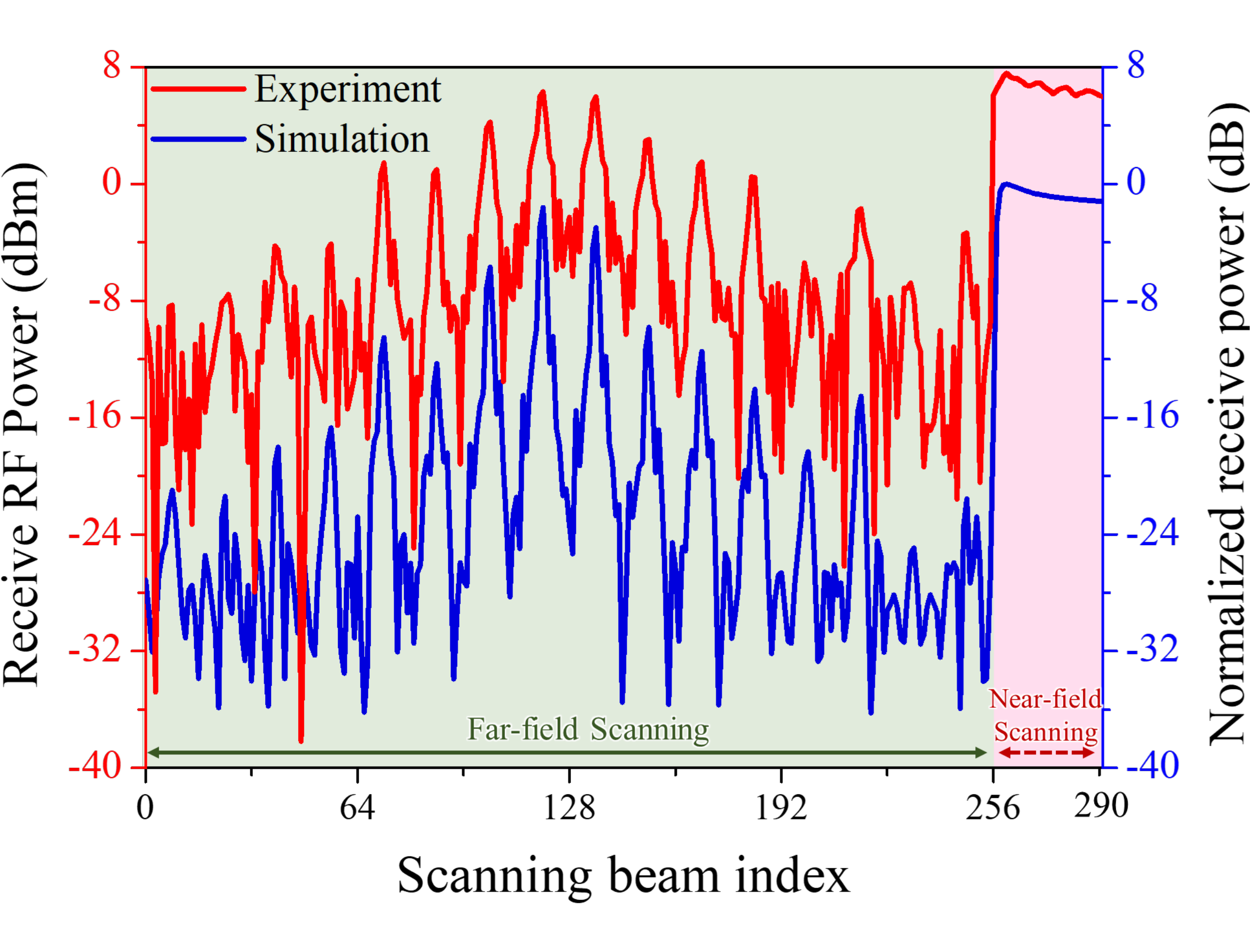}
        \caption{Receive power according to scanning beams}
        \label{fig:scanpat_exp}
\end{figure}

Fig.~\ref{fig:scanpat_exp} presents the received power according to each scanning beam when the receiver is located  0.5 meters away from the transmitter. A very good agreement between the simulation and experimental results is observed. The optimal received power is measured at the 120th scanning beam index during the far-field scanning phase for both results.
There are around 1.61 dB and 1.26 dB improvements, respectively, for the simulation and experiment in the received power at the sensor antenna. And the scanning beam index with the highest receive power is the 260th one which corresponds to the 0.5-meter distance. 

\begin{figure}
    \centering
    \subfigure[Far-field scanning]{
        \label{fig:Txphase_0.5m_far_exp}\includegraphics[width=0.45\linewidth] {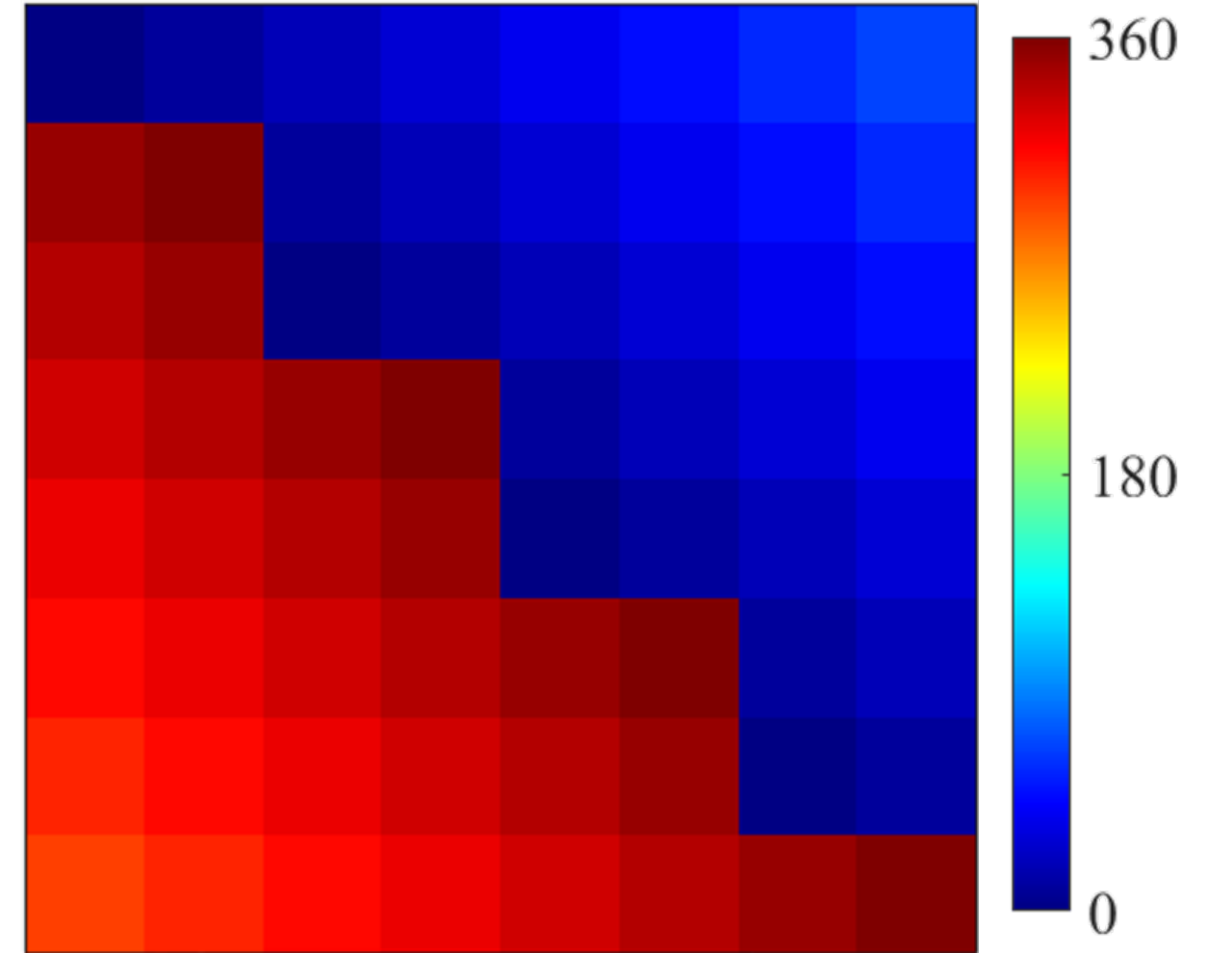}
    }
        \subfigure[Near-field scanning]{
        \label{fig:Txphase_0.5m_near_exp}\includegraphics[width=0.45\linewidth] {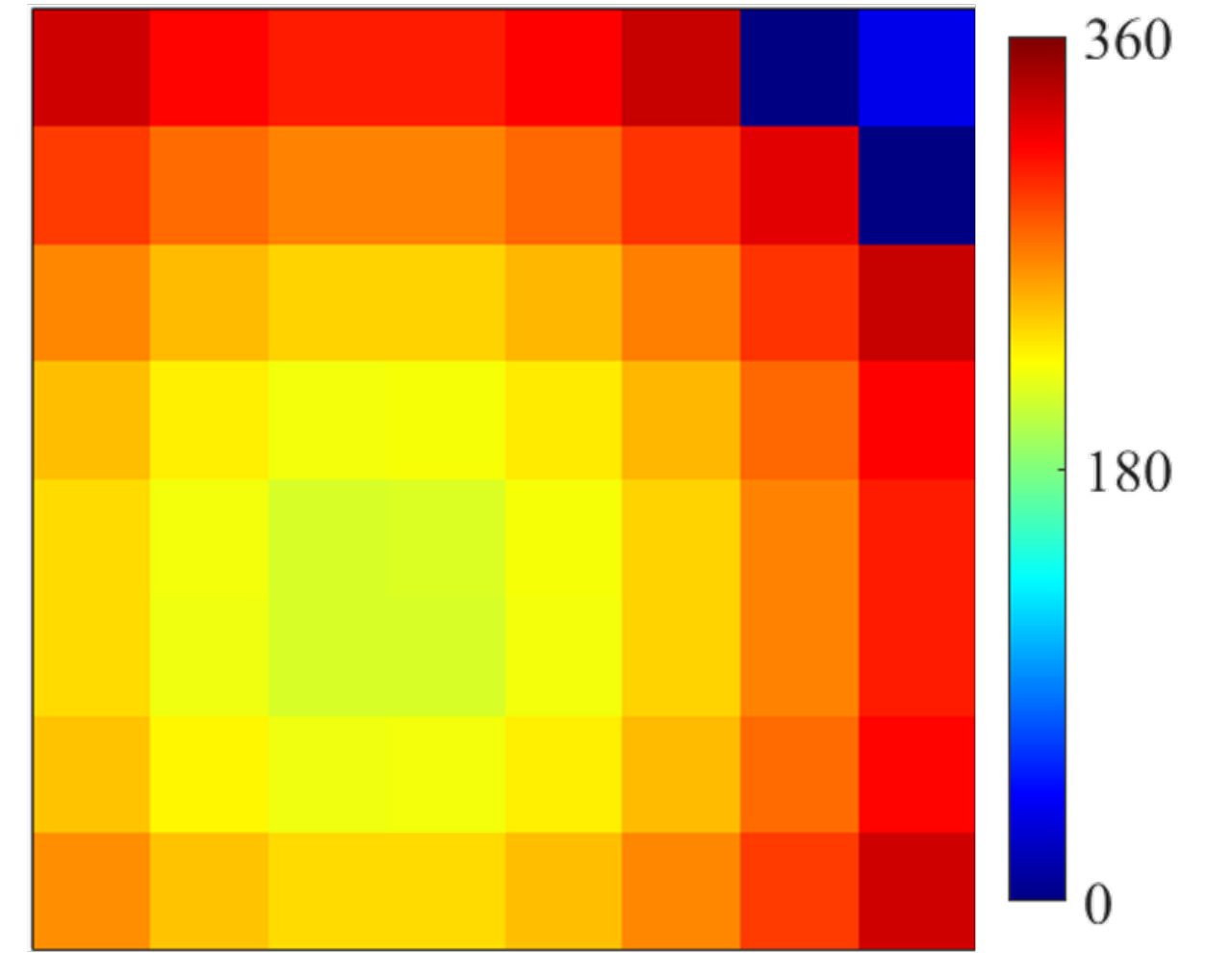}
    }\\
    \caption{Transmit antenna array optimal phase distribution in experiment}
    \label{fig:txphase_exp}
\end{figure}

Since we use an 8-by-8 transmit antenna in the real test, which is relatively small compared to the simulation scenario with a 16-by-16 transmitter array in Fig.~\ref{fig:scanpat}, the enhancement of the received power is not as great as the simulation result as seen in Fig.~\ref{fig:scanpat}. Besides, the phase shift is discrete in the experiment since the phase shifter has 5.625 degrees of LSB, which may have an impact on the degradation of the performance.
In Figs.~\ref{fig:Txphase_0.5m_far_exp} and~\ref{fig:Txphase_0.5m_near_exp}, we can see the optimal phases of the transmit array weights during the far-field scanning and near-field scanning, respectively.

\begin{figure}
    \centering
    \subfigure[Distance along z-axis]{
        \label{fig:eff_dist}\includegraphics[width=0.8\linewidth] {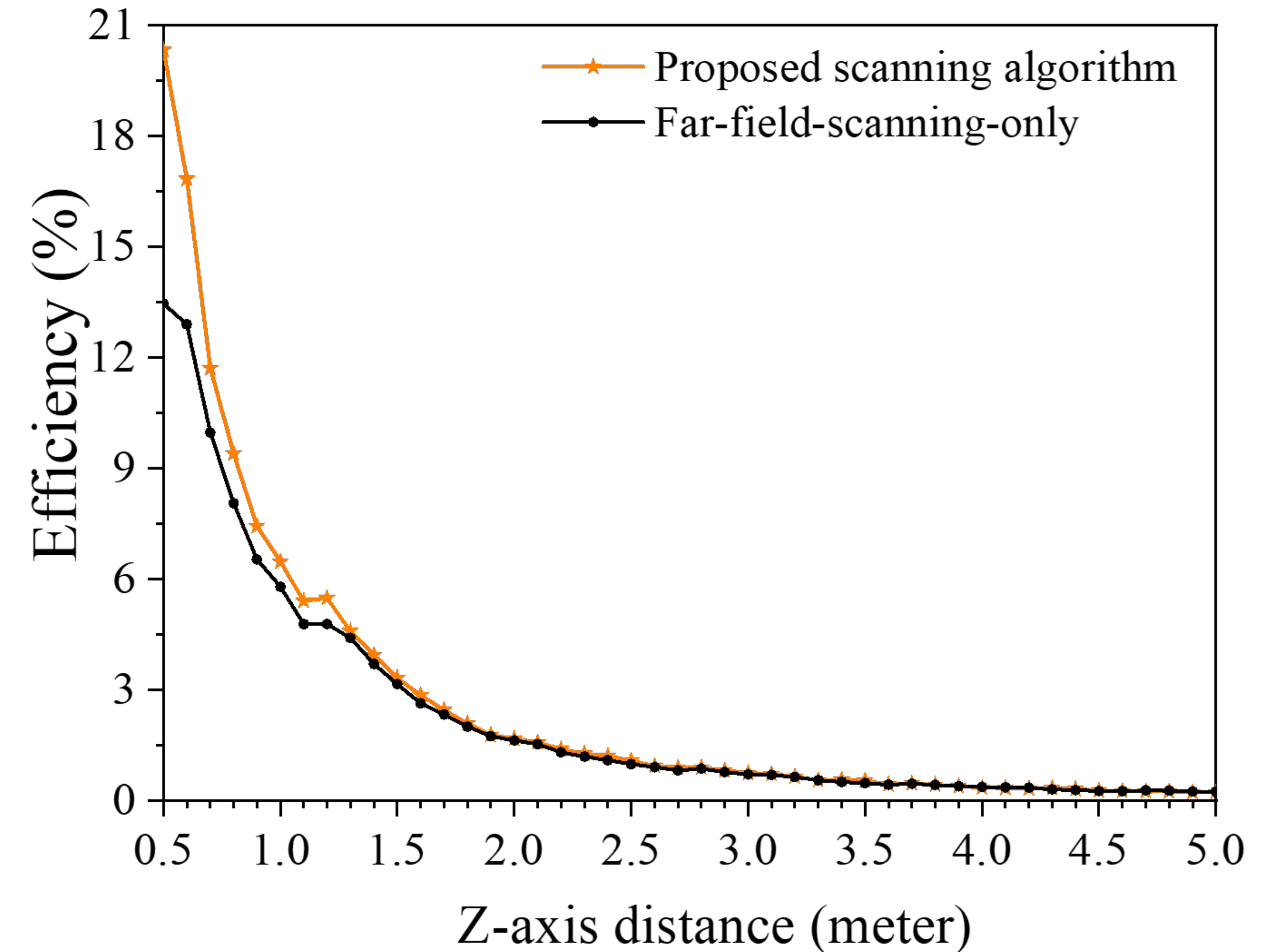}
    }\\
    \subfigure[Distance along y-axis]{
        \label{fig:offset_eff}\includegraphics[width=0.8\linewidth] {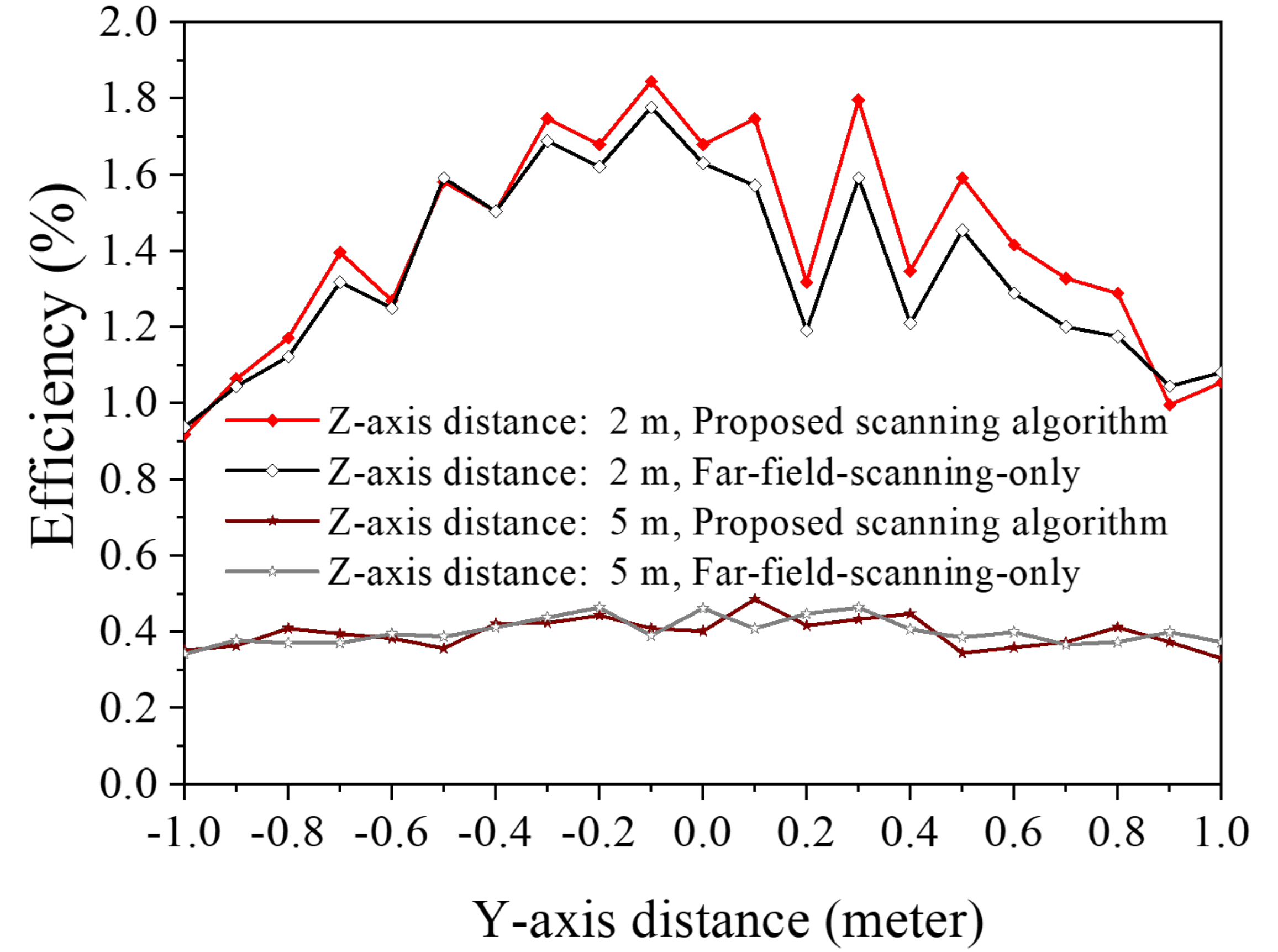}
    }
    \caption{Power transfer efficiency according to the distance}
    \label{fig:efficiency}
\end{figure}

We have conducted experiments to obtain the received power and the power transfer efficiency according to the distance between the transmitter
and receiver with the proposed beam scanning algorithm. Fig.~\ref{fig:eff_dist} shows the experimental results of a power transfer test in which the distance varies from 0.5 to 5 meters. In Fig.~\ref{fig:eff_dist}, we can see that the power transfer efficiency reaches up to 21.24 percent at the 0.5-meter distance with the proposed scanning algorithm, and gradually decreases with the distance. 
The efficiency is calculated with the combined DC power from every rectenna of the receiver.
This efficiency is comparable to that presented in \cite{Park:2021} with the consideration of rectifier efficiency. To prove the effectiveness of the proposed algorithm, which includes near-field scanning iteration, we compared the results to the far-field-scanning-only scheme that scans only over the u-v grid. 

In Fig.~\ref{fig:eff_dist}, we can see that the proposed scanning algorithm achieves much higher power transfer efficiency compared to the far-field-scanning-only scheme (i.e., 6.86 percent difference at a distance of 0.5 meters) when the receiver is located relatively close to the transmitter. This experimental results prove that by conducting the near-field scanning with the proposed algorithm, the power transfer efficiency can be greatly enhanced in the near-field region.
We can expect that much clearer power transfer efficiency difference between the two schemes can be achieved within the near-field region by increasing the number of transmit antenna elements.

Fig.~\ref{fig:offset_eff} describes the power transfer efficiency according to the offset of the receiver from the centerline. When the distance between the transmitter and receiver is 2 meters, which is in the radiative near-field region, we can see that the proposed scanning algorithm reaches significantly higher efficiency from -0.8 to 0.8 meters.

In Fig.~\ref{fig:offset_eff}, for the 5-meter scenario which is considered as the far-field region, we arbitrarily set the parameters $r_\text{b}$ and $\Upsilon^\text{d}$ to 5 and 50, respectively, in \eqref{eq:rgrid}. 
Then, for this scenario, the near-field scanning iteration is carried out from 0.1 to 5 meters with 0.1-meter step size. The reason why we slightly abuse the algorithm for this scenario is to verify that just extending the scanning distance over the radiative near-field region does not necessarily result in the higher transfer efficiency. We can see that there is no remarkable improvement and effectiveness between the two schemes in the far-field scenario. By setting the maximum distance $r_\text{b}$ based on \eqref{eq:rb}, we can reduce the radio resource and operation time.

\begin{figure}
    \centering
    \subfigure[Power transfer with the different number of transmit antenna]{
        \label{fig:Tx_num}\includegraphics[width=\linewidth] {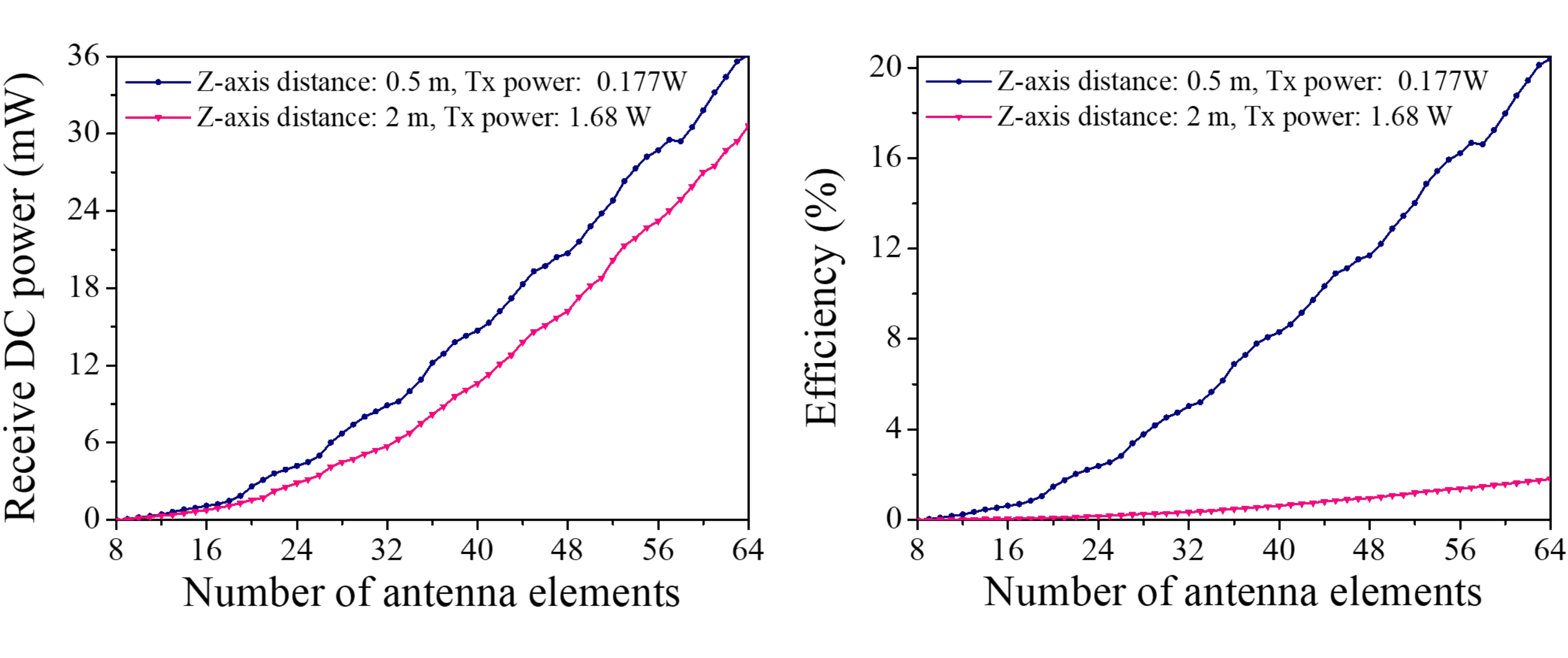}
    }\\
    \subfigure[Power transfer with the different number of receive antenna]{
        \label{fig:Rx_num}\includegraphics[width=\linewidth] {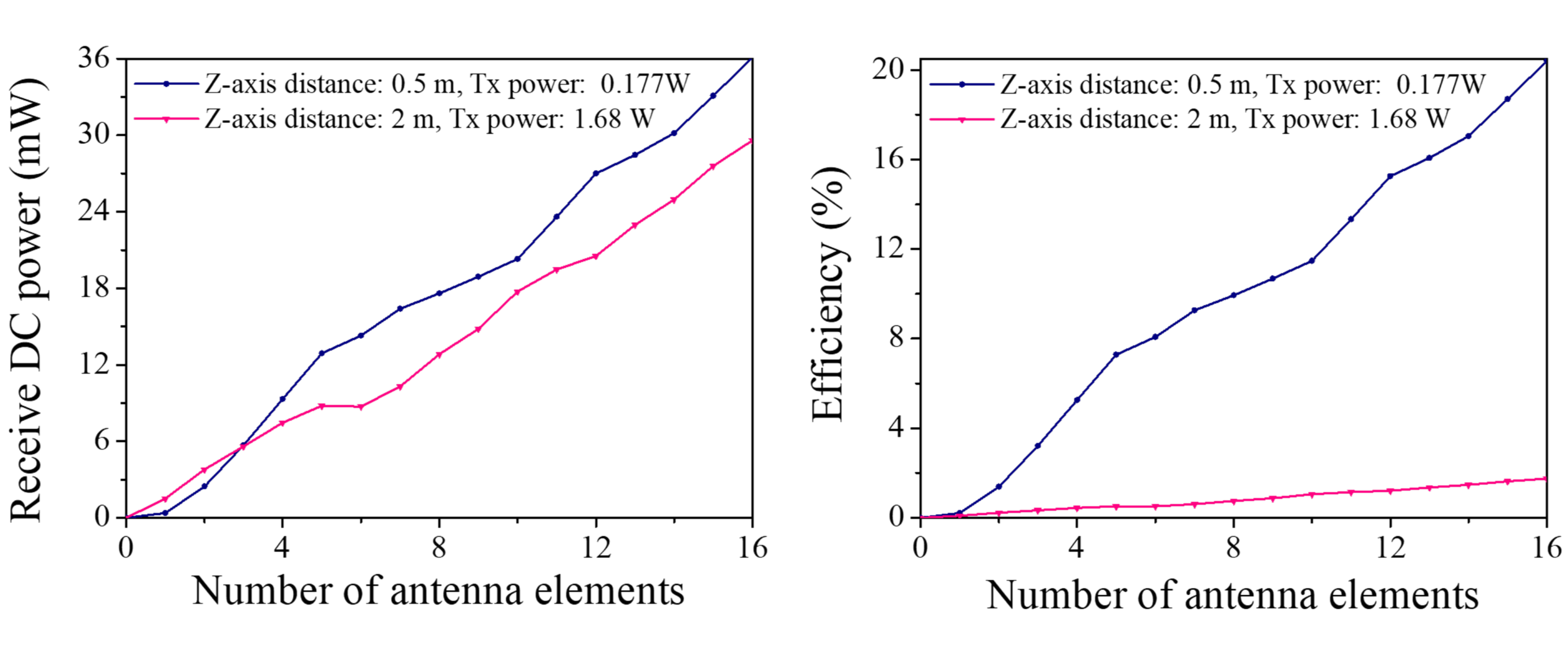}
    }
    \caption{Effect of antenna number to power transfer efficiency}
    \label{fig:ant_num}
\end{figure}

Fig.~\ref{fig:ant_num} shows the effect of the number of transmit and receive antenna elements on the received power and the power
transfer efficiency. 
We have conducted a wireless power transfer test by running the proposed scanning algorithm at 0.5-meter and 2-meter distances with different transmit powers 0.177 W and 1.68 W, respectively. 
For the results in Fig.~\ref{fig:Tx_num}, we have
activated each transmit antenna element one by one by turning on the switch of each path from the center of the antenna array. We can see the received power and power transfer efficiency linearly increase according to the number of transmit antenna elements.
In Fig.~\ref{fig:Rx_num}, we can see that the received power and power transfer efficiency both increase with the number of receive antenna elements.

\begin{figure}
    \centering
    \subfigure[Test environment A]{
        \label{fig:Exp_A}\includegraphics[width=0.45\linewidth] {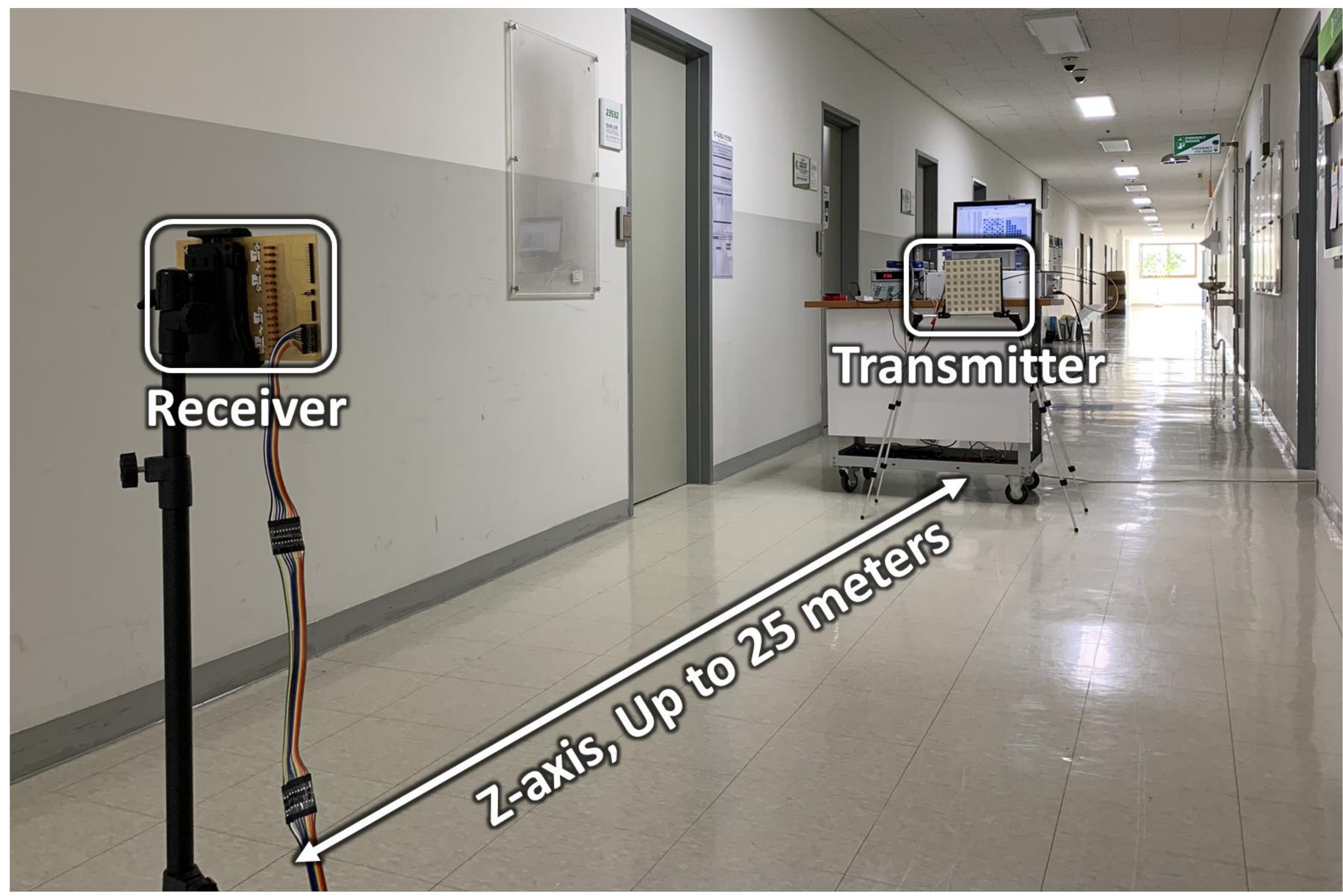}
    }
    \subfigure[Test environment B]{
        \label{fig:Exp_B}\includegraphics[width=0.45\linewidth] {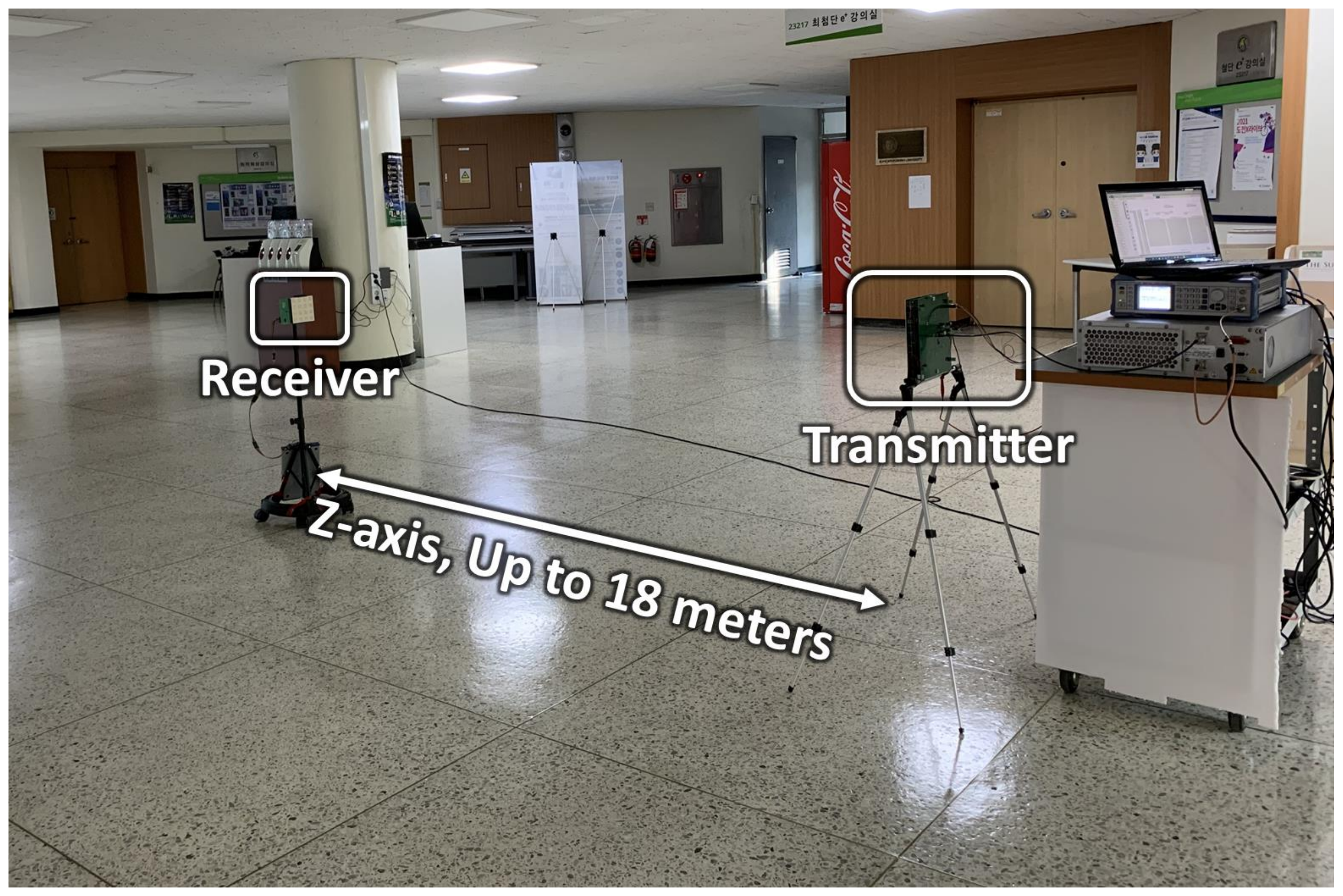}
    }\\
    \subfigure[Receive power and power transfer efficiency over distance]{
        \label{fig:Exp_outdoor}\includegraphics[width=0.9\linewidth] {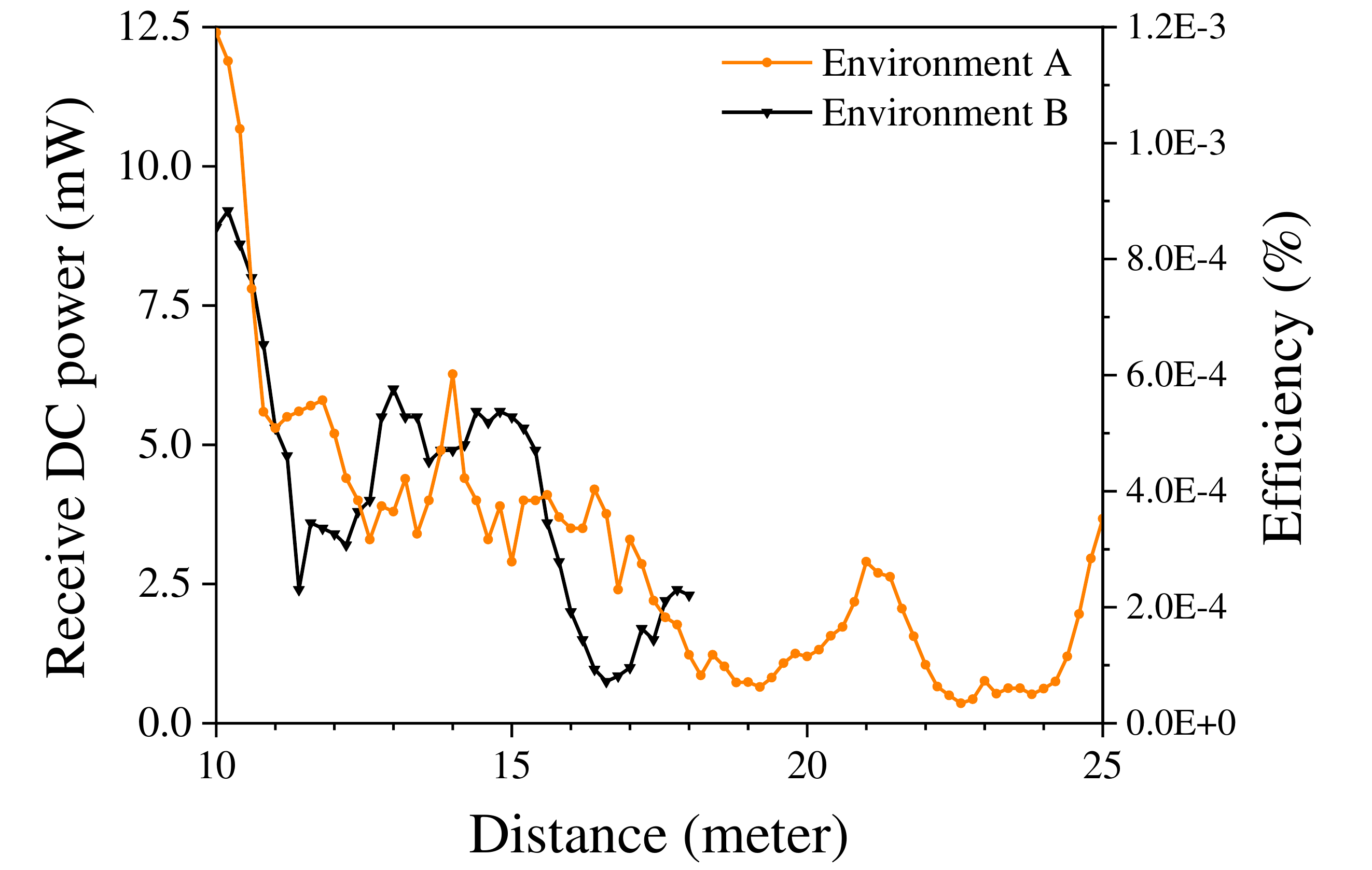}
    }
    \caption{Power transfer test}
    \label{fig:outdoor_exp}
\end{figure}

In Fig.~\ref{fig:outdoor_exp}, we show the power transfer test results for which the distance varies from 10 to 25 meters. The test environment is shown in Figs.~\ref{fig:Exp_A} and \ref{fig:Exp_B}. In this test, we use the maximum power of the implemented transmitter, which is 10.33 W. In the test environment A, due to the limited space, we have only conducted the test up to 18 meters and the received dc power of 2.3 mW was reported. The test in environment B is conducted along the corridor, which is narrow compared to environment A. At a distance of 25 meters, around 3.7 mW was reported at the receiver.

\section{Conclusion}\label{section:Conclusion}
In this paper, we have designed and implemented a 5.8 GHz RF WPT system. The transmitter of the proposed RF WPT system consists of an 8-by-8 phased antenna array, and the receiver is comprised of a 4-by-4 rectenna array. We also proposed a beam scanning algorithm that is capable of covering the radiative near-field, unlike the conventional codebook-based far-field beam scanning schemes. The proposed beam scanning algorithm is verified with the experimented results. Implemented RF WPT system can be used for charging various types of IoT devices wirelessly, which is located far away from the power transmitter.

\bibliographystyle{IEEEtran}
\bibliography{IEEEabrv,reference}

\begin{thebibliography}{10}
\providecommand{\url}[1]{#1}
\csname url@samestyle\endcsname
\providecommand{\newblock}{\relax}
\providecommand{\bibinfo}[2]{#2}
\providecommand{\BIBentrySTDinterwordspacing}{\spaceskip=0pt\relax}
\providecommand{\BIBentryALTinterwordstretchfactor}{4}
\providecommand{\BIBentryALTinterwordspacing}{\spaceskip=\fontdimen2\font plus
\BIBentryALTinterwordstretchfactor\fontdimen3\font minus
  \fontdimen4\font\relax}
\providecommand{\BIBforeignlanguage}[2]{{%
\expandafter\ifx\csname l@#1\endcsname\relax
\typeout{** WARNING: IEEEtran.bst: No hyphenation pattern has been}%
\typeout{** loaded for the language `#1'. Using the pattern for}%
\typeout{** the default language instead.}%
\else
\language=\csname l@#1\endcsname
\fi
#2}}
\providecommand{\BIBdecl}{\relax}
\BIBdecl

\bibitem{Hui:2014}
S.~Y.~R. Hui, W.~Zhong, and C.~K. Lee, ``A critical review of recent progress
  in mid-range wireless power transfer,'' \emph{{IEEE} Trans. Power Electron.},
  vol.~29, no.~9, pp. 4500--4511, Sep. 2014.

\bibitem{Clerckx:2018}
B.~Clerckx, R.~Zhang, R.~Schober, D.~W.~K. Ng, D.~I. Kim, and H.~V. Poor,
  ``Fundamentals of wireless information and power transfer: From {RF} energy
  harvester models to signal and system designs,'' \emph{{IEEE} J. Sel. Areas
  Commun.}, vol.~37, no.~1, pp. 4--33, Jan. 2018.

\bibitem{Choi:2020}
K.~W. {Choi}, S.~I. {Hwang}, A.~A. {Aziz}, H.~H. {Jang}, J.~S. {Kim}, D.~S.
  {Kang}, and D.~I. {Kim}, ``Simultaneous wireless information and power
  transfer {(SWIPT)} for internet of things: Novel receiver design and
  experimental validation,'' \emph{{IEEE} Internet Things J.}, vol.~7, no.~4,
  pp. 2996--3012, Apr. 2020.

\bibitem{Huang:2015}
K.~Huang and X.~Zhou, ``Cutting the last wires for mobile communications by
  microwave power transfer,'' \emph{{IEEE} Commun. Mag.}, vol.~53, no.~6, pp.
  86--93, Jun. 2015.

\bibitem{Shinohara:2021}
N.~Shinohara, ``History and innovation of wireless power transfer via
  microwaves,'' \emph{IEEE Journal of Microwaves}, vol.~1, no.~1, pp. 218--228,
  2021.

\bibitem{khan:2020}
D.~Khan, M.~Basim, I.~Ali, Y.~Pu, K.~C. Hwang, Y.~Yang, D.~I. Kim, and K.-Y.
  Lee, ``A survey on {RF} energy harvesting system with high efficiency {RF-DC}
  converters,'' \emph{Journal of Semiconductor Engineering}, vol.~1, no.~1, pp.
  13--30, Jun. 2020.

\bibitem{Setiawan:2017}
D.~Setiawan, A.~A. Aziz, D.~I. Kim, and K.~W. Choi, ``Experiment, modeling, and
  analysis of wireless-powered sensor network for energy neutral power
  management,'' \emph{{IEEE} Syst. J.}, vol.~12, no.~4, pp. 3381--3392, Dec.
  2017.

\bibitem{Shen:2020}
S.~Shen, J.~Kim, C.~Song, and B.~Clerckx, ``Wireless power transfer with
  distributed antennas: System design, prototype, and experiments,''
  \emph{{IEEE} Trans. Ind. Electron.}, vol.~68, no.~11, pp. 10\,868--10\,878,
  Nov. 2020.

\bibitem{Kim:2017}
J.~Kim, B.~Clerckx, and P.~D. Mitcheson, ``Prototyping and experimentation of a
  closed-loop wireless power transmission with channel acquisition and waveform
  optimization,'' in \emph{Proc. IEEE Wireless Power Transf. Conf. (WPTC)},
  Taipei, Taiwan, May 2017.

\bibitem{Clerckx:2018_2}
B.~Clerckx and J.~Kim, ``On the beneficial roles of fading and transmit
  diversity in wireless power transfer with nonlinear energy harvesting,''
  \emph{{IEEE} Trans. Wireless Commun.}, vol.~17, no.~11, pp. 7731--7743, Nov.
  2018.

\bibitem{Hui:2019}
Q.~Hui, K.~Jin, and X.~Zhu, ``Directional radiation technique for maximum
  receiving power in microwave power transmission system,'' \emph{{IEEE} Trans.
  Ind. Electron.}, vol.~67, no.~8, pp. 6376--6386, Aug. 2019.

\bibitem{Choi:2019}
K.~W. Choi, L.~Ginting, A.~A. Aziz, D.~Setiawan, J.~H. Park, S.~I. Hwang, D.~S.
  Kang, M.~Y. Chung, and D.~I. Kim, ``Toward realization of long-range
  wireless-powered sensor networks,'' \emph{{IEEE} Wireless Commun.}, vol.~26,
  no.~4, pp. 184--192, Aug. 2019.

\bibitem{Koo:2020}
H.~Koo, J.~Bae, W.~Choi, H.~Oh, H.~Lim, J.~Lee, C.~Song, K.~Lee, K.~Hwang, and
  Y.~Yang, ``Retroreflective transceiver array using a novel calibration method
  based on optimum phase searching,'' \emph{{IEEE} Trans. Ind. Electron.},
  vol.~68, no.~3, pp. 2510--2520, Mar. 2021.

\bibitem{Bae:2020}
J.~Bae, S.-H. Yi, H.~Koo, S.~Oh, H.~Oh, W.~Choi, J.~Shin, C.~M. Song, K.~C.
  Hwang, K.-Y. Lee, and Y.~Yang, ``{LUT}-based focal beamforming system using
  2-{D} adaptive sequential searching algorithm for microwave power transfer,''
  \emph{{IEEE} Access}, vol.~8, pp. 196\,024--196\,033, Nov. 2020.

\bibitem{Pabbisetty:2019}
G.~Pabbisetty, K.~Murata, K.~Taniguchi, T.~Mitomo, and H.~Mori, ``Evaluation of
  space time beamforming algorithm to realize maintenance-free {IoT} sensors
  with wireless power transfer system in 5.7-{GHz} band,'' \emph{{IEEE} Trans.
  Microw. Theory Techn.}, vol.~67, no.~12, pp. 5228--5234, Dec. 2019.

\bibitem{Arai:2021}
K.~Arai, K.~Wang, M.~Toshiya, M.~Higaki, and K.~Onizuka, ``A tile-based
  8$\times$8 triangular grid array beamformer for 5.7 {GHz} microwave power
  transmission,'' in \emph{Proc. IEEE Radio and Wireless Symp. (RWS)}, SAN,
  United States, Jan. 2021.

\bibitem{Belo:2019}
D.~Belo, D.~C. Ribeiro, P.~Pinho, and N.~B. Carvalho, ``A selective, tracking,
  and power adaptive far-field wireless power transfer system,'' \emph{{IEEE}
  Trans. Microw. Theory Techn.}, vol.~67, no.~9, pp. 3856--3866, Sep. 2019.

\bibitem{Gowda:2016}
V.~R. Gowda, O.~Yurduseven, G.~Lipworth, T.~Zupan, M.~S. Reynolds, and D.~R.
  Smith, ``Wireless power transfer in the radiative near field,'' \emph{{IEEE}
  Antennas Wireless Propag. Lett.}, vol.~15, pp. 1865--1868, Mar. 2016.

\bibitem{Xianjin:2019}
X.~Yi, X.~Chen, L.~Zhou, S.~Hao, B.~Zhang, and X.~Duan, ``A microwave power
  transmission experiment based on the near-field focused transmitter,''
  \emph{{IEEE} Antennas Wireless Propag. Lett.}, vol.~18, no.~6, pp.
  1105--1108, Jun. 2019.

\bibitem{Liu:2018}
Y.~Liu, J.~Bai, K.~D. Xu, Z.~Xu, F.~Han, Q.~H. Liu, and Y.~J. Guo, ``Linearly
  polarized shaped power pattern synthesis with sidelobe and cross-polarization
  control by using semideﬁnite relaxation,'' \emph{{IEEE} Trans. Antennas
  Propag.}, vol.~66, no.~6, pp. 3207--3212, Mar. 2018.

\bibitem{Sun:2018}
G.~Sun, Y.~Liu, Z.~Chen, S.~Liang, A.~Wang, and Y.~Zhang, ``Radiation beam
  pattern synthesis of concentric circular antenna arrays using hybrid approach
  based on cuckoo search,'' \emph{{IEEE} Trans. Antennas Propag.}, vol.~66,
  no.~9, pp. 4563--4576, Jun. 2018.

\bibitem{Aslan:2019}
Y.~Aslan, J.~Puskely, A.~Roederer, and A.~Yarovoy, ``Multiple beam synthesis of
  passively cooled {5G} planar arrays using convex optimization,'' \emph{{IEEE}
  Trans. Antennas Propag.}, vol.~68, no.~5, pp. 3557--3566, May 2020.

\bibitem{Morabito:2012}
A.~F. Morabito, A.~Massa, P.~Rocca, and T.~Isernia, ``An effective approach to
  the synthesis of phase-only reconﬁgurable linear arrays,'' \emph{{IEEE}
  Trans. Antennas Propag.}, vol.~60, no.~8, pp. 3622--3631, May 2012.

\bibitem{Yoon:2021}
H.~S. Yoon, D.~G. Jo, D.~I. Kim, and K.~W. Choi, ``On-{O}ff arbitrary beam
  synthesis and non-interactive beam management for phased antenna array
  communications,'' \emph{{IEEE} Trans. Veh. Technol.}, vol.~70, no.~6, pp.
  5959--5973, Jun. 2021.

\bibitem{Bae:2017}
J.~Bae, H.~Koo, H.~Lee, W.~Lim, W.~Lee, H.~Kang, K.~C. Hwang, K.-Y. Lee, and
  Y.~Yang, ``High-efficiency rectifier {5.2 GHz} using a {C}lass-{F} {D}ickson
  charge pump,'' \emph{Microwave and Optical Technology Letters}, vol.~59,
  no.~12, pp. 3018--3023, Sep. 2017.

\bibitem{Park:2021}
J.~H. Park, D.~I. Kim, and K.~W. Choi, ``Analysis and experiment on
  multi-antenna-to-multi-antenna {RF} wireless power transfer,'' \emph{{IEEE}
  Access}, vol.~9, pp. 2018--2031, Jan. 2021.

\end{thebibliography}

\newpage
\ \
\newpage

\begin{IEEEbiography}[{\includegraphics[width=1in,height=1.25in]{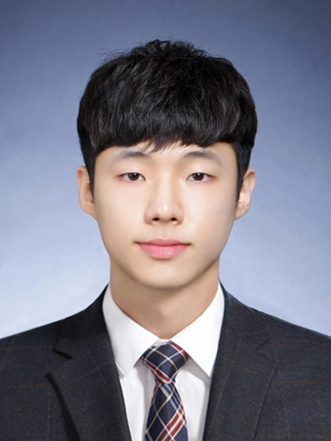}}]
{\textbf{Je Hyeon Park}} received the B.S. degree from the
School of Electronic and Electronic and Electrical
Engineering, Sungkyunkwan University, Suwon,
South Korea, in 2018, where he is currently pursuing
the Ph.D. degree.
His current research interests include radio
frequency circuit design and antenna array signal
processing.
\end{IEEEbiography}

\vskip 0pt plus -1fil

\begin{IEEEbiography}[{\includegraphics[width=1in,height=1.25in]{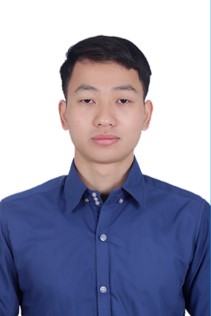}}]
	{\textbf{Nguyen Minh Tran}} received the B.S. and M.S. degrees in electronics and telecommunication technology from the VNU University of Engineering and Technology, Hanoi, Vietnam, in 2014 and 2016, respectively. He is currently pursuing the Ph.D. degree with the Department of Electrical and Computer Engineering, Sungkyunkwan University, Suwon, South Korea. His current research interests include antenna design, reconfigurable intelligent surface, radio frequency (RF) circuit design, and RF wireless power transfer.
\end{IEEEbiography}

\vskip 0pt plus -1fil

\begin{IEEEbiography}[{\includegraphics[width=1in,height=1.25in]{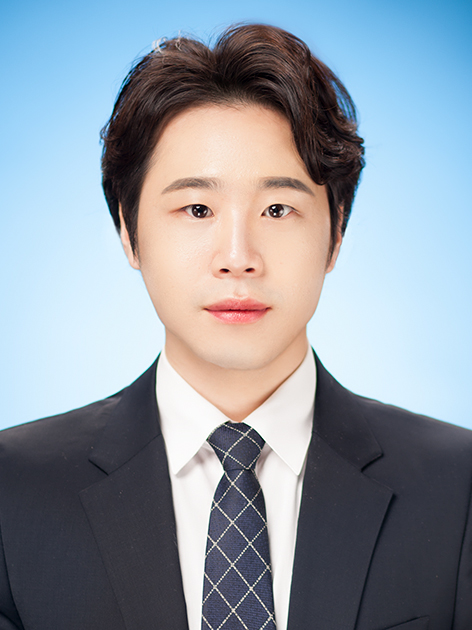}}]
{\textbf{Sa Il Hwang}} received the B.S. degree from the Department of Mechatronics Engineering, Korea Polytechnic University, Siheung, South Korea, in 2018, and the M.S. degree with the College of Information and Communication Engineering, Sungkyunkwan University, Suwon, South Korea, in 2020. Since 2021, he has been with Global Zeus, Hwaseong, South Korea. His research interests include energy harvesting, wireless power transfer and Industrial robot.
\end{IEEEbiography}

\vskip 0pt plus -1fil

\begin{IEEEbiography}[{\includegraphics[width=1in,height=1.25in]{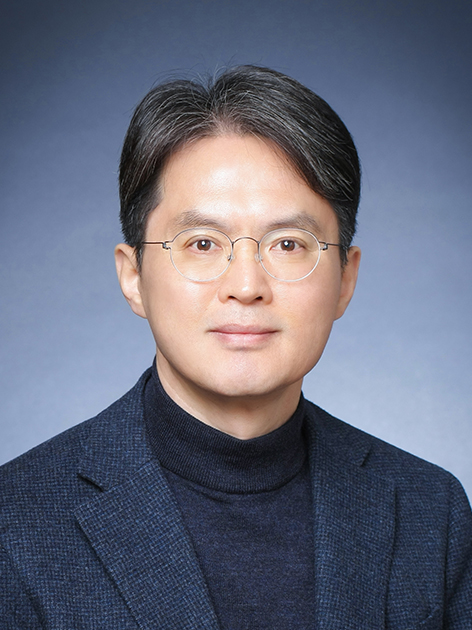}}]
{\textbf{Dong In Kim}} (S’89-M’91-SM’02-F’19) received the Ph.D. degree in electrical engineering from the University of Southern California, Los Angeles, CA, USA, in 1990. He was a tenured Professor with the School of Engineering Science, Simon Fraser University, Burnaby, BC, Canada. Since 2007, he has been an SKKU-Fellowship Professor with the College of Information \& Communication Engineering, Sungkyunkwan University (SKKU), Suwon, South Korea. He is a Fellow of the IEEE, a Fellow of the Korean Academy of Science and Technology, and a member of the National Academy of Engineering of Korea. He has been a first recipient of the NRF of Korea Engineering Research Center in Wireless Communications for RF Energy Harvesting since 2014. He has been listed as a 2020 Highly Cited Researcher by Clarivate Analytics. From 2001 to 2020, he served as an editor and an editor-at-large for Wireless Communication I for the IEEE Transaction on Communications. From 2002 to 2011, he also served as an editor and a Founding Area Editor of Cross-Layer Design and Optimization for the IEEE Transactions on Wireless Communications. From 2008 to 2011, he served as the Co-Editor-in-Chief for the IEEE/KICS Journal on Communications and Networks. He served as the Founding Editor-in-Chief for the IEEE Wireless Communications Letters, from 2012 to 2015. He was selected the 2019 recipient of the IEEE Communications Society Joseph LoCicero Award for Exemplary Service to Publications. He is the Executive Chair for IEEE ICC 2022 in Seoul.
\end{IEEEbiography}

\vskip 0pt plus -1fil

\begin{IEEEbiography}[{\includegraphics[width=1in,height=1.25in]{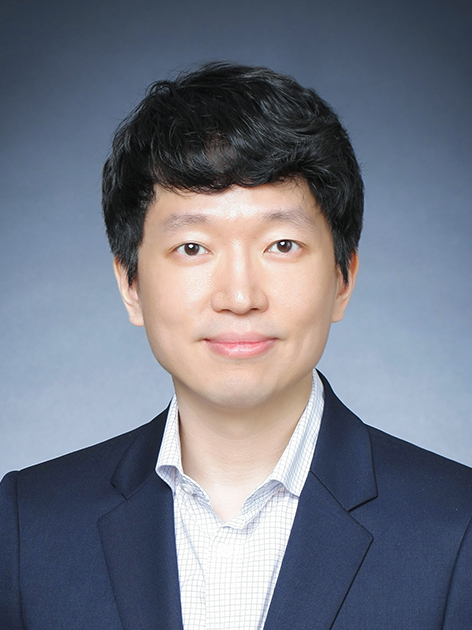}}]
{\textbf{Kae Won Choi}} (M’08-SM’15) received the B.S. degree in Civil, Urban, and Geosystem Engineering in 2001, and the M.S. and Ph.D. degrees in Electrical Engineering and Computer Science in 2003 and 2007, respectively, all from Seoul National University, Seoul, Korea. From 2008 to 2009, he was with Telecommunication Business of Samsung Electronics Co., Ltd., Korea. From 2009 to 2010, he was a postdoctoral researcher in the Department of Electrical and Computer Engineering, University of Manitoba, Winnipeg, MB, Canada. From 2010 to 2016, he was an assistant professor in the Department of Computer Science and Engineering, Seoul National University of Science and Technology, Korea. In 2016, he joined the faculty at Sungkyunkwan University, Korea, where he is currently an associate professor in the College of Information and Communication Engineering. His research interests include RF energy transfer, metasurface communication, visible light communication, cellular communication, cognitive radio, and radio resource management. He has served as an editor of IEEE Communications Surveys and Tutorials from 2014, an editor of IEEE Wireless Communications Letters from 2015, an editor of IEEE Transactions on Wireless Communications from 2017, and an editor of IEEE Transactions on Cognitive Communications and Networking from 2019.
\end{IEEEbiography}

\EOD

\end{document}